# DISPERSION CHAIN OF QUANTUM MECHANICS EQUATIONS

**E.E. Perepelkin[a,b,d], B.I. Sadovnikov[a], N.G. Inozemtseva[b,c], A.A. Korepanova[a]**

[a] *Faculty of Physics, Lomonosov Moscow State University, Moscow, 119991 Russia*
[b] *Moscow Technical University of Communications and Informatics, Moscow, 123423 Russia*
[c] *Dubna State University, Moscow region, Dubna,141980 Russia*
[d] *Joint Institute for Nuclear Research, Moscow region, Dubna,141980 Russia*

## Abstract

Based on the dispersion chain of the Vlasov equations, the paper considers the construction of a new chain of equations of quantum mechanics of high kinematical values. The proposed approach can be applied to consideration of classical and quantum systems with radiation. A number of theorems are proved on the form of extensions of the Hamilton operators, Lagrange functions, Hamilton-Jacobi equations, and Maxwell equations to the case of a generalized phase space.

In some special cases of lower dimensions, the dispersion chain of quantum mechanics is reduced to quantum mechanics in phase space (the Wigner function) and the de Broglie-Bohm «pilot wave» theory.

An example of solving the Schrödinger equation of the second rank (for the phase space) is analyzed, which, in contrast to the Wigner function, gives a positive distribution density function.

**Key words:** quantum mechanics of high kinematical values, dispersion chain of Vlasov equations, Schrödinger equation, generalized phase space, PSI-algebra, rigorous result.

## Introduction

In [1], a dispersion chain of the Vlasov equations was obtained for distribution functions $f^{n\backslash k}$ of mixed type and kinematic values $\left\langle \vec{\xi}^{\ell} \right\rangle_{n\backslash k}$ of order $\ell$, which are represented in the form of extensives/tensors of rank $R-K$:

$$f^{n\backslash k}\left(\vec{\xi}^{n\backslash k},t\right) \overset{\det}{=} \int\limits_{\Omega^{k_1}} \dots \int\limits_{\Omega^{k_K}} f^{n}\left(\vec{\xi}^{n},t\right)\prod_{s=1}^{K} d^3\xi^{k_s}, \tag{i.1}$$

$$\left\langle \vec{\xi}^{\ell} \right\rangle_{n\backslash k} = \frac{1}{f^{n\backslash k}} \int\limits_{\Omega^{k_1}} \dots \int\limits_{\Omega^{k_K}} \int\limits_{\Omega^{\ell}} f^{n\cup\{\ell\}}\left(\vec{\xi}^{n\cup\{\ell\}},t\right)\vec{\xi}^{\ell} d^3\xi^{\ell}\prod_{s=1}^{K} d^3\xi^{k_s}, \tag{i.2}$$

where $k=\{k_1,\dots,k_K\}$, $n=\{n_1,\dots,n_R\}$, $k\subset n$, $k_i\neq k_j$, $i\neq j$, $n_s\neq n_l$, $l\neq s$ and $i,j=1,\dots,K$, $s,l=1,\dots,R$; $\vec{\xi}^{\ell}$ is a kinematical value of order $\ell$; $\vec{\xi}^{\ell}\in\Omega$, where $\Omega$ is a generalized phase space [2].

The distribution functions satisfy the dispersion chain of the Vlasov equations [1]:

**1st rank**

$$\hat{\pi}_n S^n = -Q_n^n. \tag{i.3}$$

**2nd rank**

$$\hat{\pi}_{n,n+1} S^{n,n+1} = -Q^{n+1}_{n,n+1},$$
$$\hat{\pi}_{n,n+k} S^{n,n+k} = -\left(Q^{n}_{n,n+k} + Q^{n+k}_{n,n+k}\right). \tag{i.4}$$

**3rd rank**

$$\hat{\pi}_{n,n+1,n+2} S^{n,n+1,n+2} = -Q^{n+2}_{n,n+1,n+2},$$
$$\hat{\pi}_{n,n+1,n+1+k} S^{n,n+1,n+1+k} = -\left(Q^{n+1}_{n,n+1,n+1+k} + Q^{n+1+k}_{n,n+1,n+1+k}\right),$$
$$\hat{\pi}_{n,n+s,n+s+1} S^{n,n+s,n+s+1} = -\left(Q^{n}_{n,n+s,n+s+1} + Q^{n+s+1}_{n,n+s,n+s+1}\right), \tag{i.5}$$
$$\hat{\pi}_{n,n+s,n+s+k} S^{n,n+s,n+s+k} = -\left(Q^{n}_{n,n+s,n+s+k} + Q^{n+s}_{n,n+s,n+s+k} + Q^{n+s+k}_{n,n+s,n+s+k}\right),$$
$$\ldots$$

where $S^{n_1,\ldots,n_R} \overset{\det}{=} \mathrm{Ln}\, f^{n_1,\ldots,n_R}$, $Q^{p}_{n_1,\ldots,n_R} \overset{\det}{=} \mathrm{div}_{\xi^p} \left\langle \vec{\xi}^{\,p+1} \right\rangle_{n_1,\ldots,n_R}$ and differential operators $\hat{\pi}$ are defined by extensive/tensor $\hat{\Pi}^R = \left\{\hat{\pi}_{n_1,\ldots,n_R}\right\}$ of $R$ rank at $R=0$: $\quad \hat{\pi}_0 = \dfrac{d}{dt}$, at $R=1$: $\quad \hat{\pi}_n = \dfrac{\partial}{\partial t} + \left\langle \vec{\xi}^{\,n+1} \right\rangle_n \nabla_{\xi^n}$, at $R=2$ $\quad \hat{\pi}_{n,n+1} = \dfrac{\partial}{\partial t} + \vec{\xi}^{\,n+1} \nabla_{\xi^n} + \left\langle \vec{\xi}^{\,n+2} \right\rangle_{n,n+1} \nabla_{\xi^{n+1}}$,

$$\hat{\pi}_{n,n+k} = \frac{\partial}{\partial t} + \left\langle \vec{\xi}^{\,n+1} \right\rangle_{n,n+k} \nabla_{\xi^n} + \left\langle \vec{\xi}^{\,n+1+k} \right\rangle_{n,n+k} \nabla_{\xi^{n+k}}, \tag{i.6}$$

at $R=3$

$$\hat{\pi}_{n,n+1,n+2} = \frac{\partial}{\partial t} + \vec{\xi}^{\,n+1} \nabla_{\xi^n} + \vec{\xi}^{\,n+2} \nabla_{\xi^{n+1}} + \left\langle \vec{\xi}^{\,n+3} \right\rangle_{n,n+1,n+2} \nabla_{\xi^{n+2}},$$
$$\hat{\pi}_{n,n+1,n+1+k} = \frac{\partial}{\partial t} + \vec{\xi}^{\,n+1} \nabla_{\xi^n} + \left\langle \vec{\xi}^{\,n+2} \right\rangle_{n,n+1,n+1+k} \nabla_{\xi^{n+1}} + \left\langle \vec{\xi}^{\,n+2+k} \right\rangle_{n,n+1,n+1+k} \nabla_{\xi^{n+1+k}},$$
$$\hat{\pi}_{n,n+s,n+s+1} = \frac{\partial}{\partial t} + \left\langle \vec{\xi}^{\,n+1} \right\rangle_{n,n+s,n+s+1} \nabla_{\xi^n} + \vec{\xi}^{\,n+s+1} \nabla_{\xi^{n+s}} + \left\langle \vec{\xi}^{\,n+s+2} \right\rangle_{n,n+s,n+s+1} \nabla_{\xi^{n+s+1}},$$
$$\hat{\pi}_{n,n+s,n+s+k} = \frac{\partial}{\partial t} + \left\langle \vec{\xi}^{\,n+1} \right\rangle_{n,n+s,n+s+k} \nabla_{\xi^n} + \left\langle \vec{\xi}^{\,n+1+s} \right\rangle_{n,n+s,n+s+k} \nabla_{\xi^{n+s}} + \left\langle \vec{\xi}^{\,n+1+s+k} \right\rangle_{n,n+s,n+s+k} \nabla_{\xi^{n+s+k}},$$
$$\ldots.$$

where $k,s \in \{2,3,\ldots\}$. Dispersion chain (i.3)-(i.5) describes the evolution of distribution functions in the generalized phase space along the phase trajectories. On the left-hand side of the dispersion chain, in fact, there are «time derivatives» (operators $\hat{\pi}$), and on the right-hand side there are the sources of dissipations $Q$ of various kinematic values. If there are no sources of dissipation ($Q=0$), then the corresponding probability density function will be constant along the phase trajectory in the generalized phase space. It was shown in [1] that dispersion chain (i.3)-(i.5) contains three types of conservation laws, which can be conditionally called the laws of «mass», «kinetic momentum» and «energy» conservation. The dispersion equations (i.3)-(i.5) themselves act as the laws of conservation of «mass». The laws of conservation of «kinetic momentum» and «energy», for example, for equations of the first rank (i.3) are as follows:

The first group of equations

$$\hat{\pi}_n \left\langle \xi_\alpha^{n+1} \right\rangle_n = \left[ \frac{\partial}{\partial t} + \left\langle \xi_\beta^{n+1} \right\rangle_n \frac{\partial}{\partial \xi_\beta^n} \right] \left\langle \xi_\alpha^{n+1} \right\rangle_n = -\frac{1}{f^n} \frac{\partial P_{\alpha\beta}^{n+1}}{\partial \xi_\beta^n} + \left\langle \xi_\alpha^{n+2} \right\rangle_n , \tag{i.7}$$

$$\frac{\partial}{\partial t}\left[ \frac{f^n}{2} \left\langle \xi^{n+1} \right\rangle_n^2 + \frac{1}{2} \mathrm{Tr}\, P_{\alpha\alpha}^{n+1} \right] + \frac{\partial}{\partial \xi_\beta^n}\left[ \frac{f^n}{2} \left\langle \xi^{n+1} \right\rangle_n^2 \left\langle \xi_\beta^{n+1} \right\rangle_n + \frac{1}{2} \left\langle \xi_\beta^{n+1} \right\rangle_n \mathrm{Tr}\, P_{\alpha\alpha}^{n+1} + \left\langle \xi_\alpha^{n+1} \right\rangle_n P_{\alpha\beta}^{n+1} + \frac{1}{2} \mathrm{Tr}\, P_{\alpha\alpha\beta}^{n+1} \right] =$$

$$= \int\limits_{\Omega_{n+1}} f^{n,n+1} \left\langle \xi_\alpha^{n+2} \right\rangle_{n,n+1} \xi_\alpha^{n+1}\, d^3 \xi^{n+1}, \tag{i.8}$$

where $P_{\alpha\beta}^{n+1}$ and $P_{\alpha\beta\mu}^{n+1}$ are momenta of the second and third orders respectively for kinematic value $\vec{\xi}^{n+1}$:

$$P_{\alpha\beta}^{n+1} \overset{\det}{=} \int\limits_{\Omega_{n+1}} \left( \xi_\alpha^{n+1} - \left\langle \xi_\alpha^{n+1} \right\rangle_n \right)\left( \xi_\beta^{n+1} - \left\langle \xi_\beta^{n+1} \right\rangle_n \right) f^{n,n+1} d^3 \xi^{n+1}, \tag{i.9}$$

$$P_{\alpha\beta\mu}^{n+1} \overset{\det}{=} \int\limits_{\Omega_{n+1}} \left( \xi_\alpha^{n+1} - \left\langle \xi_\alpha^{n+1} \right\rangle_n \right)\left( \xi_\beta^{n+1} - \left\langle \xi_\beta^{n+1} \right\rangle_n \right)\left( \xi_\mu^{n+1} - \left\langle \xi_\mu^{n+1} \right\rangle_n \right) f^{n,n+1} d^3 \xi^{n+1}. \tag{i.10}$$

The second group of equations

$$\hat{\pi}_n \left\langle \xi_\alpha^{n+k} \right\rangle_n = \left[ \frac{\partial}{\partial t} + \left\langle \xi_\beta^{n+1} \right\rangle_n \frac{\partial}{\partial \xi_\beta^n} \right] \left\langle \xi_\alpha^{n+k} \right\rangle_n = -\frac{1}{f^n} \frac{\partial P_{\beta\alpha}^{n+1,n+k}}{\partial \xi_\beta^n} + \left\langle \xi_\alpha^{n+k+1} \right\rangle_n , \tag{i.11}$$

$$\frac{\partial}{\partial t}\left[ \frac{f^n}{2} \left\langle \xi^{n+k} \right\rangle_n^2 + \frac{1}{2} \mathrm{Tr}\, P_{\alpha\alpha}^{n+k} \right] +$$

$$+ \frac{\partial}{\partial \xi_\beta^n}\left[ \frac{f^n}{2} \left\langle \xi^{n+k} \right\rangle_n^2 \left\langle \xi_\beta^{n+1} \right\rangle_n + \frac{1}{2} \left\langle \xi_\beta^{n+1} \right\rangle_n \mathrm{Tr}\, P_{\alpha\alpha}^{n+k} + P_{\beta\alpha}^{n+1,n+k} \left\langle \xi_\alpha^{n+k} \right\rangle_n + \frac{1}{2} \mathrm{Tr}\, P_{\beta\alpha\alpha}^{n+1,n+k,n+k} \right] = \tag{i.12}$$

$$= \int\limits_{\Omega_{n+k}} f^{n,n+k} \left\langle \xi_\alpha^{n+k+1} \right\rangle_{n,n+k} \xi_\alpha^{n+k}\, d^3 \xi^{n+k},$$

where $P_{\alpha\beta}^{n+1,n+k}$ and $P_{\beta\alpha\alpha}^{n+1,n+k,n+k}$ are momenta of the second and third orders respectively of kinematic values $\left( \vec{\xi}^{n+1}, \vec{\xi}^{n+k} \right)$:

$$P_{\alpha\beta}^{n+1,n+k} \overset{\det}{=} \int\limits_{\Omega_{n+1}} \int\limits_{\Omega_{n+k}} \left( \xi_\alpha^{n+1} - \left\langle \xi_\alpha^{n+1} \right\rangle_n \right)\left( \xi_\beta^{n+k} - \left\langle \xi_\beta^{n+k} \right\rangle_n \right) f^{n,n+1,n+k} d^3 \xi^{n+1} d^3 \xi^{n+k}, \tag{i.13}$$

$$P_{\beta\alpha\alpha}^{n+1,n+k,n+k} \overset{\det}{=} \int\limits_{\Omega_{n+1}} \int\limits_{\Omega_{n+k}} \left( \xi_\beta^{n+1} - \left\langle \xi_\beta^{n+1} \right\rangle_n \right)\left( \xi_\alpha^{n+k} - \left\langle \xi_\alpha^{n+k} \right\rangle_n \right)^2 f^{n,n+1,n+k} d^3 \xi^{n+1} d^3 \xi^{n+k}. \tag{i.14}$$

The kinetic momentum conservation laws for equations of the second rank (i.4) have the form:

$$\hat{\pi}_{n+1,n+2} \left\langle \xi_\alpha^n \right\rangle_{n+1,n+2} = -\frac{1}{f^{n+1,n+2}} \frac{\partial P_{\alpha\beta}^{n,n+3}}{\partial \xi_\beta^{n+2}} + \xi_\alpha^{n+1}, \tag{i.15}$$

$$\hat{\pi}_{n,n+2}\left\langle \xi_{\alpha}^{n+1}\right\rangle_{n,n+2} = -\frac{1}{f^{n,n+2}}\left(\frac{\partial P_{\alpha\beta}^{n+1}}{\partial \xi_{\beta}^{n}} + \frac{\partial P_{\alpha\beta}^{n+1,n+3}}{\partial \xi_{\beta}^{n+2}}\right) + \xi_{\alpha}^{n+2}, \quad \text{(i.16)}$$

$$\hat{\pi}_{n,n+1}\left\langle \xi_{\alpha}^{n+2}\right\rangle_{n,n+1} = -\frac{1}{f^{n,n+1}}\frac{\partial P_{\alpha\beta}^{n+2}}{\partial \xi_{\beta}^{n+1}} + \left\langle \xi_{\alpha}^{n+3}\right\rangle_{n,n+1}, \quad \text{(i.17)}$$

where $P_{\alpha\beta}^{n+2}$, $P_{\alpha\beta}^{n,n+3}$, $P_{\alpha\beta}^{n+1}$, $P_{\alpha\beta}^{n+1,n+3}$ are momenta of the second order defined as $P_{\alpha\beta}^{n+2} = P_{\alpha\beta}^{n+2}\left(\vec{\xi}^{\,n}, \vec{\xi}^{\,n+1}\right) \overset{\text{det}}{=} P_{\alpha\beta}^{n+2}\left(n, n+1\right)$. Similarly, $P_{\alpha\beta}^{n,n+3}\left(n+1, n+2\right)$, $P_{\alpha\beta}^{n+1}\left(n, n+2\right)$ and $P_{\alpha\beta}^{n+1,n+3}\left(n, n+2\right)$.

Dispersion chain (i.3)-(i.5) contains the usual chain of the Vlasov equations [3], for which there is a chain of higher-order quantum mechanics equations [2]. The apparatus of the dispersion chain of the Vlasov equations allows one to consider physical systems described by higher kinematic values. For instance, based on the work of J. Larmor of 1897 on the radiating an electromagnetic wave by a charged particle moving with acceleration, H. Lorentz proposed the equation

$$\ddot{\vec{v}} = \frac{6\pi\varepsilon_0 c^3}{e^2}\left(m\dot{\vec{v}} - \vec{F}_{ext}\right), \quad \text{(i.18)}$$

where $\vec{F}_{ext}$ is an external force. Derivative $\ddot{\vec{v}}$ contains the information about the force of radiative friction and determines the third order of a differential motion equation (i.18), which is beyond the scope of classical mechanics. The power of electromagnetic radiation is proportional to $\dot{v}^2$, therefore, when describing physical systems with radiation, for example, the H. Lorentz's equation (i.18), it is necessary to use the third-rank Vlasov dispersion equations (i.5) for the distribution function $f^{1,2,3}\left(\vec{r}, \vec{v}, \dot{\vec{v}}, t\right)$. Chain of equations (i.5) can be broken at the third rank using the second Vlasov approximation of average flow $\left\langle \ddot{\vec{v}}\right\rangle$ [4]

$$\left\langle \ddot{v}_{\mu}\right\rangle_{1,2} = -\frac{1}{m}\frac{\partial N_{1,2}}{\partial x_{\mu}}, \quad N_{1,2} \overset{\text{det}}{=} \frac{\partial U}{\partial t} + v_{\lambda}\frac{\partial U}{\partial x_{\lambda}}, \quad \text{(i.19)}$$

where $N_{1,2}$ corresponds to the radiation power; $U$ is an electric potential. Approximation (i.19) leads to the Vlasov $\Psi$-equation [4]:

$$\frac{\partial f^{1,2,3}}{\partial t} + v_{\lambda}\frac{\partial f^{1,2,3}}{\partial x_{\lambda}} + \dot{v}_{\lambda}\frac{\partial f^{1,2,3}}{\partial v_{\lambda}} - \frac{1}{m}\frac{\partial N_{1,2}}{\partial x_{\lambda}}\frac{\partial f^{1,2,3}}{\partial \dot{v}_{\lambda}} = 0. \quad \text{(i.20)}$$

Note that when constructing the chain of the Vlasov equations [3], the condition for distribution functions to be positive only is not imposed. When considering the first Vlasov equation, the assumption of positivity of $f^1 = \left|\Psi\right|^2 \geq 0$ led to the introduction of wave function $\Psi$ and the Schrödinger equation corresponding to it [5]. The absence of the positivity condition for function $f^{1,2}$ leads to its connection with the Wigner function $W$ [6, 7]. The paper [2] considers the case when all distribution functions in the Vlasov chain are positive. The condition $f^{n,n+1\ldots} = \left|\Psi^{n,n+1\ldots}\right|^2 \geq 0$ leads to a chain of equations of quantum mechanics of high kinematical values [2].

The purpose of this paper is to build a dispersive chain of equations of quantum mechanics based on the chain (i.3)-(i.5), obtain the generalized Lagrange functions, extended analogues of the Hamilton-Jacobi equations, Legendre transformations, generalized Maxwell equations and equations of motion in generalized electromagnetic fields.

The paper has the following structure. In §1, based on chain of equations (i.3)-(i.5) and the assumption that the distribution function $f^{n_1\ldots n_R} = \left|\Psi^{n_1\ldots n_R}\right|^2 \geq 0$ is positive, we construct a dispersion chain of the Schrödinger equations for wave functions $\Psi^{n_1\ldots n_R}$. In §2, the Lagrangian and Hamiltonian formalism in the generalized phase space is described. The generalized Hamilton-Jacobi equations are obtained, the concept of a generalized complex action is introduced. In §3, the generalized Maxwell equations are constructed and equations of motion in generalized electromagnetic fields are obtained. Section 4 describes an example of the exact solution of the time-dependent Schrödinger equation of the second rank using the apparatus of quantum mechanics in the generalized phase space constructed in this paper. The Appendix contains the proofs of the theorems.

## §1 Dispersions chain of quantum mechanics equations

We will construct a dispersion chain of quantum mechanics equations by analogy with [2].

**Definition 1** *Let us define the extensive of operators* $\partial^R = \left\{\partial_{n_1\ldots n_R}\right\}$ *of rank* $R \in \mathbb{N}_0 = \mathbb{N} \cup \{0\}$ *as:*

$$
\begin{aligned}
&at\ R=0 \quad \partial_0 = \frac{\partial}{\partial t},\\
&at\ R=1 \quad \partial_n = \frac{\partial}{\partial t} + \vec{\xi}^{\,n+1}\nabla_{\xi^n},\\
&at\ R=2 \quad \partial_{n,n+1} = \frac{\partial}{\partial t} + \vec{\xi}^{\,n+1}\nabla_{\xi^n} + \vec{\xi}^{\,n+2}\nabla_{\xi^{n+1}},\\
&\ldots
\end{aligned}
\tag{1.1}
$$

*where operator* $\partial_{n_1\ldots n_R}$ *acts upon some sufficiently smooth function.*

**Definition 2** *Let us define the extensive of the Hamilton operators* $\hat{\mathfrak{H}}^R = \left\{\hat{\mathrm{H}}_{n_1\ldots n_R}\right\}$ *of rank* $R \in \mathbb{N}$ *as:*

$$\hat{\mathrm{H}}_{n_1\ldots n_R} = \hat{\mathrm{T}}_{n_1\ldots n_R} + U^{\hat{n}_1\ldots\hat{n}_R}, \tag{1.2}$$

$$at\ R=1 \quad \hat{\mathrm{T}}_n = \hat{\tau}_n^n, \tag{1.3}$$

$$at\ R=2 \quad \hat{\mathrm{T}}_{n,n+1} = \hat{\tau}_{n,n+1}^{n+1}, \tag{1.4}$$

$$\hat{\mathrm{T}}_{n,n+k} = \hat{\tau}_{n,n+k}^{n+k} + \frac{\beta_n}{\beta_{n+k}}\hat{\tau}_{n,n+k}^{n}, \tag{1.5}$$

$$at\ R=3 \quad \hat{\mathrm{T}}_{n,n+1,n+2} = \hat{\tau}_{n,n+1,n+2}^{n+2}, \tag{1.6}$$

$$\hat{\mathrm{T}}_{n,n+1,n+1+k} = \hat{\tau}^{n+1+k}_{n,n+1,n+1+k} + \frac{\beta_{n+1}}{\beta_{n+1+k}}\hat{\tau}^{n+k}_{n,n+k}, \tag{1.7}$$

$$\hat{\mathrm{T}}_{n,n+s,n+s+1} = \hat{\tau}^{n+s+1}_{n,n+s,n+s+1} + \frac{\beta_n}{\beta_{n+s+1}}\hat{\tau}^{n}_{n,n+s,n+s+1}, \tag{1.8}$$

$$\hat{\mathrm{T}}_{n,n+s,n+s+k} = \hat{\tau}^{n+k}_{n,n+s,n+s+k} + \frac{\beta_{n+s}}{\beta_{n+s+k}}\hat{\tau}^{n+s}_{n,n+s,n+s+k} + \frac{\beta_n}{\beta_{n+s+k}}\hat{\tau}^{n}_{n,n+s,n+s+k}, \tag{1.9}$$

…

*where* $\hat{\tau}^{\ell}_{n_1\ldots n_R} \overset{\det}{=} -\alpha_\ell\beta_\ell\left(\hat{\mathrm{p}}_\ell - \frac{\gamma_\ell}{2\alpha_\ell\beta_\ell}\vec{A}^{\ell}_{n_1\ldots n_R}\right)^2$ *is a kinematic energy operator of order* $\ell$*;* $U^{\hat{n}_1\ldots\hat{n}_R}$, $\vec{A}^{\ell}_{n_1\ldots n_R} \in \mathbb{R}$, $\ell \in \{n_1,\ldots,n_R\}$, $\hat{\mathrm{p}}_n \overset{\det}{=} -\frac{i}{\beta_n}\nabla_{\xi^n}$*;* $\alpha_n$, $\beta_n$ *and* $\gamma_n$ *are constant;* $n,k,s \in \mathbb{N}$, $k,s > 1$.

**Remark**

Let us point out the features of introduced Definitions 1 and 2. Operator $\partial_{n_1\ldots n_R}$ (1.1) is a «fragment» of operator $\hat{\pi}_{n_1\ldots n_R}$ (i.6) and determines an analogue of the partial derivative with respect to time along the phase trajectory defined by kinematical values $\vec{\xi}^{n_1},\ldots,\vec{\xi}^{n_R}$, for example,

$$\hat{\pi}_n = \partial_0 + \left\langle \vec{\xi}^{n+1} \right\rangle_n \nabla_{\xi^n},$$
$$\hat{\pi}_{n,n+1} = \partial_n + \left\langle \vec{\xi}^{n+2} \right\rangle_{n,n+1} \nabla_{\xi^{n+1}},$$
$$\hat{\pi}_{n,n+1,n+2} = \partial_{n,n+1} + \left\langle \vec{\xi}^{n+3} \right\rangle_{n,n+1,n+2} \nabla_{\xi^{n+2}}.$$

Operators $\hat{\mathrm{H}}^{n_1\ldots n_R}$ (1.2)-(1.9) are a generalization of the Hamilton operator ($\hat{\mathrm{H}}^1$) of a quantum system to a generalized phase space. As can be seen from Definition (1.2)-(1.9) operator $\hat{\mathrm{H}}^{n_1\ldots n_R}$ is represented as the sum of kinematic energy operator $\hat{\mathrm{T}}_{n_1\ldots n_R}$ and the «potential» energy $U$. Operator $\hat{\mathrm{T}}_{n_1\ldots n_R}$ is represented as a linear combination of operators $\hat{\tau}^{\ell}_{n_1\ldots n_R}$, that is, depending on rank $R$, it contains the contributions of various kinematical values (1.2)-(1.9).

<u>**Theorem 1**</u> *Let* $f^{n_1\ldots n_R}$ *be a positive distribution function of rank* $R$ *that satisfies the dispersion chain of Vlasov equations (i.3)-(i.5) and there are such functions* $\Phi^{n_1\ldots n_R}$, $\vec{A}^{\ell}_{n_1\ldots n_R}$, *that the Helmholtz theorem on the representation of a vector field* $\left\langle \vec{\xi}^{\ell+1} \right\rangle_{n_1\ldots n_R}$ *is fulfilled in the form:*

$$\left\langle \vec{\xi}^{\ell+1} \right\rangle_{n_1\ldots n_R} = -\alpha_\ell \nabla_{\xi^\ell}\Phi^{n_1\ldots n_R} + \gamma_\ell \vec{A}^{\ell}_{n_1\ldots n_R}, \tag{1.10}$$

*where* $\alpha_\ell$ *and* $\gamma_\ell$ *are constant vlues;* $\mathrm{div}_{\xi^\ell}\vec{A}^{\ell}_{n_1\ldots n_R} = 0$, $\mathrm{curl}_{\xi^\ell}\vec{A}^{\ell}_{n_1\ldots n_R} \overset{\det}{=} \vec{B}^{\ell}_{n_1\ldots n_R}$*;* $\ell \in \mathrm{n} = \{n_1,\ldots,n_R\}$, $\ell+1 \notin \mathrm{n}$.

*Then the dispersion chain of equations of quantum mechanics is true for wave functions* $\Psi^{n_1\ldots n_R}$*:*

**1st rank**

$$\frac{i}{\beta_n}\partial_0\Psi^n = \hat{\mathrm{H}}_n \Psi^n. \tag{1.11}$$

**2nd rank**

$$\frac{i}{\beta_{n+1}}\partial_n\Psi^{n,n+1} = \hat{\mathrm{H}}_{n,n+1} \Psi^{n,n+1}, \tag{1.12}$$

$$\frac{i}{\beta_{n+k}}\partial_0\Psi^{n,n+k} = \hat{\mathrm{H}}_{n,n+k} \Psi^{n,n+k}. \tag{1.13}$$

**3rd rank**

$$\frac{i}{\beta_{n+2}}\partial_{n,n+1}\Psi^{n,n+1,n+2} = \hat{\mathrm{H}}_{n,n+1,n+2} \Psi^{n,n+1,n+2}, \tag{1.14}$$

$$\frac{i}{\beta_{n+1+k}}\partial_n\Psi^{n,n+1,n+1+k} = \hat{\mathrm{H}}_{n,n+1,n+1+k} \Psi^{n,n+1,n+1+k}, \qquad \frac{i}{\beta_{n+s+1}}\partial_{n+s}\Psi^{n,n+s,n+s+1} = \hat{\mathrm{H}}_{n,n+s,n+s+1} \Psi^{n,n+s,n+s+1}, \tag{1.15}$$

$$\frac{i}{\beta_{n+s+k}}\partial_0\Psi^{n,n+s,n+s+k} = \hat{\mathrm{H}}_{n,n+s,n+s+k} \Psi^{n,n+s,n+s+k}, \tag{1.16}$$

$$\ldots$$

*where* $\Psi^{n_1\ldots n_R} = \sqrt{f^{n_1\ldots n_R}}\exp\left(i\varphi^{n_1\ldots n_R}\right)$*;* $\Phi^{n_1\ldots n_R} = 2\varphi^{n_1\ldots n_R} + 2\pi j,\ j\in\mathbb{Z}$*;* $\beta_n$ *is constant;* $n,k\in\mathbb{N}$, $k>1$.

The proof of Theorem 1 is given in Appendix A.

In [2], the following values were considered as constants $\alpha_n$, $\beta_n$ and $\gamma_n$:

$$\alpha_n = -\frac{\hbar_n}{2m}, \qquad \beta_n = \frac{1}{\hbar_n}, \qquad \gamma_n = -\frac{q}{m}, \tag{1.17}$$

where $\hbar_1 = \hbar$ is a Planck constant; $q,m$ are charge and mass of the particle respectively. In the case of a quantum harmonic oscillator in the generalized phase space, the conditions of the generalized uncertainty principle [8, 2] are satisfied:

$$\sigma_n\sigma_{n+1} = \left|\alpha_n\right| = \frac{\hbar_n}{2m}, \qquad \frac{\sigma_{n+1}}{\sigma_n} = \omega, \tag{1.18}$$

where $\omega$ is oscillator frequency and $\sigma_n^2$ is root-mean-square deviation of kinematical value of order $n$, for example:

$$\sigma_4^2 = \int_{-\infty}^{+\infty}\int_{-\infty}^{+\infty}\int_{-\infty}^{+\infty}\int_{-\infty}^{+\infty} f_s^{1,2,3,4}\left(\ddot{v} - \left\langle\ddot{v}\right\rangle_0\right)^2 dx dv d\dot{v} d\ddot{v}, \tag{1.19}$$

$$f_s^{1,2,3,4}(x,v,\dot v,\ddot v)=\frac{(-1)^s}{2\pi\sigma_1\sigma_2}e^{-\frac{\ddot v^2}{2\sigma_4^2}-\frac{\dot v^2}{2\sigma_3^2}}L_s\left(\frac{\ddot v^2}{\sigma_4^2}+\frac{\dot v^2}{\sigma_3^2}\right)\delta\left(\ddot v+\omega^2 v\right)\times$$
$$\times\delta\left(\dot v+\omega^2 x+\frac{\ddot v+\omega^2 v}{\omega}\arcsin\frac{\ddot v}{\sqrt{\ddot v^2+\omega^2\dot v^2}}\right), \tag{1.20}$$

where $s$ is the number of quantum state of the harmonic oscillator; $L_s$ are Laguerre polynomials and $\delta$ is a delta function of Dirac. Probability density function (1.20) satisfies the Vlasov dispersion equation of the fourth rank and the following relations are true for it:

$$\begin{aligned}
&\int_{-\infty}^{+\infty} f_s^{1,2,3,4}(x,v,\dot v,\ddot v)\,d\ddot v = f_s^{1,2,3}(x,v,\dot v),\\
&\int_{-\infty}^{+\infty} f_s^{1,2,3}(x,v,\dot v)\,d\dot v = f_s^{1,2}(x,v)=\frac{(-1)^s}{2\pi\sigma_1\sigma_2}e^{-\frac{v^2+\omega^2x^2}{2\sigma_2^2}}L_s\left(\frac{v^2+\omega^2x^2}{\sigma_2^2}\right),\\
&\int_{-\infty}^{+\infty} f_s^{1,2}(x,v)\,dv = f_s^{1}(x)=\frac{1}{2^s s!\sqrt{2\pi}\sigma_1}e^{-\frac{x^2}{2\sigma_1^2}}H_s^2\left(\frac{x}{\sigma_1\sqrt{2}}\right),\\
&\int_{-\infty}^{+\infty} f_s^{1}(x)\,dx=1,
\end{aligned} \tag{1.21}$$

where $H_s$ are Hermitian polynomials and function $f_s^{1,2}$ corresponds to the Wigner function [6] $f_s^{1,2}(x,v)=mW_s(x,mv)$ of the quantum harmonic oscillator. Note that Wigner function $W$ satisfies the Moyal equation [9], which coincides with the second Vlasov equation (i.4) with the Vlasov-Moyal approximation [7].

**Remark**

The choice of constant values (1.17) is due to the fact that for the equation of the first rank (1.11) at $n=1$, there is obtained a Schrödinger equation corresponding to a scalar particle in an electromagnetic field:

$$i\hbar\frac{\partial\Psi^1}{\partial t}=\frac{1}{2m}\left(\hat{\mathrm{p}}_1-q\vec{A}_1^1\right)^2\Psi^1+U^1\Psi^1, \tag{1.22}$$

In which $U^1$ corresponds to the potential energy of the electric field and vortex component $\vec{A}_1^1$ of the vector field of probability flow velocity $\langle\vec v\rangle_1$ is the vector potential of the magnetic induction, that is $\mathrm{curl}_r\,\vec{A}_1^1=\vec{B}_1^1$.

In the general case, the constants (1.17) can be arbitrary. For $n>1$, the values of $\hbar_n$ can be chosen by expressions (1.18) as $\hbar_n=\hbar_{n-1}\omega^2=\hbar\omega^{2(n-1)}$. Frequency $\omega$ in the framework of de Broglie theory can be related to Compton wavelength $\lambda_C$, for example, by setting the energy of the ground state of a harmonic oscillator $\hbar\omega/2$ equal to the rest energy $mc^2$, then

$$\omega=\frac{2mc^2}{\hbar}=\frac{4\pi c}{\lambda_C},\qquad \hbar^{2n-3}\hbar_n=\left(2mc^2\right)^{2(n-1)}, \tag{1.23}$$

from here

$$\left(\frac{\hbar}{2m}\right)^{2n-3}\frac{\hbar_n}{2m}=c^{4(n-1)}=\alpha_1^{2n-3}\alpha_n\,.$$

## §2 Lagrangian and Hamiltonian representation

Dispersion chain of equations (1.11)-(1.16) allows one to construct analogues of the Hamilton and Lagrange functions in the generalized phase space.

**Definition 3** *Let us define the extensive of Hamilton functions* $\mathfrak{H}^R=\left\{\mathrm{H}^{n_1\dots n_R}\right\}$ *of rank* $R\in\mathbb{N}$ *as the sum of the kinetic energy* $\mathfrak{T}^R=\left\{\mathrm{T}^{n_1\dots n_R}\right\}$ *and potential energy* $\mathfrak{V}^R=\left\{\mathrm{V}^{n_1\dots n_R}\right\}$ *extensives as*

$$\mathrm{H}^{n_1\dots n_R}=\mathrm{T}^{n_1\dots n_R}+\mathrm{V}^{n_1\dots n_R},\tag{2.1}$$

*where*

$$\mathrm{V}^{n_1\dots n_R}=U^{n_1\dots n_R}+\mathrm{Q}^{n_1\dots n_R},$$

*at* $R=1$ $\quad \mathrm{T}^n=\tau_n^n,\qquad \mathrm{Q}^n\overset{\det}{=}\mathrm{q}_n^n,$ (2.2)

*at* $R=2$ $\quad \mathrm{T}^{n,n+1}=\tau_{n+1}^{n,n+1},\ \mathrm{Q}^{n,n+1}=\mathrm{q}_{n+1}^{n,n+1},$ (2.3)

$$\mathrm{T}^{n,n+k}=\tau_{n+k}^{n,n+k}+\frac{\beta_n}{\beta_{n+k}}\tau_n^{n,n+k},\tag{2.4}$$

$$\mathrm{Q}^{n,n+k}=\mathrm{q}_{n+k}^{n,n+k}+\frac{\beta_n}{\beta_{n+k}}\mathrm{q}_n^{n,n+k},\tag{2.5}$$

*at* $R=3$ $\quad \mathrm{T}^{n,n+1,n+2}=\tau_{n+2}^{n,n+1,n+2},$ (2.6)

$$\mathrm{Q}^{n,n+1,n+2}=\mathrm{q}_{n+2}^{n,n+1,n+2},\tag{2.7}$$

$$\mathrm{T}^{n,n+1,n+1+k}=\tau_{n+1+k}^{n,n+1,n+1+k}+\frac{\beta_{n+1}}{\beta_{n+1+k}}\tau_{n+1}^{n,n+1,n+1+k},\tag{2.8}$$

$$\mathrm{Q}^{n,n+1,n+1+k}=\mathrm{q}_{n+1+k}^{n,n+1,n+1+k}+\frac{\beta_{n+1}}{\beta_{n+1+k}}\mathrm{q}_{n+1}^{n,n+1,n+1+k},\tag{2.9}$$

$$\mathrm{T}^{n,n+s,n+s+1}=\tau_{n+s+1}^{n,n+s,n+s+1}+\frac{\beta_n}{\beta_{n+s+1}}\tau_n^{n,n+s,n+s+1},\tag{2.10}$$

$$\mathrm{Q}^{n,n+s,n+s+1}=\mathrm{q}_{n+s+1}^{n,n+s,n+s+1}+\frac{\beta_n}{\beta_{n+s+1}}\mathrm{q}_n^{n,n+s,n+s+1},\tag{2.11}$$

$$\mathrm{T}^{n,n+s,n+s+k}=\tau_{n+s+k}^{n,n+s,n+s+k}+\frac{\beta_{n+s}}{\beta_{n+s+k}}\tau_{n+s}^{n,n+s,n+s+k}+\frac{\beta_n}{\beta_{n+s+k}}\tau_n^{n,n+s,n+s+k},\tag{2.12}$$

$$\mathrm{Q}^{n,n+s,n+s+k}=\mathrm{q}_{n+s+k}^{n,n+s,n+s+k}+\frac{\beta_{n+s}}{\beta_{n+s+k}}\mathrm{q}_{n+s}^{n,n+s,n+s+k}+\frac{\beta_n}{\beta_{n+s+k}}\mathrm{q}_n^{n,n+s,n+s+k},\tag{2.13}$$

....

*where* $\tau_\ell^{n_1\dots n_R}\overset{\det}{=}-\dfrac{1}{4\alpha_\ell\beta_\ell}\left|\left\langle\vec{\xi}^{\,\ell+1}\right\rangle_{n_1\dots n_R}\right|^2$ *is kinematical energy of order* $\ell$, $\mathrm{q}_\ell^{n_1\dots n_R}\overset{\det}{=}\dfrac{\alpha_\ell}{\beta_\ell}\dfrac{\Delta_{\xi^\ell}\left|\Psi^{n_1\dots n_R}\right|}{\left|\Psi^{n_1\dots n_R}\right|}$. *In this case, extensive* $\mathfrak{Q}^R=\left\{\mathrm{Q}^{n_1\dots n_R}\right\}$ *is a generalization of quantum potential* ($\mathrm{Q}^1$) *from the de Broglie-Bohm «pilot wave» theory [10-12].*

**Theorem 2** *Let the conditions of Theorem 1 be satisfied, then the extensives of the Hamilton-Jacobi equations correspond to dispersion chain of equations of quantum mechanics (1.11)-(1.16):*

**1st rank**

$$-\frac{1}{\beta_n}\partial_0\varphi^n = \mathrm{H}^n, \qquad (2.14)$$

**2nd rank**

$$-\frac{1}{\beta_{n+1}}\partial_n\varphi^{n,n+1} = \mathrm{H}^{n,n+1}, \qquad (2.15)$$

$$-\frac{1}{\beta_{n+k}}\partial_0\varphi^{n,n+k} = \mathrm{H}^{n,n+k}, \qquad (2.16)$$

**3rd rank**

$$-\frac{1}{\beta_{n+2}}\partial_{n,n+1}\varphi^{n,n+1,n+2} = \mathrm{H}^{n,n+1,n+2}, \qquad (2.17)$$

$$-\frac{1}{\beta_{n+1+k}}\partial_n\varphi^{n,n+1,n+1+k} = \mathrm{H}^{n,n+1,n+1+k}, \quad -\frac{1}{\beta_{n+s+1}}\partial_{n+s}\varphi^{n,n+s,n+s+1} = \mathrm{H}^{n,n+s,n+s+1}, \qquad (2.18)$$

$$-\frac{1}{\beta_{n+s+k}}\partial_0\varphi^{n,n+s,n+s+k} = \mathrm{H}^{n,n+s,n+s+k}, \qquad (2.19)$$

....

*where extensive* $\{\varphi^{n_1 \ldots n_R}\}$ *determines the phases of the extensives of wave functions* $\{\Psi^{n_1 \ldots n_R}\}$.

The proof of Theorem 2 is given in Appendix B.

**Remark**

Hamilton functions (2.1) consist of «kinetic» energy $\mathrm{T}^{n_1 \ldots n_R}$ and generalized «potential» energy $\mathrm{V}^{n_1 \ldots n_R}$. Value $\mathrm{T}^{n_1 \ldots n_R}$ contains not only usual kinetic energy $\mathrm{T}^1$ (2.2) but also energies related to high order kinematical values $(n>1)$ $\mathrm{T}^n$ (2.2), $\mathrm{T}^{n,n+1}$ (2.3), $\mathrm{T}^{n,n+1,n+2}$ (2.6) as well as their various combinations $(k,s>1)$ $\mathrm{T}^{n,n+k}$ (2.4), $\mathrm{T}^{n,n+1,n+1+k}$ (2.8), $\mathrm{T}^{n,n+s,n+s+1}$ (2.10), $\mathrm{T}^{n,n+s,n+s+k}$ (2.12).

Generalized «potential energy» $\mathrm{V}^{n_1 \ldots n_R}$ (2.1) in addition to energy $U^{n_1 \ldots n_R}$ (1.2)-(1.6) contains generalized quantum potential $\mathrm{Q}^{n_1 \ldots n_R}$, which in a particular case at $n=1$ coincides with known quantum potential $\mathrm{Q}^1 = -\frac{\hbar^2}{2m}\frac{\Delta_r|\Psi^2|}{|\Psi^2|}$ from the de Broglie-Bohm «pilot wave» theory [10-12]. In the general case, the generalized quantum potential, as can be seen from expressions $\mathrm{Q}^n$ (2.2), $\mathrm{Q}^{n,n+1}$ (2.3), $\mathrm{Q}^{n,n+1,n+2}$ (2.7), is associated with high order kinematical values $(n>1)$ as well as with their various combinations $(k,s>1)$ $\mathrm{Q}^{n,n+k}$ (2.5), $\mathrm{Q}^{n,n+1,n+1+k}$ (2.9), $\mathrm{Q}^{n,n+s,n+s+1}$ (2.11), $\mathrm{Q}^{n,n+s,n+s+k}$ (2.13).

The values of the form $\frac{1}{\beta_{n_R}}\varphi^{n_1\ldots n_R}=\hbar_{n_R}\varphi^{n_1\ldots n_R}$ present in equations (2.14)-(2.19) correspond to the generalized action [2]. Thus, equations (2.14)-(2.19) are generalizations of the Hamilton-Jacobi equation ($n=1$) to the case of a generalized phase space ($n\geq 1$).

Comparing expressions for operators $\hat{\mathrm{H}}^{n_1\ldots n_R}$ (1.2)-(1.7) with expressions for functions $\mathrm{H}^{n_1\ldots n_R}$ (2.1)-(2.3) one can see a general pattern. First rank equations (1.11) and (2.14) contain one kinematic summand each: operator $-\alpha_n\beta_n\left(\hat{\mathrm{p}}_n-\frac{\gamma_n}{2\alpha_n\beta_n}\vec{A}_n^n\right)^2$ and function $-\frac{1}{4\alpha_n\beta_n}\left|\left\langle\vec{\xi}^{n+1}\right\rangle_n\right|^2$, respectively. In this case, in equation (1.9), quantum potential $\mathrm{Q}^n$ (2.2) contains only one summand, which, due to (1.8), is due to the kinematical value $\left\langle\vec{\xi}^{n+1}\right\rangle_n$. Such situation is determined by the initial equation of the dispersion chain (i.3), which contains only one source $Q_n^n$ of the kinematic dissipation of field $\left\langle\vec{\xi}^{n+1}\right\rangle_n$. Equations of the second rank ($R=2$) (1.12)-(1.13) and (2.15)-(2.16) are divided into two groups. Equations from the first group (1.12) and (2.15) in the kinetic part contain only one summand each: $\hat{\mathrm{T}}_{n,n+1}$ (1.4) and $\mathrm{T}^{n,n+1}$ (2.3), respectively. In the second group, equations (1.13) and (2.16) each contain two summands determined by kinematical values $\left\langle\vec{\xi}^{n+k+1}\right\rangle_{n,n+k}$ and $\left\langle\vec{\xi}^{n+1}\right\rangle_{n,n+k}$ (see $\hat{\mathrm{T}}_{n,n+k}$ (1.5), $\mathrm{T}^{n,n+k}$ (2.4) and quantum potential $\mathrm{Q}^{n,n+k}$ (2.5)). This fact is a consequence of equations (i.4) of the dispersion chain. Indeed, the equation for function $S^{n,n+1}$ contains only one source of dissipation $Q_{n,n+1}^{n+1}$, while the equation for $S^{n,n+k}$ has two sources $Q_{n,n+k}^{n}$ and $Q_{n,n+k}^{n+k}$. Similar remarks are also true for equations of the third rank (1.14)-(1.16) and (2.17)-(2.19), which may contain information from three sources of dissipation of kinematical values (i.5).

Let us construct analogues of the Lagrange functions for the Hamilton functions (2.1). In classical mechanics, the Lagrangian and the Hamiltonian are related by the Legendre transformation. This relation is not always one-to-one. For example, Lagrange function $\mathrm{L}$ is used in the principle of least action (PLA), in which the relation $\vec{v}=\nabla_r\mathrm{S}$ is true, where $\mathrm{S}$ is the action. Thus, vector velocity field $\vec{v}$ is potential and does not contain a vortex component, which in the general case should be present according to the Helmholtz theorem (1.10). The Hamilton function (2.1) contains the total (potential and vortex) kinetic energy. From the above said, it follows that in the general case (the presence of a vortex component of the field) it is possible to construct only a certain analogue of the Lagrange function for the Hamilton function.

We will start from the classical relation for PLA: $\frac{d}{dt}\mathrm{S}=\mathrm{L}$. In the generalized phase space, the extensive of operators $\hat{\Pi}^R=\left\{\hat{\pi}_{n_1,\ldots,n_R}\right\}$ (i.6) corresponds to the total time derivative. Action $\mathrm{S}$, according to quantum mechanics, will be associated with value $\hbar_{n_R}\varphi^{n_1\ldots n_R}$. Thus, the following theorem can be formulated.

**<u>Definition 4</u>** *Let us define the extensive of Lagrange functions* $\mathfrak{L}^R=\left\{\mathrm{L}^{n_1\ldots n_R}\right\}$ *of rank* $R\in\mathbb{N}$ *in the form*

$$\mathrm{L}^{n_1\ldots n_R}=\mathrm{L}_p^{n_1\ldots n_R}-\mathrm{T}_s^{n_1\ldots n_R},$$

$$\mathrm{L}_p^{n_1\dots n_R} = \mathrm{T}_p^{n_1\dots n_R} - \mathrm{V}^{n_1\dots n_R},$$

*at* $R=1$ $$\mathrm{T}_\lambda^{n} = \tau_{n,\lambda}^{n}, \tag{2.20}$$

*at* $R=2$ $$\mathrm{T}_\lambda^{n,n+k} = \tau_{n+1,\lambda}^{n,n+1}, \tag{2.21}$$

$$\mathrm{T}_\lambda^{n,n+k} = \tau_{n+k,\lambda}^{n,n+k} + \frac{\beta_n}{\beta_{n+k}}\tau_{n,\lambda}^{n,n+k}, \tag{2.22}$$

*at* $R=3$

$$\mathrm{T}_\lambda^{n,n+1,n+2} = \tau_{n+2,\lambda}^{n,n+1,n+2}, \tag{2.23}$$

$$\mathrm{T}_\lambda^{n,n+1,n+1+k} = \tau_{n+k+1,\lambda}^{n,n+1,n+1+k} + \frac{\beta_{n+1}}{\beta_{n+1+k}}\tau_{n+1,\lambda}^{n,n+1,n+1+k}, \tag{2.24}$$

$$\mathrm{T}_\lambda^{n,n+s,n+s+1} = \tau_{n+s+1,\lambda}^{n,n+s,n+s+1} + \frac{\beta_n}{\beta_{n+s+1}}\tau_{n,\lambda}^{n,n+s,n+s+1}, \tag{2.25}$$

$$\mathrm{T}_\lambda^{n,n+s,n+s+k} = \tau_{n+s+k,\lambda}^{n,n+s,n+s+k} + \frac{\beta_{n+s}}{\beta_{n+s+k}}\tau_{n+s,\lambda}^{n,n+s,n+s+k} + \frac{\beta_n}{\beta_{n+s+k}}\tau_{n,\lambda}^{n,n+s,n+s+k}, \tag{2.26}$$

*where, according to representation (1.10) for field* $\left\langle \vec{\xi}^{\,\ell+1}\right\rangle_{n_1\dots n_R}$ *:*

$$\left\langle \vec{\xi}_p^{\,\ell+1}\right\rangle_{n_1\dots n_R} \overset{\det}{=} -\alpha_\ell \nabla_{\xi^\ell}\Phi^{n_1\dots n_R}, \quad \left\langle \vec{\xi}_s^{\,\ell+1}\right\rangle_{n_1\dots n_R} \overset{\det}{=} \gamma_\ell \vec{A}^{\ell}_{n_1\dots n_R}. \tag{2.27}$$

$$\tau_{\ell,\lambda}^{n_1\dots n_R} \overset{\det}{=} -\frac{1}{4\alpha_\ell\beta_\ell}\left|\left\langle \vec{\xi}_\lambda^{\,\ell+1}\right\rangle_{n_1\dots n_R}\right|^2, \ \lambda = \{p,s\}.$$

**Theorem 3** *Let the conditions of Theorem 2 be satisfied, then the equations for the generalized Lagrange functions correspond to the generalized Hamilton-Jacobi equations (2.14)-(2.19)*

$$\hat{\pi}_{n_1\dots n_R}\mathrm{S}^{n_1\dots n_R} = \mathrm{L}^{n_1\dots n_R}, \tag{2.28}$$

*as well as the generalized Legendre transformations do:*

*at* $R=1$ $$\mathrm{H}^n + \mathrm{L}^n = -\frac{1}{2\alpha_n\beta_n}\left\langle \vec{\xi}^{\,n+1}\right\rangle_n \left\langle \vec{\xi}_p^{\,n+1}\right\rangle_n, \tag{2.29}$$

*at* $R=2$ $$\mathrm{H}^{n,n+1} + \mathrm{L}^{n,n+1} = -\frac{1}{2\alpha_{n+1}\beta_{n+1}}\left\langle \vec{\xi}^{\,n+2}\right\rangle_{n,n+1} \left\langle \vec{\xi}_p^{\,n+2}\right\rangle_{n,n+1}, \tag{2.30}$$

$$\mathrm{H}^{n,n+k} + \mathrm{L}^{n,n+k} = -\frac{1}{2\alpha_{n+k}\beta_{n+k}}\left\langle \vec{\xi}^{\,n+1+k}\right\rangle_{n,n+k} \left\langle \vec{\xi}_p^{\,n+k+1}\right\rangle_{n,n+k} - \\ -\frac{1}{2\alpha_n\beta_{n+k}}\left\langle \vec{\xi}^{\,n+1}\right\rangle_{n,n+k} \left\langle \vec{\xi}_p^{\,n+1}\right\rangle_{n,n+k}, \tag{2.31}$$

*at* $R=3$ $$\mathrm{H}^{n,n+1,n+2} + \mathrm{L}^{n,n+1,n+2} = -\frac{1}{2\alpha_{n+2}\beta_{n+2}}\left\langle \vec{\xi}^{\,n+3}\right\rangle_{n,n+1,n+2} \left\langle \vec{\xi}_p^{\,n+3}\right\rangle_{n,n+1,n+2}, \tag{2.32}$$

$$\mathrm{H}^{n,n+1,n+1+k} + \mathrm{L}^{n,n+1,n+1+k} = -\frac{1}{2\alpha_{n+1}\beta_{n+1+k}}\left\langle \vec{\xi}^{\,n+2}\right\rangle_{n,n+1,n+1+k} \left\langle \vec{\xi}_p^{\,n+2}\right\rangle_{n,n+1,n+1+k} - \\ -\frac{1}{2\alpha_{n+1+k}\beta_{n+1+k}}\left\langle \vec{\xi}^{\,n+2+k}\right\rangle_{n,n+1,n+1+k} \left\langle \vec{\xi}_p^{\,n+2+k}\right\rangle_{n,n+1,n+1+k}, \tag{2.33}$$

$$\mathrm{H}^{n,n+s,n+s+1}+\mathrm{L}^{n,n+s,n+s+1}=-\frac{1}{2\alpha_n\beta_{n+s+1}}\left\langle\vec{\xi}^{\,n+1}\right\rangle_{n,n+s,n+s+1}\left\langle\vec{\xi}_p^{\,n+1}\right\rangle_{n,n+s,n+s+1}-$$
$$-\frac{1}{2\alpha_{n+s+1}\beta_{n+s+1}}\left\langle\vec{\xi}^{\,n+s+2}\right\rangle_{n,n+s,n+s+1}\left\langle\vec{\xi}_p^{\,n+s+2}\right\rangle_{n,n+s,n+s+1}, \tag{2.34}$$

$$\mathrm{H}^{n,n+s,n+s+k}+\mathrm{L}^{n,n+s,n+s+k}=-\frac{1}{2\alpha_n\beta_{n+s+k}}\left\langle\vec{\xi}^{\,n+1}\right\rangle_{n,n+s,n+s+k}\left\langle\vec{\xi}_p^{\,n+1}\right\rangle_{n,n+s,n+s+k}-$$
$$-\frac{1}{2\alpha_{n+s}\beta_{n+s+k}}\left\langle\vec{\xi}^{\,n+1+s}\right\rangle_{n,n+s,n+s+k}\left\langle\vec{\xi}_p^{\,n+1+s}\right\rangle_{n,n+s,n+s+k}- \tag{2.35}$$
$$-\frac{1}{2\alpha_{n+s+k}\beta_{n+s+k}}\left\langle\vec{\xi}^{\,n+1+s+k}\right\rangle_{n,n+s,n+s+k}\left\langle\vec{\xi}_p^{\,n+1+s+k}\right\rangle_{n,n+s,n+s+k},$$
$$\ldots.$$

*where* $\left\{\mathrm{S}^{n_1\ldots n_R}\right\}=\hbar_{n_R}\left\{\varphi^{n_1\ldots n_R}\right\}$ *determines the extensive of the generalized actions.*

The proof of Theorem 3 is given in Appendix C.

**Remark**

The index «p» in expression (2.27) corresponds to the potential component of field $\left\langle\vec{\xi}^{\,\ell+1}\right\rangle_{n_1\ldots n_R}$ and the index «s» corresponds to the vortex (solenoidal) component. In the absence of the vortex component ($\mathrm{T}_s^{n_1\ldots n_R}=0$), the Lagrange function $\mathrm{L}^{n_1\ldots n_R}$ transforms into function $\mathrm{L}_p^{n_1\ldots n_R}$, which is an analogue of the Lagrange function in the classical sense, since it is represented as the difference between «kinetic» energy $\mathrm{T}_p^{n_1\ldots n_R}$ and «potential» energy $\mathrm{V}^{n_1\ldots n_R}$. In this case, the generalized Legendre transformations (2.29)-(2.35) also take the «classical» form. For example, at $R=1$, transformation (2.29) taking into account (1.17) has the form:

$$\mathrm{H}^1+\mathrm{L}^1=\left\langle\vec{p}\right\rangle_1\left\langle\vec{v}\right\rangle_1,\ \left\langle\vec{p}\right\rangle_1=m\left\langle\vec{v}\right\rangle_1. \tag{2.36}$$

In the classical limit (at $\hbar\ll 1$), taking into account (1.17), we can consider the approximation of smallness of generalized quantum potential $\mathrm{Q}^{n_1\ldots n_R}$, i.e. (2.1) $\mathrm{V}^{n_1\ldots n_R}\approx U^{n_1\ldots n_R}$. Such an approximation is often used at $R=1$ and $n=1$.

Note that in the presence of vortex field component $\left\langle\vec{\xi}^{\,\ell+1}\right\rangle_{n_1\ldots n_R}$ ($\mathrm{T}_s^{n_1\ldots n_R}\neq 0$), generalized kinematic energy $\mathrm{T}^{n_1\ldots n_R}$ is not equal to the sum of $\mathrm{T}_s^{n_1\ldots n_R}$ and $\mathrm{T}_p^{n_1\ldots n_R}$. In this case, the Lagrange function $\mathrm{L}^{n_1\ldots n_R}$ has an additional «uncharacteristic» summand in the form of kinematic energy of the vortex field $-\mathrm{T}_s^{n_1\ldots n_R}$.

The wave function as a complex one is characterized by its modulus and phase. The modulus is related to the probability density $f^{n_1\ldots n_R}$ (or $S^{n_1\ldots n_R}$), and the phase is related to the scalar potential (1.10) $\Phi^{n_1\ldots n_R}$ (or action $\mathrm{S}^{n_1\ldots n_R}$). By analogy with papers [13, 2], we construct generalized compound action $\mathcal{Z}^{n_1\ldots n_R}$. Using representation (1.10) and expressions (A.2), (A.3), we obtain

$$i\Phi^{n_1\dots n_R} = \mathrm{Ln}\frac{\Psi^{n_1\dots n_R}}{\bar{\Psi}^{n_1\dots n_R}} = \mathrm{Ln}\,\Psi^{n_1\dots n_R} - \mathrm{Ln}\,\bar{\Psi}^{n_1\dots n_R}, \tag{2.37}$$

$$S^{n_1,\dots,n_R} = \mathrm{Ln}\left|\Psi^{n_1,\dots,n_R}\right|^2 = \mathrm{Ln}\,\Psi^{n_1\dots n_R} + \mathrm{Ln}\,\bar{\Psi}^{n_1\dots n_R},$$

where $S^{n_1,\dots,n_R}$ satisfies the dispersion chain of the Vlasov equations (i.3)-(i.5).

**Definition 5** *Let us define the extensive of a compound generalized action* $Z^R = \left\{\mathcal{Z}^{n_1\dots n_R}\right\}$ *of rank* $R$ *as*

$$\mathcal{Z}^{n_1\dots n_R} \stackrel{\det}{=} \frac{1}{2}M^{n_1\dots n_R} = \mathrm{Ln}\,\Psi^{n_1\dots n_R}, \quad M^{n_1\dots n_R} \stackrel{\det}{=} S^{n_1\dots n_R} + i\Phi^{n_1\dots n_R}, \tag{2.38}$$

*that is*

$$\Psi^{n_1\dots n_R} = \exp\left\{\mathcal{Z}^{n_1\dots n_R}\right\}. \tag{2.39}$$

**Theorem 4** *The relations are true:*

$$\hat{\pi}_{n_1\dots n_R}\mathcal{Z}^{n_1\dots n_R} \stackrel{\det}{=} \mathcal{L}^{n_1\dots n_R} \quad or \quad \hat{\pi}_{n_1\dots n_R}\Psi^{n_1\dots n_R} = \mathcal{L}^{n_1\dots n_R}\Psi^{n_1\dots n_R}, \tag{2.40}$$

*where the complex generalized Lagrangians* $\mathcal{L}^{n_1\dots n_R}$ *are related to Hamiltonians* $\mathcal{H}^{n_1\dots n_R}$ *by the following equations:*

*at* $R=1$

$$-\partial_0\mathcal{Z}^n = \mathcal{H}^n, \tag{2.41}$$

$$\mathcal{L}^n = -\frac{1}{2}Q_n^n + i\beta_n \mathrm{L}^n, \quad \mathcal{H}^n = \frac{1}{2}\mathcal{Q}_n^n + i\beta_n \mathrm{H}^n, \tag{2.42}$$

$$\mathcal{L}^n + \mathcal{H}^n = \left\langle \vec{\xi}^{\,n+1} \right\rangle_n \nabla_{\xi^n}\mathcal{Z}^n, \tag{2.43}$$

*at* $R=2$

$$-\partial_n\mathcal{Z}^{n,n+1} = \mathcal{H}^{n,n+1}, \tag{2.44}$$

$$\mathcal{L}^{n,n+1} = -\frac{1}{2}Q_{n,n+1}^{n+1} + i\beta_{n+1}\mathrm{L}^{n,n+1}, \quad \mathcal{H}^{n,n+1} = \frac{1}{2}\mathcal{Q}_{n,n+1}^{n+1} + i\beta_{n+1}\mathrm{H}^{n,n+1}, \tag{2.45}$$

$$\mathcal{L}^{n,n+1} + \mathcal{H}^{n,n+1} = \left\langle \vec{\xi}^{\,n+2} \right\rangle_{n,n+1} \nabla_{\xi^{n+1}}\mathcal{Z}^{n,n+1}, \tag{2.46}$$

$$-\partial_0\mathcal{Z}^{n,n+k} = \mathcal{H}^{n,n+k}, \tag{2.47}$$

$$\mathcal{L}^{n,n+k} = -\frac{1}{2}\left(Q_{n,n+k}^{n} + Q_{n,n+k}^{n+k}\right) + i\beta_{n+k}\mathrm{L}^{n,n+k}, \quad \mathcal{H}^{n,n+k} = \frac{1}{2}\left(\mathcal{Q}_{n,n+k}^{n} + \mathcal{Q}_{n,n+k}^{n+k}\right) + i\beta_{n+k}\mathrm{H}^{n,n+k}, \tag{2.48}$$

$$\mathcal{L}^{n,n+k} + \mathcal{H}^{n,n+k} = \left\langle \vec{\xi}^{\,n+1} \right\rangle_{n,n+k} \nabla_{\xi^n}\mathcal{Z}^{n,n+k} + \left\langle \vec{\xi}^{\,n+k+1} \right\rangle_{n,n+k} \nabla_{\xi^{n+k}}\mathcal{Z}^{n,n+k}, \tag{2.49}$$

*at* $R=3$

$$-\partial_{n,n+1}\mathcal{Z}^{n,n+1,n+2} = \mathcal{H}^{n,n+1,n+2}, \tag{2.50}$$

$$\mathcal{L}^{n,n+1,n+2} = -\frac{1}{2}Q_{n,n+1,n+2}^{n+2} + i\beta_{n+2}\mathrm{L}^{n,n+1,n+2}, \quad \mathcal{H}^{n,n+1,n+2} = \frac{1}{2}\mathcal{Q}_{n,n+1,n+2}^{n+2} + i\beta_{n+2}\mathrm{H}^{n,n+1,n+2}, \tag{2.51}$$

$$\mathcal{L}^{n,n+1,n+2} + \mathcal{H}^{n,n+1,n+2} = \left\langle \vec{\xi}^{\,n+3} \right\rangle_{n,n+1,n+2} \nabla_{\xi^{n+2}}\mathcal{Z}^{n,n+1,n+2}, \tag{2.52}$$

$$-\partial_n \mathcal{Z}^{n,n+1,n+1+k} = \mathcal{H}^{n,n+1,n+1+k}, \tag{2.53}$$

$$\mathcal{L}^{n,n+1,n+1+k} = -\frac{1}{2}\left(Q^{n+1}_{n,n+1,n+1+k} + Q^{n+1+k}_{n,n+1,n+1+k}\right) + i\beta_{n+1+k}\,\mathrm{L}^{n,n+1,n+1+k}, \tag{2.54}$$

$$\mathcal{H}^{n,n+1,n+1+k} = \frac{1}{2}\left(\mathcal{Q}^{n+1}_{n,n+1,n+1+k} + \mathcal{Q}^{n+1+k}_{n,n+1,n+1+k}\right) + i\beta_{n+1+k}\,\mathrm{H}^{n,n+1,n+1+k},$$

$$\mathcal{L}^{n,n+1,n+1+k} + \mathcal{H}^{n,n+1,n+1+k} = \left\langle \vec{\xi}^{\,n+2} \right\rangle_{n,n+1,n+1+k} \nabla_{\xi^{n+1}} \mathcal{Z}^{n,n+1,n+1+k} + \left\langle \vec{\xi}^{\,n+k+2} \right\rangle_{n,n+1,n+1+k} \nabla_{\xi^{n+k+1}} \mathcal{Z}^{n,n+1,n+1+k}, \tag{2.55}$$

$$-\partial_{n+s} \mathcal{Z}^{n,n+s,n+s+1} = \mathcal{H}^{n,n+s,n+s+1}, \tag{2.56}$$

$$\mathcal{L}^{n,n+s,n+s+1} = -\frac{1}{2}\left(Q^{n}_{n,n+s,n+s+1} + Q^{n+s+1}_{n,n+s,n+s+1}\right) + i\beta_{n+s+1}\,\mathrm{L}^{n,n+s,n+s+1}, \tag{2.57}$$

$$\mathcal{H}^{n,n+s,n+s+1} = \frac{1}{2}\left(\mathcal{Q}^{n}_{n,n+s,n+s+1} + \mathcal{Q}^{n+s+1}_{n,n+s,n+s+1}\right) + i\beta_{n+s+1}\,\mathrm{H}^{n,n+s,n+s+1},$$

$$\mathcal{L}^{n,n+s,n+s+1} + \mathcal{H}^{n,n+s,n+s+1} = \left\langle \vec{\xi}^{\,n+1} \right\rangle_{n,n+s,n+s+1} \nabla_{\xi^{n}} \mathcal{Z}^{n,n+s,n+s+1} + \left\langle \vec{\xi}^{\,n+s+2} \right\rangle_{n,n+s,n+s+1} \nabla_{\xi^{n+s+1}} \mathcal{Z}^{n,n+s,n+s+1}, \tag{2.58}$$

$$-\partial_0 \mathcal{Z}^{n,n+s,n+s+k} = \mathcal{H}^{n,n+s,n+s+k}, \tag{2.59}$$

$$\mathcal{L}^{n,n+s,n+s+k} = -\frac{1}{2}\left(Q^{n}_{n,n+s,n+s+k} + Q^{n+s}_{n,n+s,n+s+k} + Q^{n+s+k}_{n,n+s,n+s+k}\right) + i\beta_{n+s+k}\,\mathrm{L}^{n,n+s,n+s+k}, \tag{2.60}$$

$$\mathcal{H}^{n,n+s,n+s+k} = \frac{1}{2}\left(\mathcal{Q}^{n}_{n,n+s,n+s+k} + \mathcal{Q}^{n+s}_{n,n+s,n+s+k} + \mathcal{Q}^{n+s+k}_{n,n+s,n+s+k}\right) + i\beta_{n+s+k}\,\mathrm{H}^{n,n+s,n+s+k},$$

$$\begin{aligned}\mathcal{L}^{n,n+s,n+s+k} + \mathcal{H}^{n,n+s,n+s+k} &= \left\langle \vec{\xi}^{\,n+1} \right\rangle_{n,n+s,n+s+k} \nabla_{\xi^{n}} \mathcal{Z}^{n,n+s,n+s+k} + \\ &+ \left\langle \vec{\xi}^{\,n+s+1} \right\rangle_{n,n+s,n+s+k} \nabla_{\xi^{n+s}} \mathcal{Z}^{n,n+s,n+s+k} + \\ &+ \left\langle \vec{\xi}^{\,n+s+k+1} \right\rangle_{n,n+s,n+s+k} \nabla_{\xi^{n+s+k}} \mathcal{Z}^{n,n+s,n+s+k},\end{aligned} \tag{2.61}$$

*where* $\mathcal{Q}^{\ell}_{n_1 \dots n_R}$ *is determined by the sources of probability current densities* $\vec{J}^{\,n_1 \dots n_R}_{\ell} \overset{\text{det}}{=} f^{\,n_1 \dots n_R} \left\langle \vec{\xi}^{\,\ell+1} \right\rangle_{n_1 \dots n_R}$ *of order* $\ell$

$$f^{\,n_1 \dots n_R} \mathcal{Q}^{\ell}_{n_1 \dots n_R} \overset{\text{det}}{=} \operatorname{div}_{\xi^{\ell}} \vec{J}^{\,n_1 \dots n_R}_{\ell}, \tag{2.62}$$

*And indices* $\ell \in \{n_1, \dots, n_R\}$; $n, k, s \in \mathbb{N}$, $k, s > 1$.

The proof of Theorem 4 is given in Appendix D.

Equations (2.40) for compound action $\mathcal{Z}^{n_1 \dots n_R}$ are a generalization of equations (2.28) and the dispersion chain of the Vlasov equations. Indeed, equations (2.40) are divided into two types of equations – for the real and imaginary parts. The equations for the imaginary part (2.40) coincide with equations (2.28). The equations for the real part of (2.40) lead to dispersion chain of the Vlasov equations (i.3)-(i.5).

The imaginary parts of equations (2.41), (2.44), (2.47), (2.50), (2.53), (2.56), (2.59) for the compound action $\mathcal{Z}^{n_1 \dots n_R}$ define the generalized Hamilton-Jacobi equations (2.14)-(2.19) for action $\mathrm{S}^{n_1 \dots n_R} = 2\operatorname{Im} \mathcal{Z}^{n_1 \dots n_R} = \hbar_{n_R} \varphi^{n_1 \dots n_R}$. The real parts of equations (2.41), (2.44), (2.47), (2.50), (2.53), (2.56), (2.59) contain the dispersion chain of the Vlasov equations (i.3)-(i.5) for functions $S^{n_1 \ldots n_R} = 2\operatorname{Re}\mathcal{Z}^{n_1 \ldots n_R}$.

Expressions (2.42), (2.45), (2.48), (2.51), (2.54), (2.57), (2.60) define the generalized complex Lagrange functions $\mathcal{L}^{n_1 \ldots n_R}$ and the Hamilton functions $\mathcal{H}^{n_1 \ldots n_R}$. Relations (2.43), (2.46), (2.49), (2.52), (2.55), (2.58), (2.61) are analogues of the Legendre transformations for the functions $\mathcal{L}^{n_1 \ldots n_R}$, $\mathcal{H}^{n_1 \ldots n_R}$.

In the absence of sources $Q^{\ell}_{n_1 \ldots n_R}$ and $\mathcal{Q}^{\ell}_{n_1 \ldots n_R}$ functions $\mathcal{L}^{n_1 \ldots n_R}$, $\mathcal{H}^{n_1 \ldots n_R}$ transform into $\mathrm{L}^{n_1 \ldots n_R}$ and $\mathrm{H}^{n_1 \ldots n_R}$ respectively.

## §3 Motion equations and field equations

Based on the generalized Hamilton-Jacobi equations (2.14)-(2.19), one can obtain the equations of motion for the flows of kinematical values.

**Definition 6** *Let the representation (1.10) be true and functions (2.1) be given, then we define the extensive of vector functions* $\vec{E}^{\ell}_{n_1 \ldots n_R}$ *of order* $\ell$ *and rank* $R$ *as:*

*at* $R = 1$
$$\vec{E}^n_n \stackrel{\det}{=} -\partial_0 \vec{A}^n_n - \frac{2\alpha_n \beta_n}{\gamma_n} \nabla_{\xi^n} \mathrm{V}^n, \tag{3.1}$$

*at* $R = 2$
$$\vec{E}^{n+1}_{n,n+1} \stackrel{\det}{=} -\partial_n \vec{A}^{n+1}_{n,n+1} - \frac{\alpha_{n+1}}{\gamma_{n+1}} \left( \nabla_{\xi^n} \Phi^{n,n+1} + 2\beta_{n+1} \nabla_{\xi^{n+1}} \mathrm{V}^{n,n+1} \right), \tag{3.2}$$
$$\vec{E}^{n+k}_{n,n+k} \stackrel{\det}{=} -\partial_0 \vec{A}^{n+k}_{n,n+k} - \frac{2\alpha_{n+k} \beta_{n+k}}{\gamma_{n+k}} \nabla_{\xi^{n+k}} \left( \frac{\beta_n}{\beta_{n+k}} \tau^{n,n+k}_n + \mathrm{V}^{n,n+k} \right), \tag{3.3}$$
$$\vec{E}^{n}_{n,n+k} \stackrel{\det}{=} -\partial_0 \vec{A}^{n}_{n,n+k} - \frac{2\alpha_{n} \beta_{n+k}}{\gamma_{n}} \nabla_{\xi^{n}} \left( \tau^{n,n+k}_{n+k} + \mathrm{V}^{n,n+k} \right), \tag{3.4}$$

*at* $R = 3$
$$\vec{E}^{n+2}_{n,n+1,n+2} \stackrel{\det}{=} -\partial_{n,n+1} \vec{A}^{n+2}_{n,n+1,n+2} - \frac{\alpha_{n+2}}{\gamma_{n+2}} \left( \nabla_{\xi^{n+1}} \Phi^{n,n+1,n+2} + 2\beta_{n+2} \nabla_{\xi^{n+2}} \mathrm{V}^{n,n+1,n+2} \right), \tag{3.5}$$

$$\vec{E}^{n+1+k}_{n,n+1,n+1+k} \stackrel{\det}{=} -\partial_n \vec{A}^{n+1+k}_{n,n+1,n+1+k} - \frac{2\alpha_{n+1+k} \beta_{n+1+k}}{\gamma_{n+1+k}} \nabla_{\xi^{n+1+k}} \left( \frac{\beta_{n+1}}{\beta_{n+1+k}} \tau^{n,n+1,n+1+k}_{n+1} + \mathrm{V}^{n,n+1,n+1+k} \right), \tag{3.6}$$
$$\vec{E}^{n+1}_{n,n+1,n+1+k} \stackrel{\det}{=} -\partial_n \vec{A}^{n+1}_{n,n+1,n+1+k} - \frac{\alpha_{n+1}}{\gamma_{n+1}} \left[ \nabla_{\xi^n} \Phi^{n,n+1,n+1+k} + 2\beta_{n+1+k} \nabla_{\xi^{n+1}} \left( \tau^{n,n+1,n+1+k}_{n+1+k} + \mathrm{V}^{n,n+1,n+1+k} \right) \right], \tag{3.7}$$

$$\vec{E}^{n}_{n,n+s,n+s+1} \stackrel{\det}{=} -\partial_{n+s} \vec{A}^{n}_{n,n+s,n+s+1} - \frac{2\alpha_{n} \beta_{n+s+1}}{\gamma_{n}} \nabla_{\xi^{n}} \left( \tau^{n,n+s,n+s+1}_{n+s+1} + \mathrm{V}^{n,n+s,n+s+1} \right), \tag{3.8}$$
$$\begin{aligned} \vec{E}^{n+s+1}_{n,n+s,n+s+1} &\stackrel{\det}{=} -\partial_{n+s} \vec{A}^{n+s+1}_{n,n+s,n+s+1} - \\ &- \frac{\alpha_{n+s+1}}{\gamma_{n+s+1}} \left[ \nabla_{\xi^{n+s}} \Phi^{n,n+s,n+s+1} + 2\beta_{n+s+1} \nabla_{\xi^{n+s+1}} \left( \frac{\beta_n}{\beta_{n+s+1}} \tau^{n,n+s,n+s+1}_{n} + \mathrm{V}^{n,n+s,n+s+1} \right) \right], \end{aligned} \tag{3.9}$$

$$\vec{E}^{n}_{n,n+s,n+s+k} \stackrel{\det}{=} -\partial_0 \vec{A}^{n}_{n,n+s,n+s+k} - \frac{2\alpha_{n} \beta_{n+s+k}}{\gamma_{n}} \nabla_{\xi^{n}} \left( \tau^{n,n+s,n+s+k}_{n+s+k} + \frac{\beta_{n+s}}{\beta_{n+s+k}} \tau^{n,n+s,n+s+k}_{n+s} + \mathrm{V}^{n,n+s,n+s+k} \right), \tag{3.10}$$

$$\vec{E}^{n+s}_{n,n+s,n+s+k} \overset{\text{det}}{=} -\partial_0 \vec{A}^{n+s}_{n,n+s,n+s+k} -$$
$$-\frac{2\alpha_{n+s}\beta_{n+s+k}}{\gamma_{n+s}} \nabla_{\xi^{n+s}} \left( \tau^{n,n+s,n+s+k}_{n+s+k} + \frac{\beta_n}{\beta_{n+s+k}} \tau^{n,n+s,n+s+k}_{n} + \mathrm{V}^{n,n+s,n+s+k} \right), \tag{3.11}$$

$$\vec{E}^{n+s+k}_{n,n+s,n+s+k} \overset{\text{det}}{=} -\partial_0 \vec{A}^{n+s+k}_{n,n+s,n+s+k} -$$
$$-\frac{2\alpha_{n+s+k}\beta_{n+s+k}}{\gamma_{n+s+k}} \nabla_{\xi^{n+s+k}} \left( \frac{\beta_{n+s}}{\beta_{n+s+k}} \tau^{n,n+s,n+s+k}_{n+s} + \frac{\beta_n}{\beta_{n+s+k}} \tau^{n,n+s,n+s+k}_{n} + \mathrm{V}^{n,n+s,n+s+k} \right), \tag{3.12}$$

…

**Theorem 5** *Let representation (1.10) be true and the conditions of Theorem 2 be satisfied, then the following generalized equations of motion correspond to equations (2.14)-(2.15):*

*at* $R=1$

$$\hat{\pi}_n \left\langle \vec{\xi}^{\,n+1} \right\rangle_n = \vec{F}^n_n, \tag{3.13}$$

*at* $R=2$

$$\hat{\pi}_{n,n+1} \left\langle \vec{\xi}^{\,n+2} \right\rangle_{n,n+1} = \vec{F}^{n+1}_{n,n+1}, \tag{3.14}$$

$$\hat{\pi}^{n,n+k}_{0,n+k} \left\langle \vec{\xi}^{\,n+k+1} \right\rangle_{n,n+k} = \vec{F}^{n+k}_{n,n+k}, \tag{3.15}$$

$$\hat{\pi}^{n,n+k}_{0,n} \left\langle \vec{\xi}^{\,n+1} \right\rangle_{n,n+k} = \vec{F}^{n}_{n,n+k}, \tag{3.16}$$

*at* $R=3$

$$\hat{\pi}_{n,n+1,n+2} \left\langle \vec{\xi}^{\,n+3} \right\rangle_{n,n+1,n+2} = \vec{F}^{n+2}_{n,n+1,n+2}, \tag{3.17}$$

$$\hat{\pi}^{n,n+1,n+1+k}_{n,n+1+k} \left\langle \vec{\xi}^{\,n+k+2} \right\rangle_{n,n+1,n+k+1} = \vec{F}^{n+1+k}_{n,n+1,n+k+1}, \tag{3.18}$$

$$\hat{\pi}^{n,n+1,n+1+k}_{n,n+1} \left\langle \vec{\xi}^{\,n+2} \right\rangle_{n,n+1,n+k+1} = \vec{F}^{n+1}_{n,n+1,n+k+1}, \tag{3.19}$$

$$\hat{\pi}^{n,n+s,n+s+1}_{n+s,n} \left\langle \vec{\xi}^{\,n+1} \right\rangle_{n,n+s,n+s+1} = \vec{F}^{n}_{n,n+s,n+s+1}, \tag{3.20}$$

$$\hat{\pi}^{n,n+s,n+s+1}_{n+s,n+s+1} \left\langle \vec{\xi}^{\,n+s+2} \right\rangle_{n,n+s,n+s+1} = \vec{F}^{n+s+1}_{n,n+s,n+s+1}, \tag{3.21}$$

$$\hat{\pi}^{n,n+s,n+s+k}_{0,n} \left\langle \vec{\xi}^{\,n+1} \right\rangle_{n,n+s,n+s+k} = \vec{F}^{n}_{n,n+s,n+s+k}, \tag{3.22}$$

$$\hat{\pi}^{n,n+s,n+s+k}_{0,n+s} \left\langle \vec{\xi}^{\,n+s+1} \right\rangle_{n,n+s,n+s+k} = \vec{F}^{n+s}_{n,n+s,n+s+k}, \tag{3.23}$$

$$\hat{\pi}^{n,n+s,n+s+k}_{0,n+s+k} \left\langle \vec{\xi}^{\,n+s+k+1} \right\rangle_{n,n+s,n+s+k} = \vec{F}^{n+s+k}_{n,n+s,n+s+k}, \tag{3.24}$$

…

*where*

$$\hat{\pi}^{n_1 \dots n_R}_{q,\ell} \overset{\text{det}}{=} \partial_q + \left\langle \vec{\xi}^{\,\ell+1} \right\rangle_{n_1 \dots n_R} \nabla_{\xi^\ell}, \tag{3.25}$$

$$\vec{F}^{\ell}_{n_1 \dots n_R} \overset{\text{det}}{=} -\gamma_\ell \left( \vec{E}^{\ell}_{n_1 \dots n_R} + \left\langle \vec{\xi}^{\,\ell+1} \right\rangle_{n_1 \dots n_R} \times \vec{B}^{\ell}_{n_1 \dots n_R} \right), \tag{3.26}$$

*fields* $\vec{E}^{\ell}_{n_1 \dots n_R}$ *have the form (3.1)-(3.12);* $\vec{B}^{\ell}_{n_1 \dots n_R} = \mathrm{curl}_{\xi^\ell} \vec{A}^{\ell}_{n_1 \dots n_R}$, $\ell, q \in \mathrm{n} = \{n_1, \dots, n_R\}$, $\ell + 1 \notin \mathrm{n}$*;* $n, k, s \in \mathbb{N}$, $k, s > 1$.

The proof of Theorem 5 is given in Appendix D.

**Remark**

The relation $\hat{\pi}^{n}_{0,n} = \hat{\pi}_n$, $\hat{\pi}^{n,n+1}_{n,n+1} = \hat{\pi}_{n,n+1}$ is true for operator (3.25). Equation (3.14) is obtained from equation (2.15), and equations (3.15) and (3.16) are obtained from equation (2.16). Equations (3.18) and (3.19) correspond to the first equation in (2.18). The second equation from (2.18) gives the equations of motion (3.20) and (3.21). Equation (2.19) has three sources of dissipation and leads to three equations (3.22)-(3.24).

The generalized equations of motion (3.13)-(3.24) have a «classical» form, that is, on the left there is an analogue of the total time derivative in the generalized phase subspace of kinematical value $\hat{\pi}\left\langle \vec{\xi}^{\,\ell+1} \right\rangle$, and on the right there is some analogue of kinematic «force» $\vec{F}^{\,\ell}$. Field (3.26) consists of two components: a kinematic analogue of the «Lorentz force» $\left\langle \vec{\xi}^{\,\ell+1} \right\rangle_{n_1 \dots n_R} \times \vec{B}^{\ell}_{n_1 \dots n_R}$ and «electric force» $\vec{E}^{\ell}_{n_1 \dots n_R}$. Equations (3.13) transform into the classical equation of motion of a charged particle in an electromagnetic field only in the particular case ( $R=1$ and $n=1$).

The following theorem holds true for the extensive of fields $\vec{E}^{\ell}_{n_1 \dots n_R}$ and $\vec{B}^{\ell}_{n_1 \dots n_R}$ in generalized equations of motion (3.13)–(3.24).

**Theorem 6** *Let distribution functions $f^{n_1 \dots n_R}$ satisfy the dispersion chain of the Vlasov equations (i.22)-(i.24) and representation (1.10) is true, in which*

$$\operatorname{curl}_{\xi^{\ell}} \left\langle \vec{\xi}^{\,\ell+2} \right\rangle_{n_1 \dots n_R} = 0, \ \ell \in \mathrm{n}, \ \ell+2 \notin \mathrm{n}, \tag{3.27}$$

*therewith such an extensive of the fields $\vec{D}^{\ell}_{n_1 \dots n_R}$ is present that*

$$f^{n_1 \dots n_R} = \sum_{r=1}^{R} \operatorname{div}_{\xi^{n_r}} \vec{D}^{n_r}_{n_1 \dots n_R}, \tag{3.28}$$

$$\partial_{\ell} \vec{D}^{\ell}_{n_1 \dots n_R} = \vec{D}^{\ell+1}_{n_1 \dots n_R}, \ \ell, \ell+1 \in \mathrm{n}, \tag{3.29}$$

$$\partial_{\ell_1 \ell_2} \vec{D}^{\ell_1}_{n_1 \dots n_R} = \vec{D}^{\ell_1+1}_{n_1 \dots n_R}, \ \ell_1, \ell_1+1, \ell_2 \in \mathrm{n}, \tag{3.30}$$

$$\partial_{\ell_1 \ell_2} \vec{D}^{\ell_2}_{n_1 \dots n_R} = \vec{D}^{\ell_2+1}_{n_1 \dots n_R}, \ \ell_2, \ell_2+1, \ell_1 \in \mathrm{n}, \tag{3.31}$$

$$\dots$$

*Then the following field equations are true:*

$$\operatorname{div}_{\xi^{\ell}} \vec{B}^{\ell}_{n_1 \dots n_R} = 0, \tag{3.32}$$

*at* $R=1$

$$\operatorname{curl}_{\xi^{n}} \vec{E}^{n}_{n} = -\partial_0 \vec{B}^{n}_{n}, \tag{3.33}$$

$$\partial_0 \vec{D}^{n}_{n} + \vec{J}^{n}_{n} = \operatorname{curl}_{\xi^{n}} \vec{H}^{n}_{n}, \tag{3.34}$$

*at* $R=2$

$$\operatorname{curl}_{\xi^{n+1}} \vec{E}^{n+1}_{n,n+1} = -\partial_n \vec{B}^{n+1}_{n,n+1}, \tag{3.35}$$

$$\partial_n \vec{D}^{n+1}_{n,n+1} + \vec{J}^{n+1}_{n,n+1} = \operatorname{curl}_{\xi^{n+1}} \vec{H}^{n+1}_{n,n+1}, \tag{3.36}$$

$$\operatorname{curl}_{\xi^{n+k}} \vec{E}^{n+k}_{n,n+k} = -\partial_0 \vec{B}^{n+k}_{n,n+k}, \tag{3.37}$$

$$\operatorname{curl}_{\xi^n} \vec{E}^n_{n,n+k} = -\partial_0 \vec{B}^n_{n,n+k}, \tag{3.38}$$

$$\partial_0 \vec{D}^n_{n,n+k} + \vec{J}^n_{n,n+k} = \operatorname{curl}_{\xi^n} \vec{H}^n_{n,n+k}, \tag{3.39}$$

$$\partial_0 \vec{D}^{n+k}_{n,n+k} + \vec{J}^{n+k}_{n,n+k} = \operatorname{curl}_{\xi^{n+k}} \vec{H}^{n+k}_{n,n+k}, \tag{3.40}$$

*at* $R = 3$

$$\operatorname{curl}_{\xi^{n+2}} \vec{E}^{n+2}_{n,n+1,n+2} \overset{\det}{=} -\partial_{n,n+1} \vec{B}^{n+2}_{n,n+1,n+2}, \tag{3.41}$$

$$\partial_{n,n+1} \vec{D}^{n+2}_{n,n+1,n+2} + \vec{J}^{n+2}_{n,n+1,n+2} = \operatorname{curl}_{\xi^{n+2}} \vec{H}^{n+2}_{n,n+1,n+2}, \tag{3.42}$$

$$\operatorname{curl}_{\xi^{n+1+k}} \vec{E}^{n+1+k}_{n,n+1,n+1+k} = -\partial_n \vec{B}^{n+1+k}_{n,n+1,n+1+k}, \tag{3.43}$$

$$\operatorname{curl}_{\xi^{n+1}} \vec{E}^{n+1}_{n,n+1,n+1+k} = -\partial_n \vec{B}^{n+1}_{n,n+1,n+1+k}, \tag{3.44}$$

$$\partial_n \vec{D}^{n+1}_{n,n+1,n+1+k} + \vec{J}^{n+1}_{n,n+1,n+1+k} = \operatorname{curl}_{\xi^{n+1}} \vec{H}^{n+1}_{n,n+1,n+1+k}, \tag{3.45}$$

$$\partial_n \vec{D}^{n+1+k}_{n,n+1,n+1+k} + \vec{J}^{n+1+k}_{n,n+1,n+1+k} = \operatorname{curl}_{\xi^{n+1+k}} \vec{H}^{n+1+k}_{n,n+1,n+1+k}, \tag{3.46}$$

$$\operatorname{curl}_{\xi^{n+s+1}} \vec{E}^{n+s+1}_{n,n+s,n+s+1} = -\partial_{n+s} \vec{B}^{n+s+1}_{n,n+s,n+s+1}, \tag{3.47}$$

$$\operatorname{curl}_{\xi^n} \vec{E}^n_{n,n+s,n+s+1} = -\partial_{n+s} \vec{B}^n_{n,n+s,n+s+1}, \tag{3.48}$$

$$\partial_{n+s} \vec{D}^n_{n,n+s,n+s+1} + \vec{J}^n_{n,n+s,n+s+1} = \operatorname{curl}_{\xi^n} \vec{H}^n_{n,n+s,n+s+1}, \tag{3.49}$$

$$\partial_{n+s} \vec{D}^{n+s+1}_{n,n+s,n+s+1} + \vec{J}^{n+s+1}_{n,n+s,n+s+1} = \operatorname{curl}_{\xi^{n+s+1}} \vec{H}^{n+s+1}_{n,n+s,n+s+1}, \tag{3.50}$$

$$\operatorname{curl}_{\xi^n} \vec{E}^n_{n,n+s,n+s+k} \overset{\det}{=} -\partial_0 \vec{B}^n_{n,n+s,n+s+k}, \tag{3.51}$$

$$\operatorname{curl}_{\xi^{n+s}} \vec{E}^{n+s}_{n,n+s,n+s+k} \overset{\det}{=} -\partial_0 \vec{B}^{n+s}_{n,n+s,n+s+k}, \tag{3.52}$$

$$\operatorname{curl}_{\xi^{n+s+k}} \vec{E}^{n+s+k}_{n,n+s,n+s+k} \overset{\det}{=} -\partial_0 \vec{B}^{n+s+k}_{n,n+s,n+s+k}, \tag{3.53}$$

$$\partial_0 \vec{D}^n_{n,n+s,n+s+k} + \vec{J}^n_{n,n+s,n+s+k} = \operatorname{curl}_{\xi^n} \vec{H}^n_{n,n+s,n+s+k}, \tag{3.54}$$

$$\partial_0 \vec{D}^{n+s}_{n,n+s,n+s+k} + \vec{J}^{n+s}_{n,n+s,n+s+k} = \operatorname{curl}_{\xi^{n+s}} \vec{H}^{n+s}_{n,n+s,n+s+k}, \tag{3.55}$$

$$\partial_0 \vec{D}^{n+s+k}_{n,n+s,n+s+k} + \vec{J}^{n+s+k}_{n,n+s,n+s+k} = \operatorname{curl}_{\xi^{n+s+k}} \vec{H}^{n+s+k}_{n,n+s,n+s+k}, \tag{3.56}$$

$\ldots$

*where*

$$\vec{J}^{\ell}_{n_1 \ldots n_R} \overset{\det}{=} f^{n_1 \ldots n_R} \left\langle \vec{\xi}^{\ell+1} \right\rangle_{n_1 \ldots n_R}, \qquad \ell \in \mathrm{n},\ \ell+1 \notin \mathrm{n}, \tag{3.57}$$

*and* $\vec{H}^{\ell}_{n_1 \ldots n_R}$ *is an extensive of some sufficiently soft fields;* $n, k, s \in \mathbb{N}$, $k, s > 1$; $\mathrm{n} = \{n_1, \ldots, n_R\}$.

The proof of Theorem 6 is given in Appendix E.

**Theorem 7** *Let field extensives* $\vec{E}^{\ell}_{n_1\ldots n_R}$ *and* $\vec{B}^{\ell}_{n_1\ldots n_R}$ *satisfying generalized field equations (3.35)-(3.59) be defined and the following conditions be satisfied:*

$$\Phi^{n,n+1,n+2,\ldots,n+r} = \int\limits_{(\infty)} \Phi^{n,n+1,n+2,\ldots,n+r,n+r+k} d^3\xi^{n+r+k}, \quad \vec{A}^{\ell}_{n_1\ldots n_R} = \int\limits_{(\infty)} \vec{A}^{\ell}_{n_1\ldots n_R,n_R+k} d^3\xi^{n_R+k}, \tag{3.58}$$

$$\tau_{\ell}^{n_1\ldots n_R} = \int\limits_{(\infty)} \tau_{\ell}^{n_1\ldots n_R,n_R+k} d^3\xi^{n_R+k}, \quad \mathrm{V}^{n_1\ldots n_R} = \frac{\beta_{n_R+k}}{\beta_{n_R}} \int\limits_{(\infty)} \left(\tau_{n_R+k}^{n_1\ldots n_R,n_R+k} + \mathrm{V}^{n_1\ldots n_R,n_R+k}\right) d^3\xi^{n_R+k}, \tag{3.59}$$

*then the relations are true:*

$$\vec{J}^{\ell}_{n_1\ldots n_R} = \int\limits_{(\infty)} \vec{J}^{\ell}_{n_1\ldots n_R,n_R+k} d^3\xi^{n_R+k}, \tag{3.60}$$

$$\vec{E}^{\ell}_{n_1\ldots n_R} = \int\limits_{(\infty)} \vec{E}^{\ell}_{n_1\ldots n_R,n_R+k} d^3\xi^{n_R+k}, \quad \vec{D}^{\ell}_{n_1\ldots n_R} = \int\limits_{(\infty)} \vec{D}^{\ell}_{n_1\ldots n_R,n_R+k} d^3\xi^{n_R+k}, \tag{3.61}$$

$$\vec{B}^{\ell}_{n_1\ldots n_R} = \int\limits_{(\infty)} \vec{B}^{\ell}_{n_1\ldots n_R,n_R+k} d^3\xi^{n_R+k}, \quad \vec{H}^{\ell}_{n_1\ldots n_R} = \int\limits_{(\infty)} \vec{H}^{\ell}_{n_1\ldots n_R,n_R+k} d^3\xi^{n_R+k}, \tag{3.62}$$

*where* $n,k,r \in \mathbb{N}$, $k,n > 1$; $\ell \in \mathrm{n}$, $\mathrm{n} = \{n_1,\ldots,n_R\}$

The proof of Theorem 7 is given in Appendix F.

The obtained field equations (3.28), (3.32)-(3.56) are a generalization of the known Maxwell equations for an electromagnetic field to the case of a generalized phase space. In a special case for equations of the first rank at $n=1$, generalized field equations (3.28), (3.32)-(3.56) are transformed into the four well-known Maxwell equations (3.28), (3.32)-(3.34). In the general case, analogues of electric and magnetic fields have dependencies on high order kinematical values (3.1)-(3.12) and are interrelated through averaging over corresponding phase subspaces (3.60)-(3.62). It is these generalized analogues of electromagnetic fields that are present in generalized equations of motion (3.13)-(3.24). (3.13)-(3.24).

## §4 Exact solution

Let us consider second-rank equations (1.12), (2.15), (3.14) at $n=1$. Equation (1.12) for function $\Psi^{1,2}(x,v,t)$ will take the form:

$$\frac{i}{\beta_2}\partial_1 \Psi^{1,2} = \hat{\mathrm{H}}_{1,2}\Psi^{1,2},$$
$$i\hbar_2\left(\frac{\partial}{\partial t} + v\frac{\partial}{\partial x}\right)\Psi^{1,2} = -\frac{\hbar_2^2}{2m}\frac{\partial^2}{\partial v^2}\Psi^{1,2} + U^{1,2}\Psi^{1,2}, \tag{4.1}$$

where the potential is $U^{1,2} = U^{1,2}(x,v,t)$. We consider particular solutions of equation (4.1). The solution of equation (4.1) will be sought in the form

$$\Psi^{1,2}(x,v,t) = \Psi(x-vt)\exp\left(-i\frac{E^{1,2}}{3\hbar_2}t^3\right), \tag{4.2}$$

where $\Psi = \Psi(\eta)$ is some function, $E^{1,2}$ is constant. Substituting representation (4.2) into equation (4.1), we obtain:

$$i\hbar_2\left(-i\frac{E^{1,2}}{\hbar_2}t^2\Psi - v\Psi' + v\Psi'\right)e^{-i\frac{E^{1,2}}{3\hbar_2}t^3} = -\frac{\hbar_2^2}{2m}e^{-i\frac{E^{1,2}}{3\hbar_2}t^3}t^2\Psi'' + e^{-i\frac{E^{1,2}}{3\hbar_2}t^3}U^{1,2}\Psi,$$

$$-\frac{\hbar_2^2}{2m}\Psi'' + \frac{1}{t^2}U^{1,2}\Psi = E^{1,2}\Psi. \tag{4.3}$$

As a potential $U^{1,2}$, we consider

$$U^{1,2}(x,v,t) = t^2U(x-vt), \tag{4.4}$$

then equation (4.3) takes the form:

$$\Psi''(\eta) - \frac{2m}{\hbar_2^2}\left[U(\eta) - E^{1,2}\right]\Psi(\eta) = 0. \tag{4.5}$$

Equation (4.5) is an ordinary differential equation with variable coefficients. Function $\eta(x,v,t) = x - vt$ defines a characteristic surface in space $(x,v,t)$. For a single-valued solution of equation (4.5), we set the boundary conditions on two characteristic surfaces:

$$\eta = -\eta_0, \qquad \eta = \eta_0, \tag{4.6}$$

$$\Psi(-\eta_0) = \Psi(\eta_0) = 0, \tag{4.7}$$

where $\eta_0 = const$. Fig. 1 shows graphs of surfaces (4.6). Region $-\eta_0 < \eta < \eta_0$ between surfaces (4.6) is the area in which the solution of equation (4.5) is sought. Each surface $\eta$ has set of values $(x,v,t)$, while for any point $p = (x,v,t) \in \eta$ the value of wave function $\Psi$ will be the same. At each instant of time $t$, phase region $(x,v)$, in which the system is considered, changes, but its area remains constant.

Indeed, the phase region has the shape of a parallelogram that changes its slope over time. At the initial instant of time ($t = 0$), the phase region is rectangle $ABCD$ (see Fig. 1), and after an instant of time $\Delta t$ it is parallelogram $A'B'C'D'$ with the same area (see Fig. 1), since the base and height remain unchanged $\Delta x \cdot \Delta v = const$. In this formulation of the problem, we will consider the evolution of the system in a finite phase region of size $\Delta x \cdot \Delta v$, so the normalization condition for the wave function will have the form (see Fig. 1):

$$\int_{-\Delta v/2}^{\Delta v/2} dv \int_{-\eta_0+vt}^{\eta_0+vt} \left|\Psi(x-vt)\right|^2 dx = 1. \tag{4.8}$$

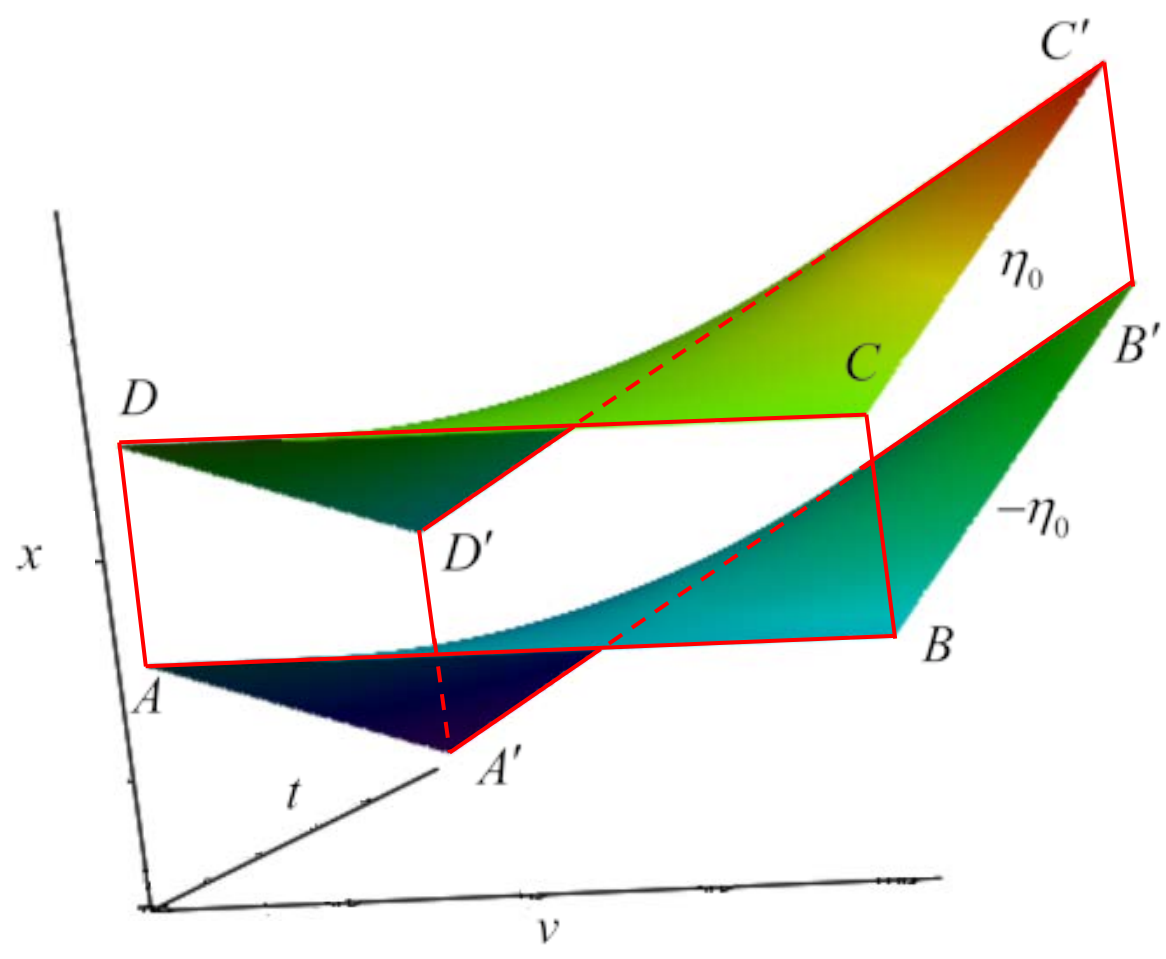

Fig. 1 Evolution of the phase region $\Omega^{1,2}$

The form of boundary conditions (4.7) is chosen as the simplest example. To solve equation (4.5), it is necessary to set potential $U(\eta)$. Consider the simplest form of potential

$$U(\eta)=\begin{cases}0, & |\eta|<\eta_0,\\ +\infty, & |\eta|=\eta_0.\end{cases} \tag{4.9}$$

Note that, despite expression (4.9), potential $U^{1,2}$ remains non-stationary, since walls of the potential $(x,v)$ in the phase space change their position over time (see Fig. 1). The solution of equation (4.5) with potential (4.9) contains even $\Psi_n^{(+)}(\eta)$ and odd $\Psi_k^{(-)}(\eta)$ wave functions:

$$\begin{aligned}
&\Psi_n^{(+)}(\eta)=\sqrt{\frac{2}{\Delta x\cdot\Delta v}}\cos\left[\frac{\pi(2n+1)}{2\eta_0}\eta\right],. \qquad E_n^{(+)}=\frac{\hbar_2^2\pi^2(2n+1)^2}{8m\eta_0^2}, \quad n=0,1,...\\
&\Psi_k^{(-)}(\eta)=\sqrt{\frac{2}{\Delta x\cdot\Delta v}}\sin\left(\frac{\pi k}{\eta_0}\eta\right), \qquad E_k^{(-)}=\frac{\hbar_2^2\pi^2k^2}{2m\eta_0^2}, \qquad k=1,2...
\end{aligned} \tag{4.10}$$

Without loss of generality, we consider only even wave functions $\Psi_n^{(+)}(\eta)$, for which function $\Psi_n^{1,2}$ takes the form:

$$\Psi_n^{1,2}(x,v,t)=\sqrt{\frac{2}{\Delta x\cdot\Delta v}}\cdot\cos\left[\frac{\pi(2n+1)}{2\eta_0}(x-vt)\right]\exp\left(-i\frac{E_n^{1,2}}{3\hbar_2}t^3\right), \tag{4.11}$$

where $E_n^{1,2}=E_n^{(+)}$. Substituting resulting wave function (4.11) into equation (4.1) with potential $U^{1,2}$ (4.4), (4.9) gives the correct identity. Thus, distribution function $f_n^{1,2}$ has the form:

$$f_n^{1,2}(x,v,t)=\begin{cases}\frac{2}{\Delta x\cdot\Delta v}\cos^2\left[\frac{\pi(2n+1)}{2\eta_0}(x-vt)\right], & |x-vt|<\frac{\Delta x}{2},\\ 0, \ \ otherwise.\end{cases} \tag{4.12}$$

The obtained distribution density function $f_n^{1,2}$ (4.12) is positive and different from the Gaussian distribution. *This fact, in accordance with the Hudson theorem [14], indicates that function (4.12) is not the Wigner function.*

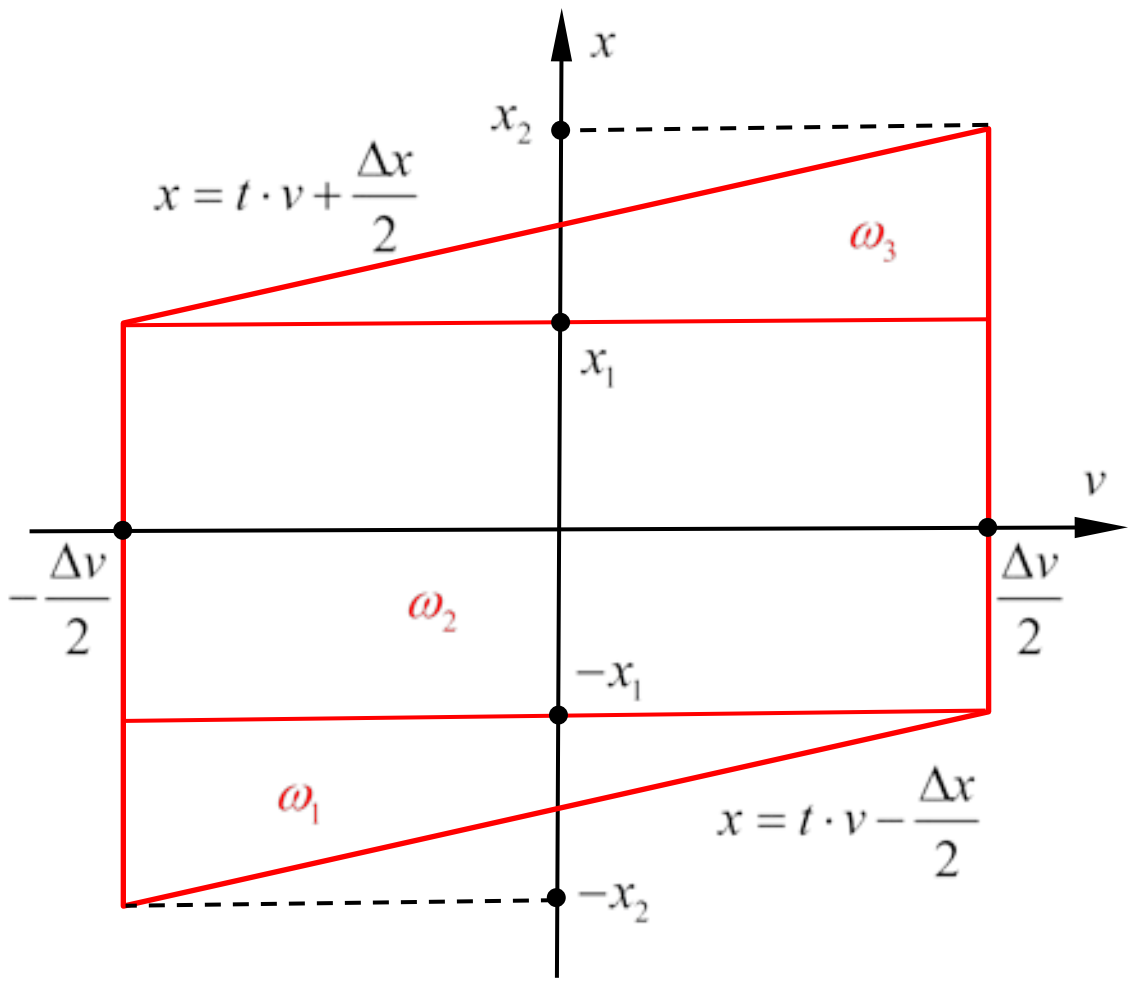

Fig. 2 Phase region at $0 < t < \frac{\Delta x}{\Delta v}$.

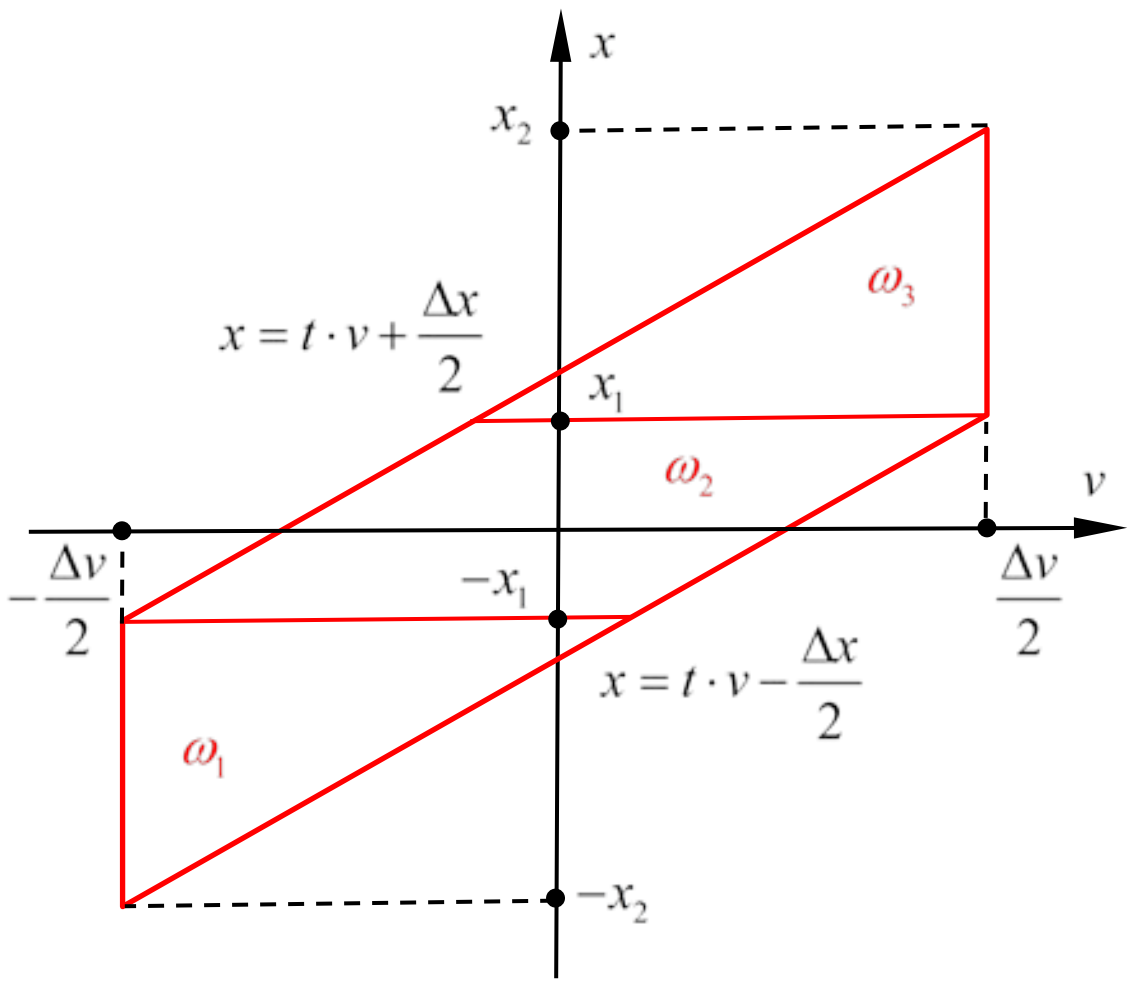

Fig. 3 Phase region at $t > \frac{\Delta x}{\Delta v}$.

**Remark**

Index $n$ of function (4.11), (4.12) will be formally called the state number. All functions (4.11) correspond to the same potential (4.9). In the further presentation, when passing to equations of the first rank, for example, (1.11), it turns out that each wave function $\Psi_n^{1,2}(x,v,t)$ determines wave function $\Psi_n^1(x,t)$, to which its potential $U_n^1(x,t)$ corresponds. Thus, wave functions $\Psi_n^1(x,t)$ will satisfy the different Schrödinger equations with different potential $U_n^1$. As a result, index $n$ for function $\Psi_n^1(x,t)$ will not have the meaning of the quantum state number.

Note that when setting potential $U^{1,2}(x,v,t)$ in the phase space, it is not known in advance what the potential will be in the coordinate $U^1(x,t)$ or momentum ($p = mv$) $U^2(v,t)$ space. To determine potentials $U^1(x,t)$ and $U^2(v,t)$ one needs to perform the corresponding averaging operations. If a potential is initially given in the coordinate or momentum space, then it is possible to construct some potential in the phase space using it only in a phenomenological way. This is a natural fact, since the generalized phase space of lower dimensions contains no information from spaces of higher dimensions. Such information is «lost» as a result of averaging.

Knowing distribution (4.12), we can find the probability densities $f_n^1$ and $f_n^2$. Let us find $f_n^1$. Let us perform integration over variable $v$ in phase region $\Omega^{1,2}$ (see Fig. 1). The shape of phase region $\Omega^{1,2}$ depends on time $t$. At the initial instant of time ($t = 0$), region $\Omega^{1,2}$ has the shape of a rectangle with sides $\Delta x$ and $\Delta v$ (see Fig. 1). At $t > 0$, region $\Omega^{1,2}$ is a parallelogram (see Fig. 1, 2), therefore the limits of integration with respect to variable $v$ will change depending on time $t$. Let us divide phase region $\Omega^{1,2}$ (at $t > 0$) into three subregions (see Fig. 2, 3):

$$\Omega^{1,2} = \omega_1 \cup \omega_2 \cup \omega_3, \tag{4.13}$$

where at $0 < t < \frac{\Delta x}{\Delta v}$:

$$\omega_1=\left\{(x,v):\ -t\Delta v<2x+\Delta x<t\Delta v\ ;\ -\Delta v<2v<\frac{1}{t}(2x+\Delta x)\right\}, \tag{4.14}$$
$$\omega_2=\{(x,v):t\Delta v-\Delta x<2x<-t\Delta v+\Delta x\ ;\ -\Delta v<2v<\Delta v\},$$
$$\omega_3=\left\{(x,v):-t\Delta v<2x-\Delta x<t\Delta v\ ;\ \frac{1}{t}(2x-\Delta x)<2v<\Delta v\right\},$$

and at $t>\frac{\Delta x}{\Delta v}$:

$$\omega_1=\left\{(x,v):\ -\Delta x<2x+t\Delta v<\Delta x\ ;\ -\Delta v<2v<\frac{1}{t}(2x+\Delta x)\right\}, \tag{4.15}$$
$$\omega_2=\{(x,v):-t\Delta v+\Delta x<2x<t\Delta v-\Delta x\ ;\ -\Delta x<2(vt-x)<\Delta x\},$$
$$\omega_3=\left\{(x,v):-\Delta x<2x-t\Delta v<\Delta x\ ;\ \frac{1}{t}(2x-\Delta x)<2v<\Delta v\right\}.$$

The region shown in Fig. 2 corresponds to time interval $0<t<\frac{\Delta x}{\Delta v}$ and is determined by expressions (4.14). At $t>\frac{\Delta x}{\Delta v}$, phase region (4.15) undergoes a significant change, shown in Fig. 3.

As a result of integration, we obtain the following result:

At $t=0$

$$f_n^1(x,0)=\begin{cases}\frac{2}{\Delta x}\cos^2\vartheta_n x, & |x|<\frac{\Delta x}{2},\\ 0, & otherwise.\end{cases} \tag{4.16}$$

At $0<t\le\frac{\Delta x}{\Delta v}$

$$f_n^1(x,t)=\begin{cases}g_n(x,t), & -t\Delta v<2x+\Delta x\le t\Delta v,\\ \frac{1+\operatorname{sinc}\tau_n\cdot\cos 2\vartheta_n x}{\Delta x}, & t\Delta v-\Delta x<2x\le -t\Delta v+\Delta x,\\ g_n(-x,t), & -t\Delta v<2x-\Delta x\le t\Delta v,\\ 0, & otherwise.\end{cases} \tag{4.17}$$

At $t\ge\frac{\Delta x}{\Delta v}$

$$f^1(x,t)=\begin{cases}g_n(x,t), & -\Delta x<2x+t\Delta v\le\Delta x,\\ \frac{1}{t\Delta v}, & \Delta x-t\Delta v<2x\le t\Delta v-\Delta x,\\ g_n(-x,t), & -\Delta x<2x-t\Delta v\le\Delta x,\\ 0, & otherwise.\end{cases} \tag{4.18}$$

where $\tau_n\overset{\det}{=}t\vartheta_n\Delta v$, $\vartheta_n\overset{\det}{=}\frac{\pi(2n+1)}{\Delta x}$, $\operatorname{sinc}x\overset{\det}{=}\frac{\sin x}{x}$,

$$g_n(x,t) \overset{\text{det}}{=} \frac{2x+t\Delta v+\Delta x}{2t\Delta x\Delta v} + \frac{\sin\left[\vartheta_n\left(2x+t\Delta v\right)\right]}{2\vartheta_n t\Delta x\Delta v}.$$

Potential $U^{1,2}$ at the initial instant of time, according to expressions (4.1) and (4.9), depends only on coordinate $x$ and coincides with the potential of an infinitely deep well. Thus, distribution (4.16) is a known probability density of the state (corresponding to an even wave function) of a particle in an infinitely deep potential well. At $t>0$, potential $U^{1,2}$ begins to change rapidly, and the phase region first transforms from a rectangle into a parallelogram shown in Fig. 2 ($0<t\le\frac{\Delta x}{\Delta v}$, function (4.17)), and then into a one shown in Fig. 3 ($t\ge\frac{\Delta x}{\Delta v}$, function (4.18)). The expression for function $f_n^2$ will take the form:

$$f_n^2(v,t)=\frac{1}{\Delta v}. \tag{4.19}$$

Expression (4.19) does not depend on time and differs from the known momentum distribution for an infinitely deep potential well. Such a difference is due to the time dependent of potential $U^{1,2}$.

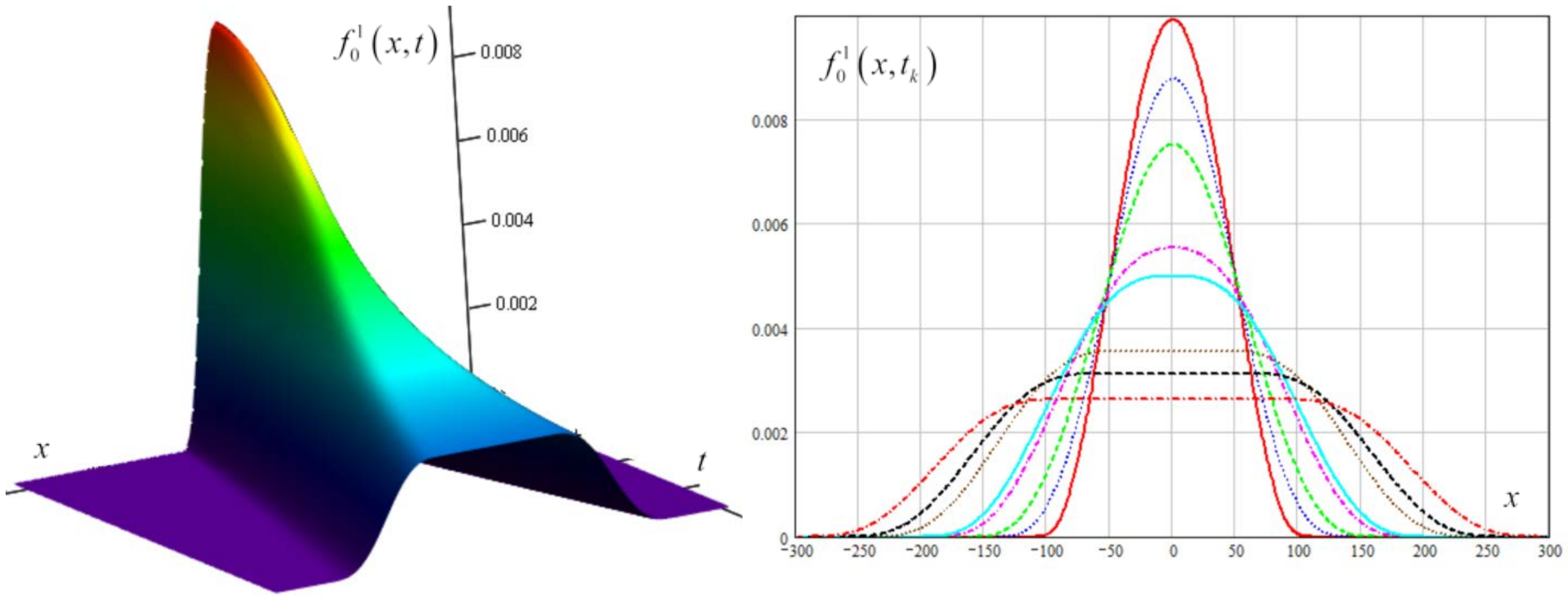

Fig. 4 Evolution of the ground state distribution density $f_0^1(x,t)$.

In Figs. 4, 5 (left), the evolution is shown of the probability density distribution function $f_0^1(x,t)$ and $f_1^1(x,t)$, respectively. Figs. 4 and 5 (on the right) show successive time «junctures» for the functions $f_0^1(x,t)$ and $f_1^1(x,t)$ at time instants $t_k=\{0;\ 0.4;\ 0.6;\ 0.9;\ 1;\ 1.4;\ 1.6;\ 1.9\}$. Note that function $f_0^1(x,t)$ corresponds to the ground «state» of the quantum system, and $f_1^1(x,t)$ corresponds to the «state» with number three (as $f_n^1(x,t)$ corresponds only to even wave functions).

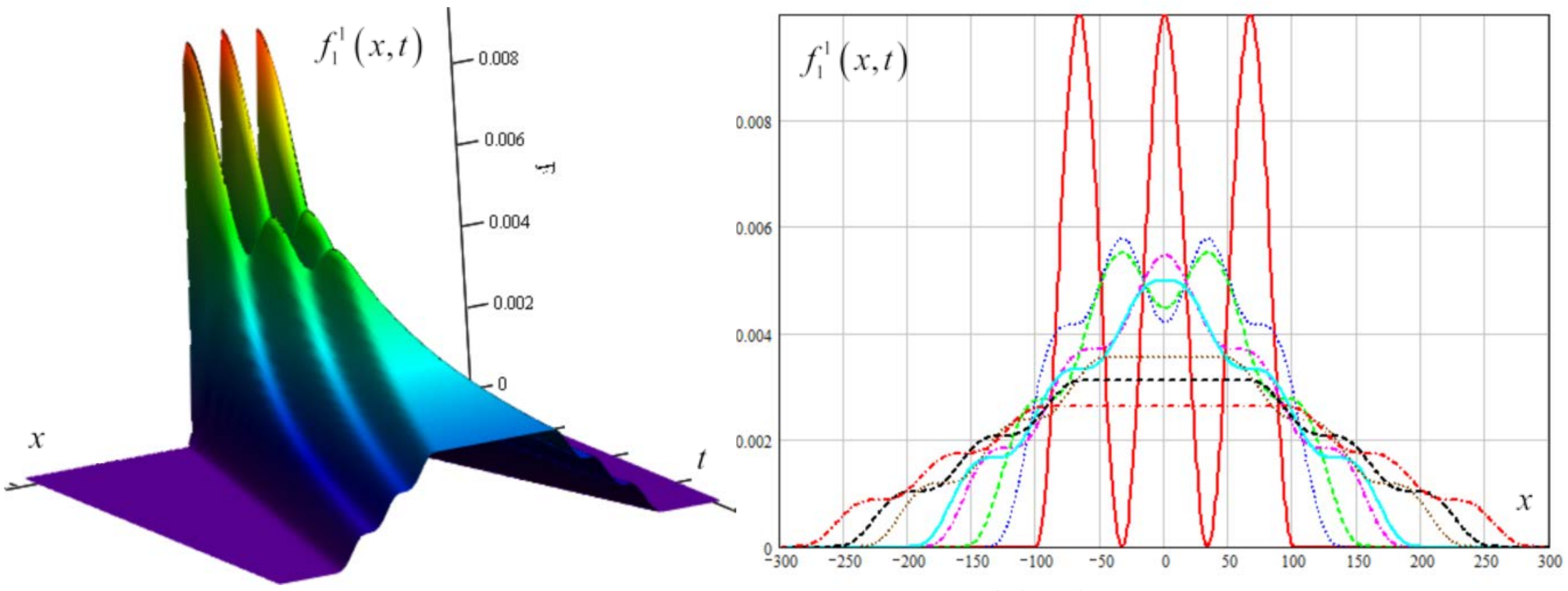

Fig. 5 Evolution of the ground state distribution density $f_1^1(x,t)$.

As an example, the ratio $\frac{\Delta x}{\Delta v}$ was taken equal to one. Over time, the distribution along the coordinate becomes blurred (see Figs. 4, 5), which corresponds to the rotation of the phase region in Figs. 2, 3. The region $(-x_2, x_2)$ increases over time from the interval $(-\Delta x/2, \Delta x/2)$ − a rectangular area at the initial instant of time and until it tends to an infinite interval. The area of the phase region remains unchanged − $\Delta x \cdot \Delta v$.

Knowledge of distribution function $f_n^{1,2}(x,v,t)$ (4.12) allows us to determine not only the functions (4.16)-(4.18) $f_n^1(x,t)$ and (4.19) $f_n^2(v,t)$, but the average field of the probability flow velocity $\langle v\rangle_1 = \langle v\rangle_1(x,t)$:

$$f_n^1(x,t)\langle v\rangle_{1|n}(x,t) = \int_{-\infty}^{+\infty} v\, f_n^{1,2}(x,v,t)\,dv. \tag{4.20}$$

where $\langle v\rangle_{1|n}$ is velocity $\langle v\rangle_1$ of the probability flow for quantum «state» number $n$. The calculation of integral (4.20) gives the following results:

At $t = 0$

$$\langle v\rangle_{1|n}(x,0) = 0. \tag{4.21}$$

At $0 < t \le \frac{\Delta x}{\Delta v}$

$$f_n^1\langle v\rangle_{1|n} = \begin{cases} h_n(x,t), & -t\Delta v < 2x + \Delta x \le t\Delta v, \\ \frac{\sin 2\vartheta_n x}{2\vartheta_n t\Delta x}\left(\operatorname{sinc}\vartheta_n t\Delta v - \cos\vartheta_n t\Delta v\right), & t\Delta v - \Delta x < 2x \le -t\Delta v + \Delta x, \\ -h_n(-x,t), & -t\Delta v < 2x - \Delta x \le t\Delta v, \\ 0, & otherwise. \end{cases} \tag{4.22}$$

At $t \geq \dfrac{\Delta x}{\Delta v}$

$$f_n^1 \langle v \rangle_{1|n} = \begin{cases} h_n(x,t), & -\Delta x < 2x + t\Delta v \leq \Delta x, \\ \dfrac{x}{t^2 \Delta v}, & \Delta x - t\Delta v < 2x \leq t\Delta v - \Delta x, \\ -h_n(-x,t), & -\Delta x < 2x - t\Delta v \leq \Delta x, \\ 0, & otherwise. \end{cases} \tag{4.23}$$

where

$$h_n(x,t) \overset{\det}{=} \frac{(2x+\Delta x)^2 - t^2\Delta v^2}{8t^2 \Delta x \Delta v} - \frac{\cos^2 \frac{\vartheta_n}{2}(2x + t\Delta v)}{2\vartheta_n^2 t^2 \Delta x \Delta v} - \frac{\sin \vartheta_n (2x + t\Delta v)}{4\vartheta_n t \Delta x}.$$

It follows from expression (4.21) that at the initial instant of time the probability flow velocity is equal to zero, which corresponds to the result of [15]. The result (4.21) is expected, since the phase of the wave function for a particle in an infinitely deep potential well does not depend on the coordinate, hence, from representation, (1.10) $\langle v \rangle_{1|n}(x,0) = 0$. For time interval $t > 0$ according to expressions (4.22) and (4.23) in Figs. 6, 7 plots of the evolution of the field of velocities $\langle v \rangle_{1|0}(x,t)$ and $\langle v \rangle_{1|1}(x,t)$ are constructed, respectively. In Figs. 6, 7 on the left there are three-dimensional distributions of velocities $\langle v \rangle_{1|0}(x,t)$ and $\langle v \rangle_{1|1}(x,t)$ from the coordinate $x$ and time $t$. In Figs. 6, 7 on the right there are successive time «junctures» of the distributions of the velocity field from the coordinate $x$ at the instants of time $t_k = \{0.01;\ 0.1;\ 0.2;\ 0.3;\ 0.4;\ 0.9;\ 1.9\}$. For the main «state» ($n = 0$) the velocity flow is directed from the center of the system to its periphery and grows in absolute value over time interval $0 < t \leq \dfrac{\Delta x}{\Delta v}$. After instant of time $t > \dfrac{\Delta x}{\Delta v}$, in the central region a general decrease in the velocity flow is observed. At large times in the central region, the distribution $\langle v \rangle_{1|0}(x,t)$ asymptotically tends to a linear dependence in the coordinate (see Fig. 6).

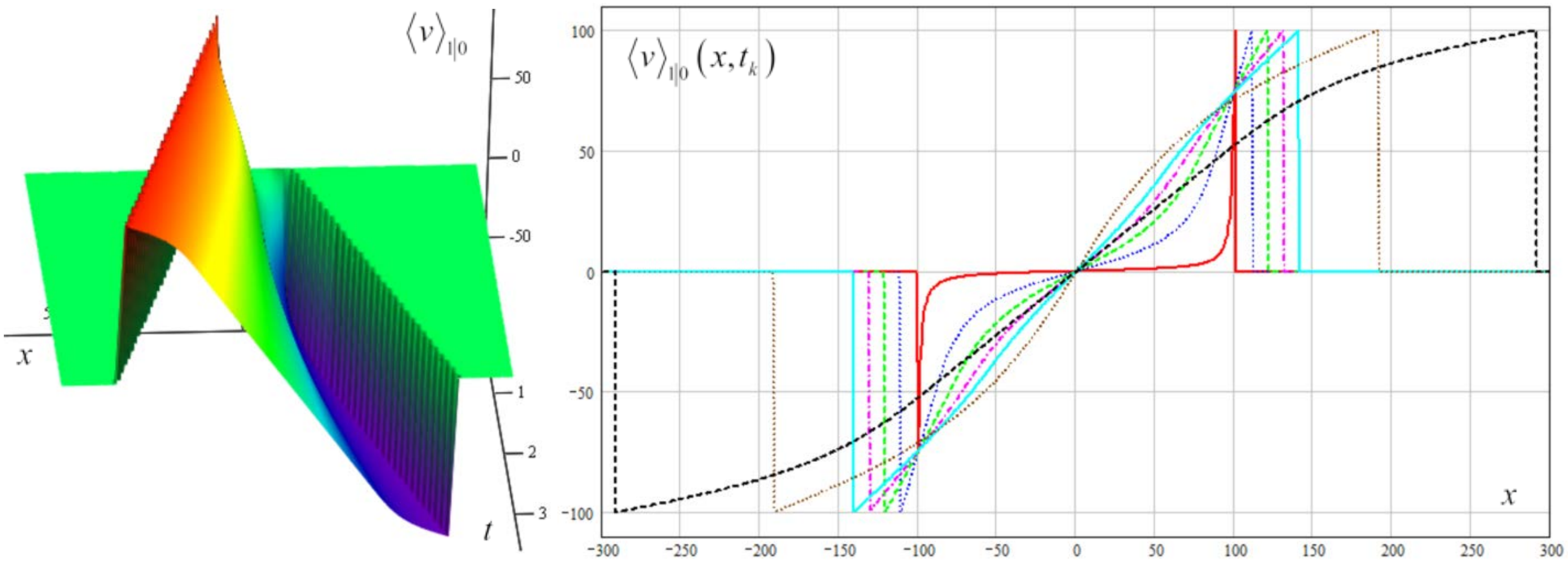

Fig. 6 Evolution of probability velocity flow $\langle v \rangle_{1|0}(x,t)$.

The state of the system with the number $n = 1$ (see Fig. 7) is similar to the «superposition» of three systems in the ground state ($n = 0$). Recall that $n = 1$ applies to the even wave function (4.10). For a quantum system containing even and odd wave functions, the state number $n=1$ (4.10) corresponds to the state number three. In general, the nature of the evolution of probability velocity flow $\langle v\rangle_{1|1}(x,t)$ in Fig. 7 has similar features with behavior of $\langle v\rangle_{1|0}(x,t)$ (see Fig. 6). At large times, velocity distribution $\langle v\rangle_{1|1}(x,t)$ (see Fig. 7), as well as $\langle v\rangle_{1|0}(x,t)$ (see Fig. 6), asymptotically tends to a linear dependence on the coordinate.

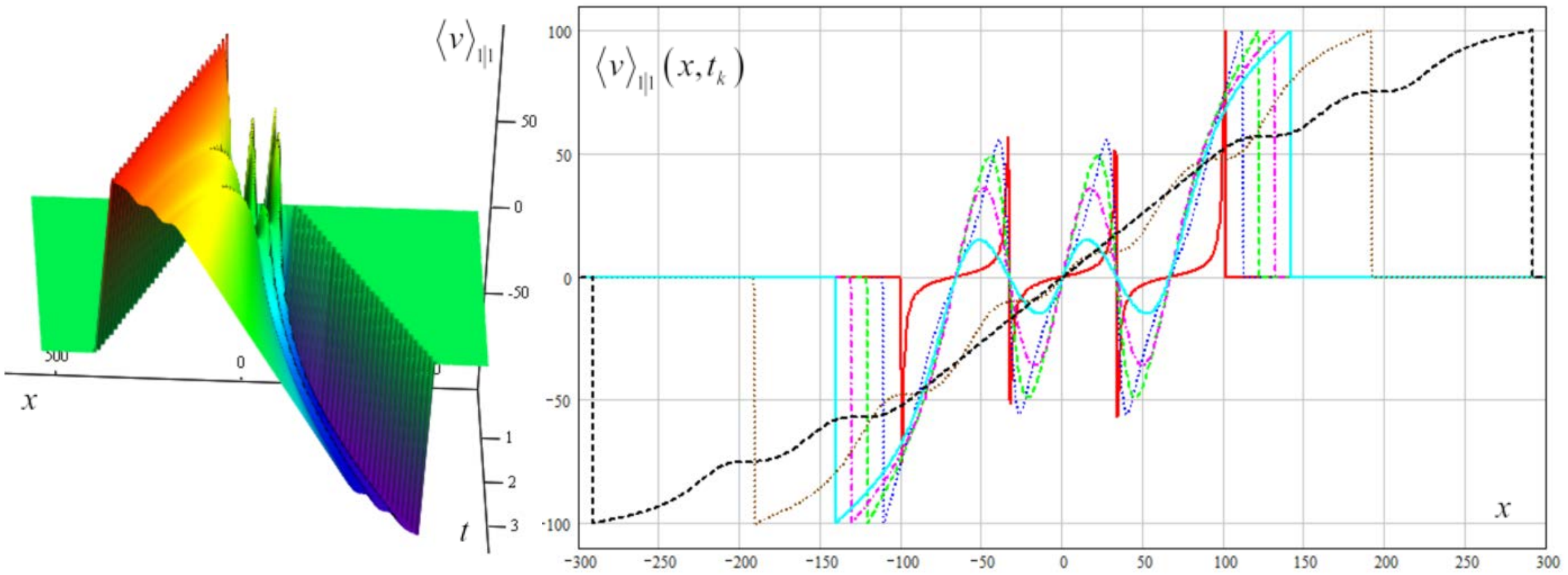

Fig. 7 Evolution of probability velocity flow $\langle v\rangle_{1|1}(x,t)$.

Based on the theorems proved above, let us consider in more detail the interpretation of the behavior of a quantum system in terms of continuum mechanics. Let us start with equations of the second rank. Phase $\varphi_n^{1,2}=-\dfrac{E_n^{1,2}}{3\hbar_2}t^3$ of wave function (4.11) does not depend on velocity $v$, therefore, vector field (1.10):

$$\langle \dot{v}\rangle_{1,2|n}=\langle \dot{v}\rangle_{1|n}=0, \tag{4.24}$$

where, due to the one-dimensional case, the vortex component is absent. Taking into account expression (4.24), the equation for distribution function $f_n^{1,2}$ according to the Vlasov dispersion chain (i.4) takes the form:

$$\frac{\partial}{\partial t}f_n^{1,2}+v\frac{\partial}{\partial x}f_n^{1,2}=0. \tag{4.25}$$

Equation (4.25) is satisfied, since according to (4.12) function $f_n^{1,2}(\eta)$ has an argument in the form of characteristic $\eta=x-vt$.

Let us find the quantum potential $\mathrm{Q}_n^{1,2}$ (2.3). Equation (4.1) implies that

$$t^2E_n^{1,2}-U^{1,2}=-\frac{\hbar_2^2}{2m}\frac{1}{\Psi_n^{1,2}}\frac{\partial^2\Psi_n^{1,2}}{\partial v^2}=-\frac{\hbar_2^2}{2m}\frac{1}{\left|\Psi_n^{1,2}\right|}\frac{\partial^2\left|\Psi_n^{1,2}\right|}{\partial v^2}=\mathrm{Q}_n^{1,2},$$
$$\mathrm{Q}_n^{1,2}=t^2E_n^{1,2}, \tag{4.26}$$

which is true almost everywhere inside the potential well. Using expressions (4.24) and (4.26) we can find the Hamilton function $\mathrm{H}^{1,2}$ (2.1):

$$\mathrm{H}_n^{1,2} = \frac{m}{2}\left|\langle \dot{v}\rangle_{1,2|n}\right|^2 + \mathrm{V}_n^{1,2} = t^2 E_n^{1,2}, \qquad (4.27)$$
$$\mathrm{V}_n^{1,2} = U^{1,2} + \mathrm{Q}_n^{1,2}.$$

A similar expression is obtained directly from the Hamilton-Jacobi equation (2.15)

$$\mathrm{H}_n^{1,2} = -\frac{1}{\beta_2}\partial_1 \varphi_n^{1,2} = \frac{1}{\beta_2}\left(\frac{\partial}{\partial t} + v\frac{\partial}{\partial x}\right)\frac{E_n^{1,2}}{3\hbar_2}t^3 = E_n^{1,2}t^2. \qquad (4.28)$$

Comparison of expressions (4.27) and (4.28) shows that they give the same representation for the Hamilton function $\mathrm{H}_n^{1,2}$. According to expressions (4.27) and (4.28), its own Hamilton function $\mathrm{H}_n^{1,2}$ corresponds to each quantum «state» $n$.

The equation of motion (3.14) is also satisfied:

$$\hat{\pi}_{1,2}\langle \dot{v}\rangle_{1,2|n} = F_{1,2|n}^2, \qquad (4.29)$$

where

$$F_{1,2|n}^2 = -\gamma_2 E_{1,2}^2 = \gamma_2 \partial_1 A_{1,2}^2 + \alpha_2\left(2\frac{\partial}{\partial x}\varphi^{1,2} + 2\beta_2\frac{\partial}{\partial v}\mathrm{V}^{1,2}\right) = 0.$$

Let us consider the equations of the first rank. Based on representation (1.10), knowing velocity field (4.21)-(4.23), one can find phase $\varphi_n^1$ of the wave function $\Psi_n^1$. We will search for the phase in the following form:

$$\langle v\rangle_{1|n}(x,t) = -2\alpha_1\frac{\partial}{\partial x}\varphi_n^1(x,t),\ \ \varphi_n^1(x,t) = -\frac{1}{2\alpha_1}\int_0^x \langle v\rangle_{1|n}(\overline{x},t)\,d\overline{x}, \qquad (4.30)$$

where, due to the one-dimensional spatial coordinate, the vortex field $\vec{A}_1^1$ is not considered. Note that representation (1.10) implies that phase (4.30) is defined up to a function independent of the coordinate. Due to the symmetry of the problem, the lower limit in integral (4.30) coincides with the coordinate origin. Taking the integral (4.30) of function $\langle v\rangle_{1|n}(\overline{x},t)$ defined by expressions (4.21)-(4.23) directly is difficult in an explicit form, so we will use numerical integration.

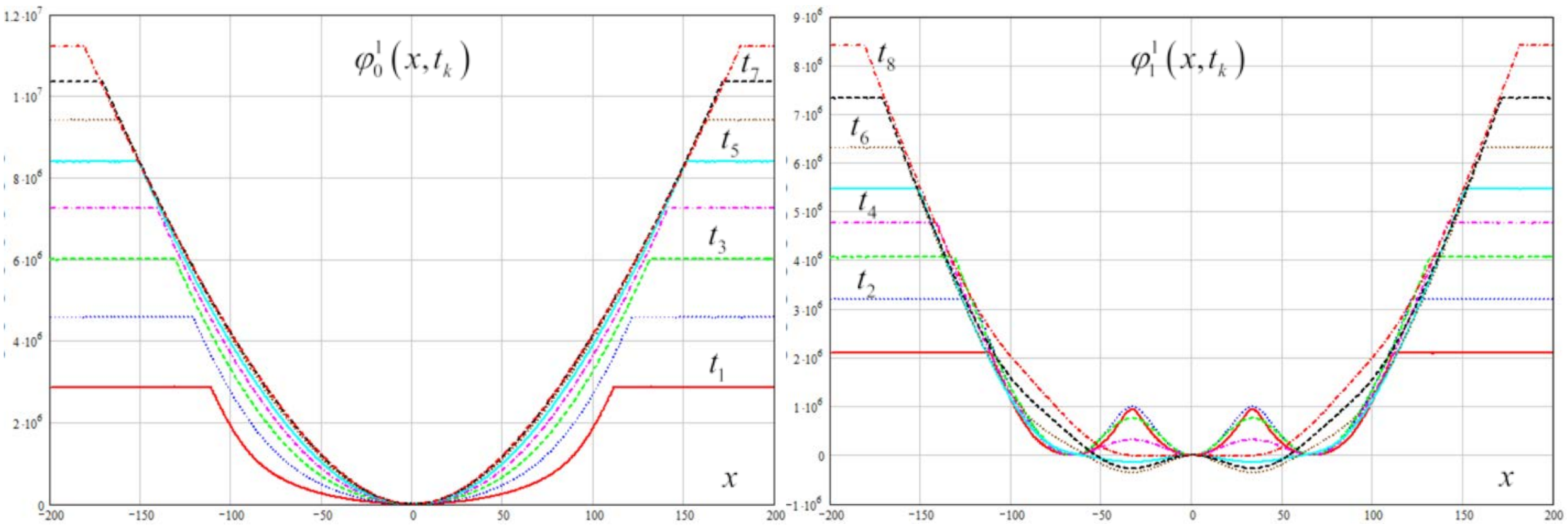

Fig. 8 Distribution of phase $\varphi_n^1$ at time instants $t_k$ .

Fig. 8 shows the distributions of phase $\varphi_n^1(x,t_k)$ for state $n=0$ (left) and state $n=1$ (right) at time instants $t_k=\{0.1;\ 0.2;\ 0.3;\ 0.4;\ 0.5;\ 0.6;\ 0.7;\ 0.8\}$. In the central region in Fig. 8, the behavior of phase $\varphi_1^1(x,t)$ is more complex than of phase $\varphi_0^1(x,t_k)$. Using the Hamilton-Jacobi equation of the first rank (2.14) and knowledge of the distribution of phase $\varphi_n^1$ (see Fig. 8), we can calculate the Hamilton function $\mathrm{H}_n^1$

$$\mathrm{H}_n^1=-\frac{1}{\beta_1}\partial_0\varphi_n^1. \tag{4.31}$$

Fig. 9 shows the distributions of the Hamilton function (4.31) for two states – $n=0$ (left) and $n=1$ (right) – at time instants $t_k=\{0.1;\ 0.2;\ 0.3;\ 0.4;\ 0.5;\ 0.6;\ 0.7;\ 0.8\}$. The comparison of distributions on the left and right in Fig. 9 gives a significant difference in the nature of the behavior of the Hamiltonians $\mathrm{H}_0^1$ and $\mathrm{H}_1^1$.

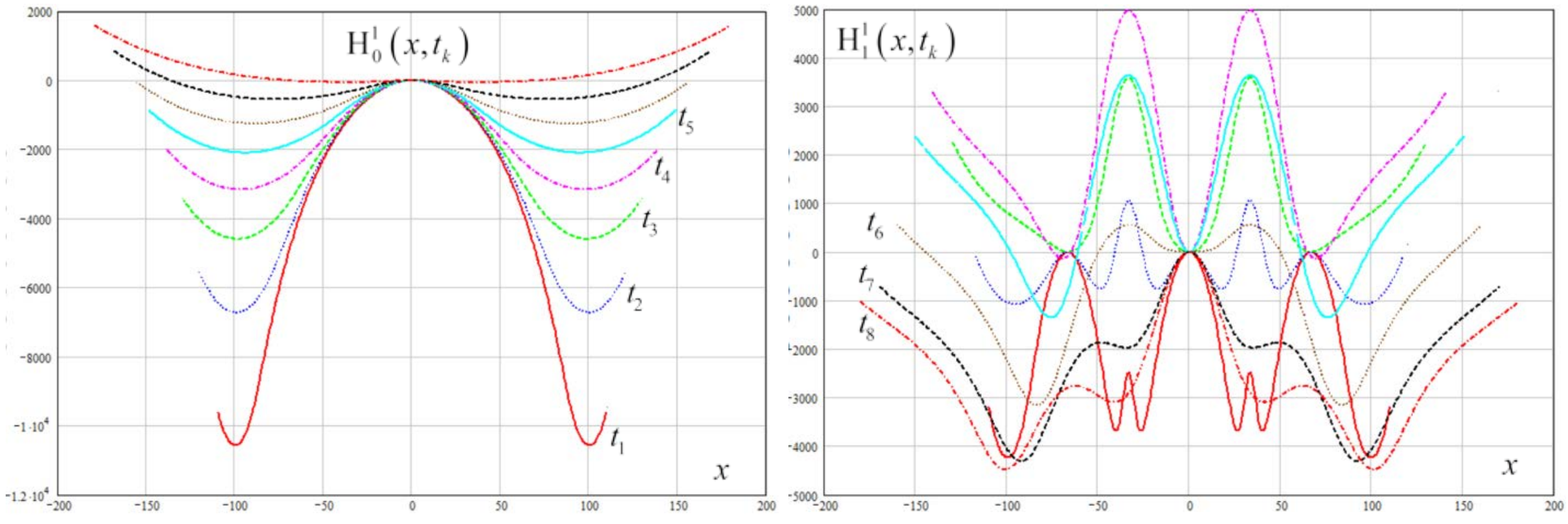

Fig. 9 Hamilton function $\mathrm{H}_n^1$ at time instants $t_k$ .

Taking into account (4.31) and expressions (4.21)-(4.23) for velocity $\langle v\rangle_{1|n}$, representation (2.1) allows us to find the classical potential $\mathrm{V}_n^1(x,t)$

$$\mathrm{V}_n^1 = \mathrm{H}_n^1 - \frac{m}{2}\left|\langle v\rangle_{1|n}\right|^2 . \quad (4.32)$$

Fig. 10 shows the distributions $\mathrm{V}_0^1$ (left) and $\mathrm{V}_1^1$ (right) at time instants $t_k$. In the central region, the distributions in Fig. 10 have a similar form to the Hamilton functions in Fig. 9. At the edges of the region, the behavior of potential (4.32) is determined by the significant contributions of the kinetic energy, which depends on distributions $\langle v\rangle_{1|n}$ (4.21)-(4.23) (see Figs. 6, 7).

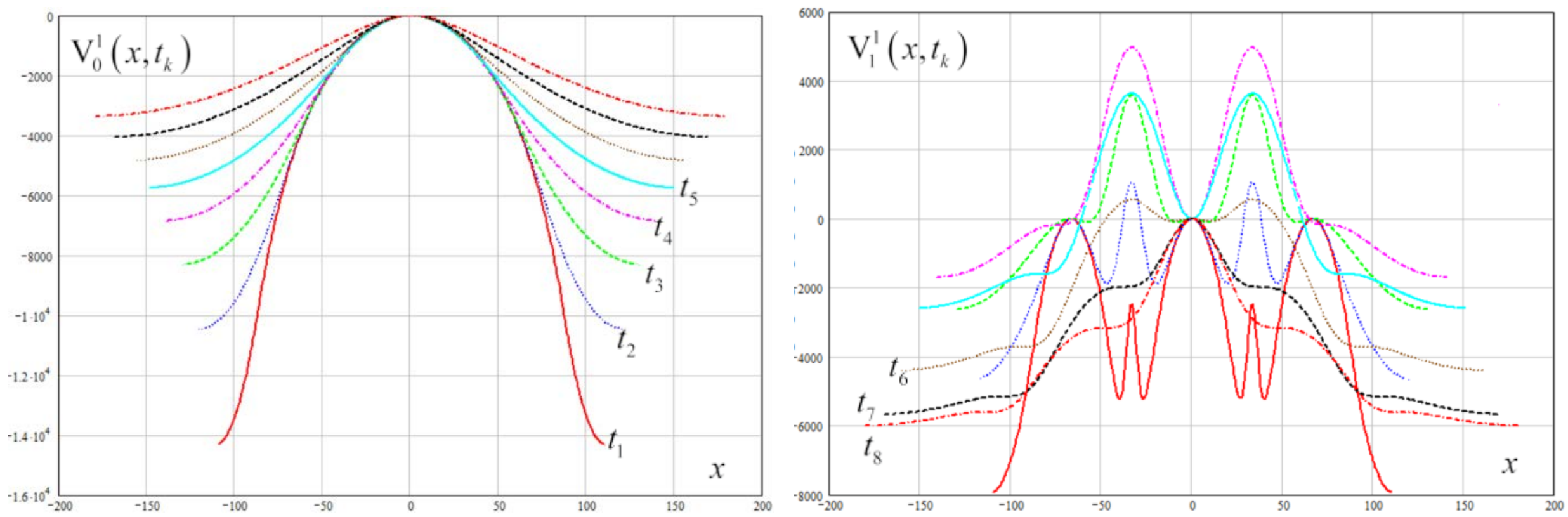

Fig. 10 Distributions of potential $\mathrm{V}_n^1$ at time instants $t_k$.

Potential $\mathrm{V}_n^1$ and velocity field $\langle v\rangle_{1|n}$ are directly related through the equation of motion of the first rank (3.13)

$$\hat{\pi}_1\langle v\rangle_{1|n} = F_1^1,$$

$$\left(\frac{\partial}{\partial t} + \langle v\rangle_{1|n}\frac{\partial}{\partial t}\right)\langle v\rangle_{1|n} = -\gamma_1 E_1^1 = \gamma_1\partial_0 A_1^1 + 2\alpha_1\beta_1\frac{\partial}{\partial x}\mathrm{V}_n^1 = 2\alpha_1\beta_1\frac{\partial}{\partial x}\mathrm{V}_n^1,$$

$$\hat{\pi}_1\langle v\rangle_{1|n} = \left(\frac{\partial}{\partial t} + \langle v\rangle_{1|n}\frac{\partial}{\partial t}\right)\langle v\rangle_{1|n} = -\frac{1}{m}\frac{\partial}{\partial x}\mathrm{V}_n^1, \quad (4.33)$$

where, due to the one-dimensionality, there is no vortex velocity component. In the hydrodynamic approximation, equation (4.33) can be represented as (i.7)

$$\hat{\pi}_1\left\langle v_{1|n}\right\rangle - \left\langle \dot{v}_{1|n}\right\rangle = -\frac{1}{f_n^1}\frac{\partial P_n^1}{\partial x}, \quad (4.34)$$

where, according to [1], the difference between derivative of the mean $\hat{\pi}_1\left\langle v_{1|n}\right\rangle$ and mean of the derivative $\left\langle \dot{v}_{1|n}\right\rangle$ is determined by pressure $P_n^1$::

$$P_n^1(x,t) = \int_{(\infty)} f_n^{1,2}(x,v,t)\left[v - \langle v\rangle_{1|n}(x,t)\right]^2 dv. \quad (4.35)$$

Note that in this case (4.35) the pressure $P_n^1$ corresponds to the distribution of the velocity field dispersion along the coordinate. Taking into account expressions (4.24), (4.27) and (4.33), equation (4.34) takes the form:

$$\frac{\partial}{\partial x}\mathrm{V}_n^1 = \frac{\partial}{\partial x}\left(U^1 + \mathrm{Q}_n^1\right) = \frac{m}{f_n^1}\frac{\partial P_n^1}{\partial x}. \tag{4.36}$$

Thus, pressure $P_n^1$ is determined by quantum potential $\mathrm{Q}_n^1$ and the Schrödinger potential $U^1$. The value $P_n^1$ can be found explicitly by expressions (4.21)-(4.23) and through the average value $\left\langle v^2 \right\rangle_{1|n}$:

$$P_n^1(x,t) = f_n^1(x,t)\left[\left\langle v^2 \right\rangle_{1|n}(x,t) - \left\langle v \right\rangle_{1|n}^2(x,t)\right], \tag{4.37}$$

$$f_n^1(x,t)\left\langle v^2 \right\rangle_{1|n}(x,t) = \int_{(\infty)} f_n^{1,2}(x,v,t)\,v^2 dv. \tag{4.38}$$

The explicit expression for the average value $\left\langle v^2 \right\rangle_{1|n}$ has the form:

At $t=0$

$$f_n^1(x,0)\left\langle v^2 \right\rangle_{1|n}(x,0) = \frac{\Delta v^2}{6\Delta x}\cos^2 \vartheta_n x. \tag{4.39}$$

At $0 < t \le \dfrac{\Delta x}{\Delta v}$

$$f_n^1\left\langle v^2 \right\rangle_{1|n} = \begin{cases} \bar{h}_n(x,t), & -t\Delta v < 2x + \Delta x \le t\Delta v, \\ \tilde{h}_n(x,t), & t\Delta v - \Delta x < 2x \le -t\Delta v + \Delta x, \\ \bar{h}_n(-x,t), & -t\Delta v < 2x - \Delta x \le t\Delta v, \\ 0, & otherwise. \end{cases} \tag{4.40}$$

At $t \ge \dfrac{\Delta x}{\Delta v}$

$$f_n^1\left\langle v^2 \right\rangle_{1|n} = \begin{cases} \bar{h}_n(x,t), & -\Delta x < 2x + t\Delta v \le \Delta x, \\ \dfrac{x^2}{2t^3\Delta v} - \dfrac{1}{2\vartheta_n^2 t^3 \Delta v}, & \Delta x - t\Delta v < 2x \le t\Delta v - \Delta x, \\ \bar{h}_n(-x,t), & -\Delta x < 2x - t\Delta v \le \Delta x, \\ 0, & otherwise. \end{cases} \tag{4.41}$$

where

$$\bar{h}_n(x,t) \overset{\det}{=} \frac{(2x+\Delta x)^3 + t^3\Delta v^3}{24t^3\Delta x\Delta v} - \frac{2x+\Delta x}{4\vartheta_n^2 t^3 \Delta x \Delta v} + \frac{1}{4\vartheta_n^2 t^2 \Delta x}\cos\vartheta_n(2x+t\Delta v) +$$
$$+\frac{\Delta v}{8\vartheta_n t\Delta x}\left(1-\frac{2}{\vartheta_n^2 t^2 \Delta v^2}\right)\sin\vartheta_n(2x+t\Delta v),$$
$$\tilde{h}_n(x,t) \overset{\det}{=} \frac{\Delta v^2}{12\Delta x} + \frac{\Delta v\cos(2\vartheta_n x)}{4\vartheta_n t\Delta x}\left[\left(1-\frac{2}{\vartheta_n^2 t^2\Delta v^2}\right)\sin(\vartheta_n t\Delta v) + \frac{2}{\vartheta_n t\Delta v}\cos(\vartheta_n t\Delta v)\right].$$

Using expressions (4.39)-(4.41) and (4.37), in Fig. 11 the distributions for pressure $P_n^1$ are shown at time instants $t_k = \{0.1;\ 0.2;\ 0.3;\ 0.4;\ 0.5;\ 0.6;\ 0.7;\ 0.8;\ 0.9\}$ at $n=0$ the (left) and at $n=1$ (right). For state $n=0$, pressure $P_0^1$ has a monotonous decrease in amplitude and blurring along the coordinate over time (see Fig. 11 on the left). The behavior of pressure $P_1^1$ (see Fig. 11 on the right) is characterized by the presence of amplitude oscillations at the initial instants of time. At long times $t \geq \frac{\Delta x}{\Delta v}$, the behavior of pressure $P_1^1$ is similar to the behavior of pressure $P_0^1$, that is, a decrease in amplitude and blurring along the coordinate is observed.

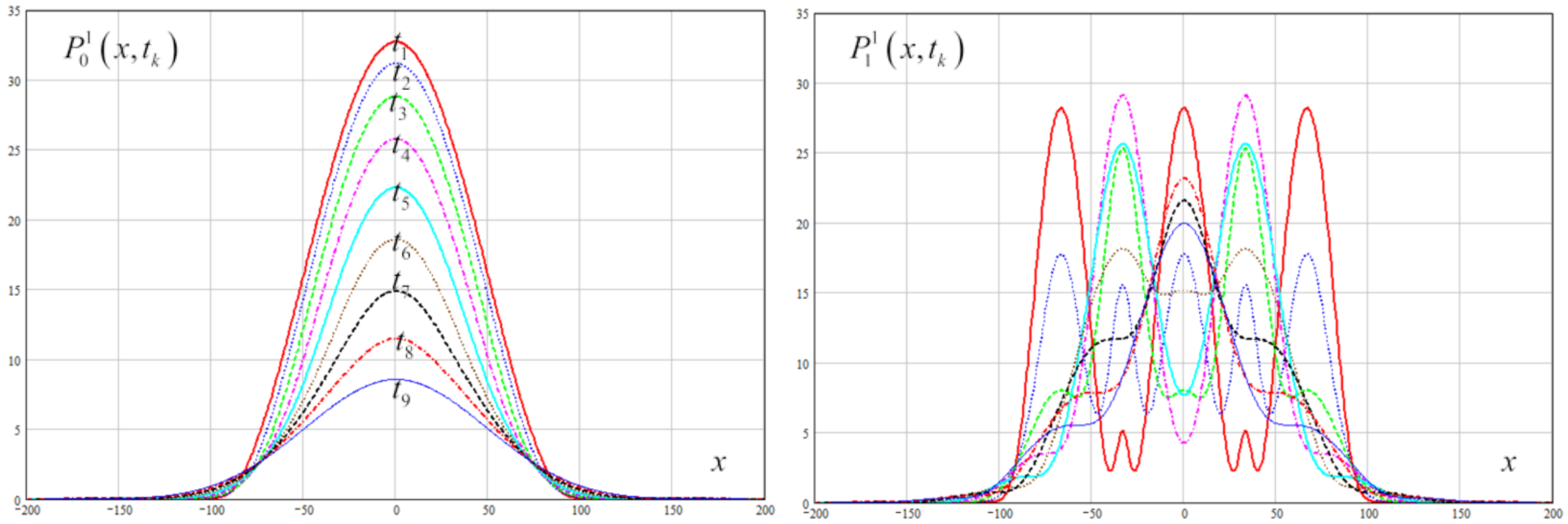

Fig. 11 Distribution of pressure $P_n^1$ at time instants $t_k$.

Fig. 12 shows the evolution of the force distribution (4.33), (4.36) $-\frac{\partial \mathrm{V}_n^1}{\partial x} = -\frac{m}{f_n^1}\frac{\partial P_n^1}{\partial x}$. Value $-\frac{m}{f_n^1}\frac{\partial P_n^1}{\partial x}$ according to expressions (4.16)-(4.18), (4.21)-(4.23) and (4.39)-(4.41) can be found explicitly. The left side of Fig. 12 shows the distribution of forces $-\frac{m}{f_n^1}\frac{\partial P_n^1}{\partial x}$ within the system at time instants $t_k$ for state $n=0$, and the right side – for state $n=1$. Fig. 12 (left) the force distribution indicates the expansion process. Indeed, as shown in Fig. 6, velocity field $\langle v\rangle_{1|0}$ indicates the movement of the flow of probabilities from the center of the system to its periphery. A similar but more complex process is observed for the state $n=1$ (see Fig. 12 on the right and Fig. 7). It was said above that, in a sense, state $n=1$ is represented as a superposition of three states with number $n=0$. As a result, Fig. 12 (on the right) shows the existence of three regions separated by potential barriers (see Fig. 10 on the right), in each of which processes characteristic of state $n=0$ occur.

Note that direct numerical differentiation of the potential $V_n^1$ (4.32) (see Fig. 10) with respect to the coordinate gives the result shown in Fig. 12. Thus, analytical expression (4.36) has a numerical confirmation.

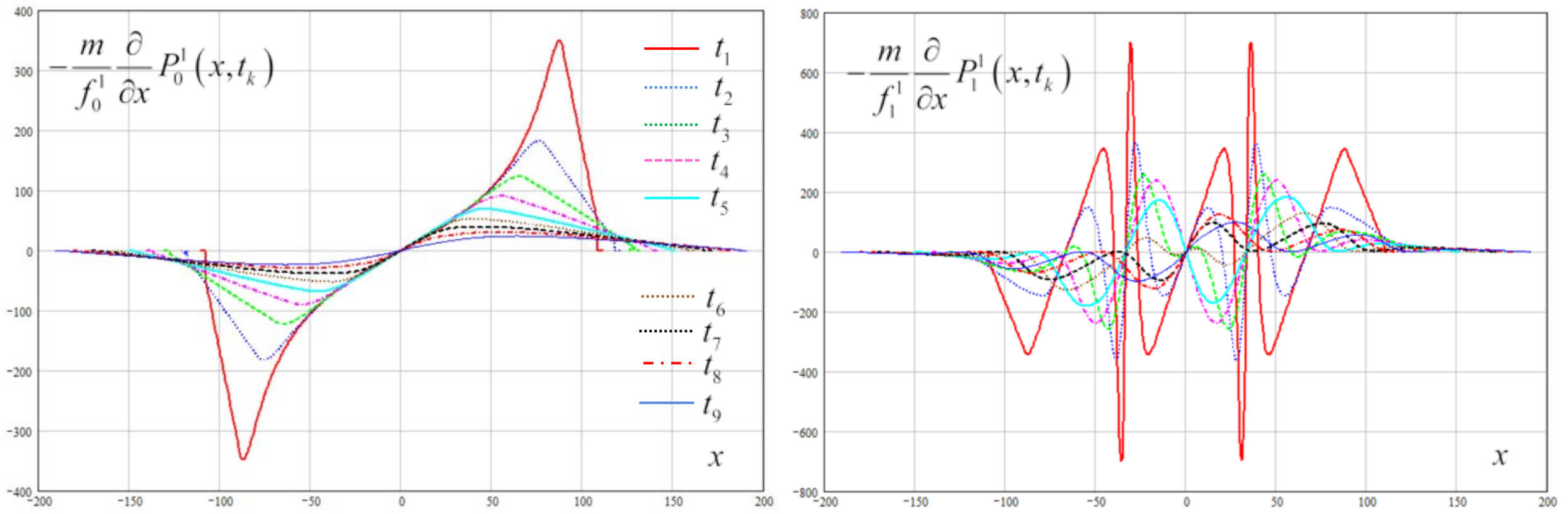

Fig. 12 Distribution of force $-\frac{m}{f_n^1}\frac{\partial P_n^1}{\partial x}$ at time instants $t_k$.

The quantum potential $Q_n^1$, according to definition (2.2), is explicitly found from expressions (4.16)-(4.18)

$$Q_n^1 = \frac{\alpha_1}{\beta_1}\frac{1}{\sqrt{f_n^1}}\frac{\partial^2}{\partial x^2}\sqrt{f_n^1}. \tag{4.42}$$

Due to the smallness of coefficient $\frac{\alpha_1}{\beta_1} = -\frac{\hbar^2}{2m}$, value (4.42) makes an insignificant contribution to potential $V_n^1$ (4.32), which leads to the closeness of the Schrödinger potential $U_n^1$ to the distributions $V_n^1$ shown in Fig. 10. The wave function $\Psi_n^1$ is represented as

$$\Psi_n^1(x,t) = \sqrt{f_n^1(x,t)}\exp\left[i\varphi_n^1(x,t)\right], \tag{4.43}$$

where distribution function $f_n^1$ (4.16)-(4.18) satisfies the Vlasov dispersion equation of the first rank (i.3)

$$\frac{\partial}{\partial t}f_n^1(x,t) + \frac{\partial}{\partial x}\left[f_n^1(x,t)\langle v\rangle_{1|n}(x,t)\right] = 0. \tag{4.44}$$

Wave function (4.43) satisfies the Schrödinger equation of the first rank (1.11)

$$i\hbar\frac{\partial}{\partial t}\Psi_n^1 = -\frac{\hbar^2}{2m}\frac{\partial^2}{\partial x^2}\Psi_n^1 + U_n^1\Psi_n^1, \tag{4.45}$$

where, as noted above, each wave function $\Psi_n^1$ has its own potential $U_n^1$.

According to expression (4.24) and representation (1.10), phase $\varphi_n^2$ does not depend on the velocity, that is, it can only depend on time. Indeed, we multiply expression (1.10) for acceleration $\langle \dot{v} \rangle_{1,2|n}$ by $f_n^{1,2}$ and integrate it over the coordinate, we obtain:

$$\langle \dot{v} \rangle_{1,2|n} = -2\alpha_2 \frac{\partial}{\partial v}\varphi_n^{1,2} = 0,$$

$$f_n^2 \langle \dot{v} \rangle_{2|n} = \int f_n^{1,2} \langle \dot{v} \rangle_{1,2|n} dx = -2\alpha_2 \int f_n^{1,2} \frac{\partial}{\partial v}\varphi_n^{1,2} dx = -2\alpha_2 \frac{\partial}{\partial v}\varphi_n^{1,2} \int f_n^{1,2} dx = -2\alpha_2 f_n^2 \frac{\partial}{\partial v}\varphi_n^{1,2},$$

$$\langle \dot{v} \rangle_{2|n} = -2\alpha_2 \frac{\partial}{\partial v}\varphi_n^{1,2}, \tag{4.45}$$

where it is taken into account that $\varphi_n^{1,2}$ depends only on time. On the other hand, due to (1.10), the representation is true:

$$\langle \dot{v} \rangle_{2|n} = -2\alpha_2 \frac{\partial}{\partial v}\varphi_n^2. \tag{4.46}$$

Comparing expressions (4.45) and (4.46), we obtain

$$\frac{\partial}{\partial v}\varphi_n^{1,2} = \frac{\partial}{\partial v}\varphi_n^2. \tag{4.47}$$

As function $\varphi_n^2$, one can take function $\varphi_n^{1,2}$ or any other function that depends only on time. Hamilton-Jacobi equation (2.14) of the first rank for phase $\varphi_n^2$ will have the form

$$-\hbar_2 \partial_0 \varphi_n^2 = \mathrm{H}_n^2 = \frac{m}{2}\left| \langle \dot{v} \rangle_{2|n} \right|^2 + \mathrm{V}_n^2 = \mathrm{V}_n^2 = U_n^2, \tag{4.48}$$

since $\mathrm{Q}_n^2 = \frac{\alpha_2}{\beta_2} \frac{1}{\sqrt{f_n^2}} \frac{\partial^2}{\partial v^2} \sqrt{f_n^2} = 0$. As a result, wave function $\Psi_n^2(v,t)$ will take the form:

$$\Psi_n^2(v,t) = \frac{1}{\sqrt{\Delta v}} \mathrm{rect}\left(\frac{v}{\Delta v}\right) \exp\left[i\varphi_n^2(t)\right], \tag{4.49}$$

where $\varphi_n^2 = \varphi_n^2(t)$ is some function of time and function $\mathrm{rect}(x)$ is defined as:

$$\mathrm{rect}(x) = \begin{cases} 0, \ if \ |x| > 1/2, \\ 1/2, \ if \ |x| = 1/2, \\ 1, \ if \ |x| < 1/2. \end{cases}$$

Function (4.49) satisfies the equation:

$$i\hbar_2 \frac{\partial}{\partial t}\Psi_n^2 = -\frac{\hbar_2^2}{2m} \frac{\partial^2}{\partial v^2}\Psi_n^2 + U_n^2 \Psi_n^2, \tag{4.50}$$

And probability density $f_n^2$ and acceleration flux (4.49) are interrelated to each other by the Vlasov equation of the first rank

$$\frac{\partial}{\partial t} f_n^2(v,t) + \frac{\partial}{\partial v}\left[ f_n^2(v,t)\langle \dot{v}\rangle_{2|n}(v,t)\right] = 0. \tag{4.51}$$

The inverse Fourier transformation of function (4.48) with respect to variable $v$ will give the expression

$$\tilde{\Psi}_n^2(x,t) \overset{\det}{=} \mathcal{F}^{-1}\left[\Psi_n^2(v,t)\right] = \frac{m}{\sqrt{2\pi\hbar}}\int_{-\infty}^{+\infty}\Psi_n^2(v,t)e^{i\frac{xmv}{\hbar}}dv,$$

$$\tilde{\Psi}_n^2(x,t) = m\sqrt{\frac{\Delta v}{2\pi\hbar}}\operatorname{sinc}\left(\frac{m\Delta v}{2\hbar}x\right)e^{i\varphi_n^2}. \tag{4.52}$$

**Remark**

Note that function (4.52) differs from function (4.43). Such a result is expected, since in the construction of the dispersion chain of equations of quantum mechanics, there was no condition of connectedness of functions $\Psi_n^1$ and $\Psi_n^2$ through the integral Fourier transformation imposed. The solution of equation (4.1) was chosen arbitrarily without imposing the conditions of «Fourier connection».

As an example, when the «Fourier connection» condition is satisfied, one can consider the wave function:

$$\Psi^{1,2}(x,v,t) = \sqrt{\frac{m}{\pi\hbar}}\cdot\exp\left[-\frac{1}{\hbar\omega}\left(\frac{mv^2}{2}+\frac{m\omega^2x^2}{2}\right) - i\left(\frac{m\omega^2}{\hbar_2}xv + \frac{\mathscr{E}_2}{\hbar_2}t\right)\right], \tag{4.53}$$

which satisfies equation (4.1) with the potential

$$U^{1,2} = \mathscr{E}^{1,2} - \frac{\hbar_2^2}{2\hbar\omega} + m\omega^2\left(1+\frac{\hbar_2^2}{2\hbar^2\omega^4}\right)v^2 - \frac{1}{2}m\omega^4x^2, \tag{4.54}$$

where $\mathscr{E}^{1,2} = \frac{\hbar_2\omega_2}{2}$ is constant. The Wigner function of ground state ($s=0$) of a quantum harmonic oscillator (1.21) corresponds to wave function (4.53):

$$f^{1,2}(x,v,t) = \left|\Psi^{1,2}\right|^2 = mW(x,mv) = \frac{m}{\pi\hbar}\cdot\exp\left[-\frac{m}{\hbar\omega}\left(v^2+\omega^2x^2\right)\right]. \tag{4.55}$$

Integration of function (4.55) over the coordinate and momentum spaces will give probability densities (1.21) $f^2(v)$ and $f^1(x)$, to which wave functions $\Psi^1(x)$ and $\Psi^2(v)$ correspond, related by the Fourier transformation $\Psi^2(v) = \mathcal{F}\left[\Psi^1(x)\right]$.

As a result, the second-rank Schrödinger equation (4.1) is satisfied by the wave functions (4.11) and (4.53) with potentials (4.4) and (4.54), respectively. In this case, wave functions of the first rank (4.43) and (4.49), which do not have the «Fourier connection», correspond to wave function (4.11). The wave functions of the first rank – the coordinate and momentum representations of the ground state of a harmonic oscillator having the «Fourier connection», correspond to wave function of the second rank (4.53).

Despite the fact that functions (4.43) and (4.49) are not related by the Fourier transformation, they remain the solutions of the first rank Schrödinger equations (4.45) and (4.51). Therefore, it is possible to construct the Fourier images and the Wigner functions for them.

## Conclusions

In this paper, an extended theory of quantum mechanics of high kinematical values is constructed on the basis of the dispersion chain of the Vlasov equations. A chain of analogues of the Schrödinger, Hamilton-Jacobi equations, Hamiltonian and Lagrangian representations, Legendre transformations, Maxwell equations and equations of motion are obtained. The presented theory of the dispersion chain of equations of quantum mechanics gives a clear interpretation and shows a deep interrelation between various branches of theoretical physics: classical mechanics, statistical physics, continuum mechanics and quantum mechanics.

In fact, the entire proposed theory is based only on the concept of the distribution function and the first principle – the law of conservation of probabilities. Surprisingly, it turns out that these concepts themselves are enough to obtain the basic fundamental equations of various branches of physics. At the same time, the equations of classical and quantum physics naturally bypassing the «limit transitions» are interrelated.

## Acknowledgements

This research has been supported by the Interdisciplinary Scientific and Educational School of Moscow University «Photonic and Quantum Technologies. Digital Medicine».

## Appendix A

### *Proof of Theorem 1*

Since, according to the condition of the theorem, the probability density function $f^{n_1 \dots n_R}$ is positive, the representation is true:

$$f^{n_1 \dots n_R} = \left| \Psi^{n_1 \dots n_R} \right|^2 = \Psi^{n_1 \dots n_R} \bar{\Psi}^{n_1 \dots n_R} \geq 0, \qquad \varphi^{n_1 \dots n_R} = \arg \Psi^{n_1 \dots n_R}, \tag{A.1}$$

where $\Psi^{n_1 \dots n_R} \in \mathbb{C}$. Due to freedom in choosing phase $\varphi^{n_1 \dots n_R}$ by analogy with [5, 2], we set it proportional to scalar potential $\Phi^{n_1 \dots n_R}$ (1.1):

$$\Phi^{n_1 \dots n_R} = 2\varphi^{n_1 \dots n_R} + 2\pi k, \;\; k \in \mathbb{Z}. \tag{A.2}$$

Taking into account the interrelation of (A.2) and (A.1), decomposition (1.1) takes the form:

$$\left\langle \vec{\xi}^{\,\ell+1} \right\rangle_{n_1 \dots n_R} = i\alpha_\ell \nabla_{\xi^\ell} \operatorname{Ln} \frac{\Psi^{n_1 \dots n_R}}{\bar{\Psi}^{n_1 \dots n_R}} + \gamma_\ell \vec{A}^{\ell}_{n_1 \dots n_R}. \tag{A.3}$$

Substituting representations (A.3) and (A.1) into the equations of the first rank of the Vlasov dispersion chain (i.22), we obtain:

$$\bar{\Psi}^n \frac{\partial \Psi^n}{\partial t} + \Psi^n \frac{\partial \bar{\Psi}^n}{\partial t} + i\alpha_n \operatorname{div}_{\xi^n} \left[ \bar{\Psi}^n \nabla_{\xi^n} \Psi^n - \Psi^n \nabla_{\xi^n} \bar{\Psi}^n \right] + \gamma_n \operatorname{div}_{\xi^n} \left[ \Psi^n \bar{\Psi}^n \vec{A}_n^n \right] = 0,$$

or, adding and subtracting the summand $\frac{i}{4\alpha_n} \left| \gamma_n \vec{A}_n^n \right|^2 \left| \Psi^n \right|^2$, we obtain

$$\begin{gathered} \bar{\Psi}^n \left[ \frac{\partial}{\partial t} + i\alpha_n \Delta_{\xi^n} + \gamma_n \vec{A}_n^n \nabla_{\xi^n} - \frac{i}{4\alpha_n} \left| \gamma_n \vec{A}_n^n \right|^2 \right] \Psi^n + \\ + \Psi^n \left[ \frac{\partial}{\partial t} - i\alpha_n \Delta_{\xi^n} + \gamma_n \vec{A}_n^n \nabla_{\xi^n} + \frac{i}{4\alpha_n} \left| \gamma_n \vec{A}_n^n \right|^2 \right] \bar{\Psi}^n = 0, \end{gathered} \tag{A.4}$$

where the condition $\operatorname{div}_{\xi^\ell} \vec{A}_{n_1 \ldots n_R}^\ell = 0$ and $\ell = n_1$, $n = n_1$ is taken into account. Expression (A.4) can be rewritten in a compact form:

$$\Lambda_n + \bar{\Lambda}_n = \operatorname{Re} \Lambda_n = 0, \tag{A.5}$$

$$\Lambda_n \stackrel{\text{det}}{=} \bar{\Psi}^n L_n \Psi^n, \qquad L_n \stackrel{\text{det}}{=} \frac{\partial}{\partial t} + i\alpha_n \Delta_{\xi^n} + \gamma_n \vec{A}_n^n \nabla_{\xi^n} - \frac{i}{4\alpha_n} \left| \gamma_n \vec{A}_n^n \right|^2. \tag{A.6}$$

It follows from expression (A.5) that value $\Lambda_n = iu_n$, where $u_n \in \mathbb{R}$, i.e.

$$\bar{\Psi}^n L_n \Psi^n = iu_n \;\Rightarrow\; L_n \Psi^n = -i\beta_n U^n \Psi^n, \tag{A.7}$$

where $-\beta_n U^n \stackrel{\text{det}}{=} u_n \Big/ \left| \Psi^n \right|^2$ is a real value. Taking into account (A.6), equation (A.7) takes the form:

$$\frac{i}{\beta_n} \frac{\partial \Psi^n}{\partial t} = \frac{\alpha_n}{\beta_n} \Delta_{\xi^n} \Psi^n - i \frac{\gamma_n}{\beta_n} \vec{A}_n^n \nabla_{\xi^n} \Psi^n - \frac{1}{4\alpha_n \beta_n} \left| \gamma_n \vec{A}_n^n \right|^2 \Psi^n + U^n \Psi^n. \tag{A.8}$$

Using the «momentum» differential operator $\hat{\mathrm{p}}_n$, equation (A.8) takes the form:

$$\frac{i}{\beta_n} \frac{\partial \Psi^n}{\partial t} = -\alpha_n \beta_n \left( \hat{\mathrm{p}}_n^2 - \frac{\gamma_n}{\alpha_n \beta_n} \vec{A}_n^n \hat{\mathrm{p}}_n \right) \Psi^n - \frac{1}{4\alpha_n \beta_n} \left| \gamma_n \vec{A}_n^n \right|^2 \Psi^n + U^n \Psi^n. \tag{A.9}$$

The relation is true:

$$\left( \hat{\mathrm{p}}^2 - \lambda \vec{A} \hat{\mathrm{p}} + \frac{\lambda^2}{4} \left| \vec{A} \right|^2 \right) \Psi = \left( \hat{\mathrm{p}} - \frac{\lambda}{2} \vec{A} \right)^2 \Psi, \tag{A.10}$$

where $\lambda$ is constant. Taking into account expression (A.10), equation (A.9) takes the form:

$$\frac{i}{\beta_n} \frac{\partial \Psi^n}{\partial t} = -\alpha_n \beta_n \left( \hat{\mathrm{p}}_n - \frac{\gamma_n}{2\alpha_n \beta_n} \vec{A}_n^n \right)^2 \Psi^n + U^n \Psi^n.$$

For equations of the second rank (i.22) belonging to the first group

$$\frac{\partial f^{n,n+1}}{\partial t}+\mathrm{div}_{\xi^n}\left[f^{n,n+1}\vec{\xi}^{n+1}\right]+\mathrm{div}_{\xi^{n+1}}\left[f^{n,n+1}\left\langle\vec{\xi}^{n+2}\right\rangle_{n,n+1}\right]=0, \qquad (A.11)$$

we obtain

$$\begin{gathered}\bar{\Psi}^{n,n+1}\left[\frac{\partial}{\partial t}+\vec{\xi}^{n+1}\nabla_{\xi^n}+i\alpha_{n+1}\Delta_{\xi^{n+1}}+\gamma_{n+1}\vec{A}^{n+1}_{n,n+1}\nabla_{\xi^{n+1}}-\frac{i}{4\alpha_{n+1}}\left|\gamma_{n+1}\vec{A}^{n+1}_{n,n+1}\right|^2\right]\Psi^{n,n+1}+\\+\Psi^{n,n+1}\left[\frac{\partial}{\partial t}+\vec{\xi}^{n+1}\nabla_{\xi^n}-i\alpha_{n+1}\Delta_{\xi^{n+1}}+\gamma_{n+1}\vec{A}^{n+1}_{n,n+1}\nabla_{\xi^{n+1}}+\frac{i}{4\alpha_{n+1}}\left|\gamma_{n+1}\vec{A}^{n+1}_{n,n+1}\right|^2\right]\bar{\Psi}^{n,n+1}=0,\end{gathered} \qquad (A.12)$$

where due to (A.3) is taken into account that $\left\langle\vec{\xi}^{n+2}\right\rangle_{n,n+1}=i\alpha_{n+1}\nabla_{\xi^{n+1}}\,\mathrm{Ln}\frac{\Psi^{n,n+1}}{\bar{\Psi}^{n,n+1}}+\gamma_{n+1}\vec{A}^{n+1}_{n,n+1}$. Using operators of the second rank $\Lambda_{n,n+1}$ and $L_{n,n+1}$ by analogy with transformations (A.5)-(A.7), we obtain

$$\begin{gathered}\Lambda_{n,n+1}+\bar{\Lambda}_{n,n+1}=\mathrm{Re}\,\Lambda_{n,n+1}=0,\\ \Lambda_{n,n+1}\overset{\det}{=}\bar{\Psi}^{n,n+1}L_{n,n+1}\Psi^{n,n+1},\\ L_{n,n+1}\overset{\det}{=}\frac{\partial}{\partial t}+\vec{\xi}^{n+1}\nabla_{\xi^n}+i\alpha_{n+1}\Delta_{\xi^{n+1}}+\gamma_{n+1}\vec{A}^{n+1}_{n,n+1}\nabla_{\xi^{n+1}}-\frac{i}{4\alpha_{n+1}}\left|\gamma_{n+1}\vec{A}^{n+1}_{n,n+1}\right|^2.\end{gathered} \qquad (A.13)$$

From expression (A.13) it follows that value $\Lambda_{n,n+1}=iu_{n,n+1}$, where $u_{n,n+1}\in\mathbb{R}$, i.e.

$$\bar{\Psi}^{n,n+1}L_{n,n+1}\Psi^{n,n+1}=iu_{n,n+1}\;\Rightarrow\;L_{n,n+1}\Psi^{n,n+1}=-i\beta_{n+1}U^{n,n+1}\Psi^{n,n+1}, \qquad (A.14)$$

where $-\beta_{n+1}U^{n,n+1}\overset{\det}{=}u_{n,n+1}\Big/\left|\Psi^{n,n+1}\right|^2$ is a real value. Given expression (A.13), equation (A.14) takes the form:

$$\begin{gathered}\frac{i}{\beta_{n+1}}\frac{\partial\Psi^{n,n+1}}{\partial t}=-\alpha_{n+1}\beta_{n+1}\left[\hat{\mathrm{p}}^2_{n+1}-\frac{\gamma_{n+1}}{\alpha_{n+1}\beta_{n+1}}\vec{A}^{n+1}_{n,n+1}\,\hat{\mathrm{p}}_{n+1}\right]\Psi^{n,n+1}-\frac{1}{4\alpha_{n+1}\beta_{n+1}}\left|\gamma_{n+1}\vec{A}^{n+1}_{n,n+1}\right|^2\Psi^{n,n+1}+\\+\frac{\beta_n}{\beta_{n+1}}\vec{\xi}^{n+1}\,\hat{\mathrm{p}}_n\,\Psi^{n,n+1}+U^{n,n+1}\Psi^{n,n+1}.\end{gathered} \qquad (A.18)$$

Taking into account expression (A.10), equation (A.18) takes the form:

$$\frac{i}{\beta_{n+1}}\left(\frac{\partial}{\partial t}+\vec{\xi}^{n+1}\nabla_{\xi^n}\right)\Psi^{n,n+1}=-\alpha_{n+1}\beta_{n+1}\left(\hat{\mathrm{p}}_{n+1}-\frac{\gamma_{n+1}}{2\alpha_{n+1}\beta_{n+1}}\vec{A}^{n+1}_{n,n+1}\right)^2\Psi^{n,n+1}+U^{n,n+1}\Psi^{n,n+1},$$

For equations of the second rank from the second group,

$$\frac{\partial f^{n,n+k}}{\partial t}+\mathrm{div}_{\xi^n}\left[f^{n,n+k}\left\langle\vec{\xi}^{n+1}\right\rangle_{n,n+k}\right]+\mathrm{div}_{\xi^{n+k}}\left[f^{n,n+k}\left\langle\vec{\xi}^{n+k+1}\right\rangle_{n,n+k}\right]=0, \qquad (A.19)$$

where

$$\left\langle \vec{\xi}^{\,n+1} \right\rangle_{n,n+k} = i\alpha_n \nabla_{\xi^n} \operatorname{Ln} \frac{\Psi^{n,n+k}}{\bar{\Psi}^{n,n+k}} + \gamma_n \vec{A}^{n}_{n,n+k},$$
$$\left\langle \vec{\xi}^{\,n+k+1} \right\rangle_{n,n+k} = i\alpha_{n+k} \nabla_{\xi^{n+k}} \operatorname{Ln} \frac{\Psi^{n,n+k}}{\bar{\Psi}^{n,n+k}} + \gamma_{n+k} \vec{A}^{n+k}_{n,n+k}, \tag{A.20}$$

we obtain

$$\begin{aligned}
&\bar{\Psi}^{n,n+k} \frac{\partial \Psi^{n,n+k}}{\partial t} + \Psi^{n,n+k} \frac{\partial \bar{\Psi}^{n,n+k}}{\partial t} + \\
&+ i\alpha_n \operatorname{div}_{\xi^n} \left[ \bar{\Psi}^{n,n+k} \Psi^{n,n+k} \nabla_{\xi^n} \operatorname{Ln} \frac{\Psi^{n,n+k}}{\bar{\Psi}^{n,n+k}} \right] + \gamma_n \operatorname{div}_{\xi^n} \left[ \bar{\Psi}^{n,n+k} \Psi^{n,n+k} \vec{A}^{n}_{n,n+k} \right] + \\
&+ i\alpha_{n+k} \operatorname{div}_{\xi^{n+k}} \left[ \bar{\Psi}^{n,n+k} \Psi^{n,n+k} \nabla_{\xi^{n+k}} \operatorname{Ln} \frac{\Psi^{n,n+k}}{\bar{\Psi}^{n,n+k}} \right] + \gamma_{n+k} \operatorname{div}_{\xi^{n+k}} \left[ \bar{\Psi}^{n,n+k} \Psi^{n,n+k} \vec{A}^{n+k}_{n,n+k} \right] = 0,
\end{aligned} \tag{A.21}$$

Adding and subtracting in expression (A.21) two terms $\frac{i}{4\alpha_n} \left| \gamma_n \vec{A}^{n}_{n,n+k} \right|^2 \left| \Psi^{n,n+k} \right|^2$ and $\frac{i}{4\alpha_{n+k}} \left| \gamma_{n+k} \vec{A}^{n+k}_{n,n+k} \right|^2 \left| \Psi^{n,n+k} \right|^2$, we obtain

$$\begin{aligned}
&\bar{\Psi}^{n,n+k} \left[ \frac{\partial}{\partial t} + i\alpha_n \Delta_{\xi^n} + \gamma_n \vec{A}^{n}_{n,n+k} \nabla_{\xi^n} - \frac{i}{4\alpha_n} \left| \gamma_n \vec{A}^{n}_{n,n+k} \right|^2 + \right. \\
&\qquad \left. + i\alpha_{n+k} \Delta_{\xi^{n+k}} + \gamma_{n+k} \vec{A}^{n+k}_{n,n+k} \nabla_{\xi^{n+k}} - \frac{i}{4\alpha_{n+k}} \left| \gamma_{n+k} \vec{A}^{n+k}_{n,n+k} \right|^2 \right] \Psi^{n,n+k} + \\
&+ \Psi^{n,n+k} \left[ \frac{\partial}{\partial t} - i\alpha_n \Delta_{\xi^n} + \gamma_n \vec{A}^{n}_{n,n+k} \nabla_{\xi^n} + \frac{i}{4\alpha_n} \left| \gamma_n \vec{A}^{n}_{n,n+k} \right|^2 - \right. \\
&\qquad \left. - i\alpha_{n+k} \Delta_{\xi^{n+k}} + \gamma_{n+k} \vec{A}^{n+k}_{n,n+k} \nabla_{\xi^{n+k}} + \frac{i}{4\alpha_{n+k}} \left| \gamma_{n+k} \vec{A}^{n+k}_{n,n+k} \right|^2 \right] \bar{\Psi}^{n,n+k} = 0.
\end{aligned} \tag{A.22}$$

Let us write relation (A.22) in the form:

$$\begin{aligned}
&\Lambda_{n,n+k} + \bar{\Lambda}_{n,n+k} = \operatorname{Re} \Lambda_{n,n+k} = 0, \\
&\Lambda_{n,n+k} \overset{\text{det}}{=} \bar{\Psi}^{n,n+k} L_{n,n+k} \Psi^{n,n+k}, \\
&L_{n,n+k} \overset{\text{det}}{=} \frac{\partial}{\partial t} + i\alpha_n \Delta_{\xi^n} + \gamma_n \vec{A}^{n}_{n,n+k} \nabla_{\xi^n} - \frac{i}{4\alpha_n} \left| \gamma_n \vec{A}^{n}_{n,n+k} \right|^2 + i\alpha_{n+k} \Delta_{\xi^{n+k}} + \gamma_{n+k} \vec{A}^{n+k}_{n,n+k} \nabla_{\xi^{n+k}} - \frac{i}{4\alpha_{n+k}} \left| \gamma_{n+k} \vec{A}^{n+k}_{n,n+k} \right|^2.
\end{aligned} \tag{A.23}$$

from here it follows

$$\bar{\Psi}^{n,n+k} L_{n,n+k} \Psi^{n,n+k} = i u_{n,n+k} \;\Rightarrow\; L_{n,n+k} \Psi^{n,n+k} = -i\beta_{n+k} U^{n,n+k} \Psi^{n,n+k}. \tag{A.24}$$

From expressions (A.23) and (A.24) follows the equation:

$$\frac{i}{\beta_{n+k}}\frac{\partial\Psi^{n,n+k}}{\partial t}=-\alpha_{n+k}\beta_{n+k}\left(\hat{\mathrm{p}}_{n+k}-\frac{\gamma_{n+k}}{2\alpha_{n+k}\beta_{n+k}}\vec{A}^{n+k}_{n,n+k}\right)^2\Psi^{n,n+k}-$$

$$-\frac{\alpha_n\beta_n^2}{\beta_{n+k}}\left(\hat{\mathrm{p}}_n-\frac{\gamma_n}{2\alpha_n\beta_n}\vec{A}^{n}_{n,n+k}\right)^2\Psi^{n,n+k}+U^{n,n+k}\Psi^{n,n+k}.$$

Equation of the third rank from the first group (i.22) is of the form:

$$\frac{\partial f^{n,n+1,n+2}}{\partial t}+\xi_\beta^{n+1}\frac{\partial f^{n,n+1,n+2}}{\partial\xi_\beta^n}+\xi_\beta^{n+2}\frac{\partial f^{n,n+1,n+2}}{\partial\xi_\beta^{n+1}}+\frac{\partial}{\partial\xi_\beta^{n+2}}\left[f^{n,n+1,n+2}\left\langle\xi_\beta^{n+3}\right\rangle_{n,n+1,n+2}\right]=0. \quad \text{(A.25)}$$

According to the condition of theorem (1.1), the average kinematical value $\left\langle\xi_\beta^{n+3}\right\rangle_{n,n+1,n+2}$ is determined by the expression:

$$\left\langle\vec{\xi}^{n+3}\right\rangle_{n,n+1,n+2}=i\alpha_{n+2}\nabla_{\xi^{n+2}}\,\mathrm{Ln}\frac{\Psi^{n,n+1,n+2}}{\bar{\Psi}^{n,n+1,n+2}}+\gamma_{n+2}\vec{A}^{n+2}_{n,n+1,n+2}. \quad \text{(A.26)}$$

Substituting (A.26) into equation (A.25), we obtain

$$\begin{aligned}&\bar{\Psi}^{n,n+1,n+2}\left[\frac{\partial}{\partial t}+\vec{\xi}^{n+1}\nabla_{\xi^n}+\vec{\xi}^{n+2}\nabla_{\xi^{n+1}}+\right.\\&\left.\quad+i\alpha_{n+2}\Delta_{\xi^{n+2}}+\gamma_{n+2}\vec{A}^{n+2}_{n,n+1,n+2}\nabla_{\xi^{n+2}}-\frac{i}{4\alpha_{n+2}}\left|\gamma_{n+2}\vec{A}^{n+2}_{n,n+1,n+2}\right|^2\right]\Psi^{n,n+1,n+2}+\\&+\Psi^{n,n+1,n+2}\left[\frac{\partial}{\partial t}+\vec{\xi}^{n+1}\nabla_{\xi^n}+\vec{\xi}^{n+2}\nabla_{\xi^{n+1}}-\right.\\&\left.\quad-i\alpha_{n+2}\Delta_{\xi^{n+2}}+\gamma_{n+2}\vec{A}^{n+2}_{n,n+1,n+2}\nabla_{\xi^{n+2}}+\frac{i}{4\alpha_{n+2}}\left|\gamma_{n+2}\vec{A}^{n+2}_{n,n+1,n+2}\right|^2\right]\bar{\Psi}^{n,n+1,n+2}=0,\end{aligned} \quad \text{(A.27)}$$

where term $\frac{i}{4\alpha_{n+2}}\left|\gamma_{n+2}\vec{A}^{n+2}_{n,n+1,n+2}\right|^2\left|\Psi^{n,n+1,n+2}\right|^2$ is added and subtracted. In operator form, expression (A.27) is of the form:

$$\Lambda_{n,n+1,n+2}+\bar{\Lambda}_{n,n+1,n+2}=\mathrm{Re}\,\Lambda_{n,n+1,n+2}=0,$$

$$\Lambda_{n,n+1,n+2}\stackrel{\det}{=}\bar{\Psi}^{n,n+1,n+2}L_{n,n+1,n+2}\Psi^{n,n+1,n+2},$$

$$L_{n,n+1,n+2}\stackrel{\det}{=}\frac{\partial}{\partial t}+\vec{\xi}^{n+1}\nabla_{\xi^n}+\vec{\xi}^{n+2}\nabla_{\xi^{n+1}}+i\alpha_{n+2}\Delta_{\xi^{n+2}}+\gamma_{n+2}\vec{A}^{n+2}_{n,n+1,n+2}\nabla_{\xi^{n+2}}-\frac{i}{4\alpha_{n+2}}\left|\gamma_{n+2}\vec{A}^{n+2}_{n,n+1,n+2}\right|^2,$$

$$L_{n,n+1,n+2}\Psi^{n,n+1,n+2}=-i\beta_{n+2}U^{n,n+1,n+2}\Psi^{n,n+1,n+2}, \quad \text{(A.28)}$$

or

$$\frac{i}{\beta_{n+2}}\left(\frac{\partial}{\partial t}+\vec{\xi}^{n+1}\nabla_{\xi^n}+\vec{\xi}^{n+2}\nabla_{\xi^{n+1}}\right)\Psi^{n,n+1,n+2}=$$

$$=-\alpha_{n+2}\beta_{n+2}\left(\hat{\mathrm{p}}_{n+2}-\frac{\gamma_{n+2}}{2\alpha_{n+2}\beta_{n+2}}\vec{A}^{n+2}_{n,n+1,n+2}\right)^2\Psi^{n,n+1,n+2}+U^{n,n+1,n+2}\Psi^{n,n+1,n+2}.$$

In the second group of equations of the third rank, there are three types of equations (i.22). Two equations have two sources of dissipation and one equation has three sources of dissipations. Let us start with an equation with two sources of dissipation:

$$\frac{\partial f^{n,n+1,n+1+k}}{\partial t}+\frac{\partial}{\partial \xi_{\beta}^{n}}\left[\xi_{\beta}^{n+1} f^{n,n+1,n+1+k}\right]+\frac{\partial}{\partial \xi_{\beta}^{n+1}}\left[f^{n,n+1,n+1+k}\left\langle \xi_{\beta}^{n+2}\right\rangle_{n,n+1,n+1+k}\right]+ \\ +\frac{\partial}{\partial \xi_{\beta}^{n+1+k}}\left[f^{n,n+1,n+1+k}\left\langle \xi_{\beta}^{n+2+k}\right\rangle_{n,n+1,n+1+k}\right]=0. \tag{A.29}$$

The corresponding average kinematical values are of the form:

$$\left\langle \vec{\xi}^{n+2}\right\rangle_{n,n+1,n+1+k}=i\alpha_{n+1}\nabla_{\xi^{n+1}}\operatorname{Ln}\frac{\Psi^{n,n+1,n+1+k}}{\overline{\Psi}^{n,n+1,n+1+k}}+\gamma_{n+1}\vec{A}^{n+1}_{n,n+1,n+1+k}, \\ \left\langle \vec{\xi}^{n+2+k}\right\rangle_{n,n+1,n+1+k}=i\alpha_{n+1+k}\nabla_{\xi^{n+1+k}}\operatorname{Ln}\frac{\Psi^{n,n+1,n+1+k}}{\overline{\Psi}^{n,n+1,n+1+k}}+\gamma_{n+1+k}\vec{A}^{n+1+k}_{n,n+1,n+1+k}. \tag{A.30}$$

Let us rewrite equation (A.29) taking into account representations (A.30) in the operator form

$$\overline{\Psi}^{n,n+1,n+1+k}L_{n,n+1,n+1+k}\Psi^{n,n+1,n+1+k}+\Psi^{n,n+1,n+1+k}\overline{L}_{n,n+1,n+1+k}\overline{\Psi}^{n,n+1,n+1+k}=0,$$

$$L_{n,n+1,n+1+k}\overset{\det}{=}\frac{\partial}{\partial t}+\vec{\xi}^{n+1}\nabla_{\xi^{n}}+i\alpha_{n+1+k}\Delta_{\xi^{n+1+k}}+\gamma_{n+1+k}\vec{A}^{n+1+k}_{n,n+1,n+1+k}\nabla_{\xi^{n+1+k}}-\frac{i}{4\alpha_{n+1+k}}\left|\gamma_{n+1+k}\vec{A}^{n+1+k}_{n,n+1,n+1+k}\right|^{2}+ \\ +i\alpha_{n+1}\Delta_{\xi^{n+1}}+\gamma_{n+1}\vec{A}^{n+1}_{n,n+1,n+1+k}\nabla_{\xi^{n+1}}-\frac{i}{4\alpha_{n+1}}\left|\gamma_{n+1}\vec{A}^{n+1}_{n,n+1,n+1+k}\right|^{2},$$

$$L_{n,n+1,n+1+k}\Psi^{n,n+1,n+1+k}=-i\beta_{n+1+k}U^{n,n+1,n+1+k}\Psi^{n,n+1,n+1+k}, \tag{A.31}$$

where summands $\frac{i}{4\alpha_{n+1+k}}\left|\gamma_{n+1+k}\vec{A}^{n+1+k}_{n,n+1,n+1+k}\right|^{2}\left|\Psi^{n,n+1,n+1+k}\right|^{2}$ and $\frac{i}{4\alpha_{n+1}}\left|\gamma_{n+1}\vec{A}^{n+1}_{n,n+1,n+1+k}\right|^{2}\left|\Psi^{n,n+1,n+1+k}\right|^{2}$ are added and subtracted. As a result, from equation (A.31), we obtain

$$\frac{i}{\beta_{n+1+k}}\left(\frac{\partial}{\partial t}+\vec{\xi}^{n+1}\nabla_{\xi^{n}}\right)\Psi^{n,n+1,n+1+k}=-\alpha_{n+1+k}\beta_{n+1+k}\left(\hat{\mathrm{p}}_{n+1+k}-\frac{\gamma_{n+1+k}}{2\alpha_{n+1+k}\beta_{n+1+k}}\vec{A}^{n+1+k}_{n,n+1,n+1+k}\right)^{2}\Psi^{n,n+1,n+1+k}- \\ -\frac{\alpha_{n+1}\beta_{n+1}^{2}}{\beta_{n+1+k}}\left(\hat{\mathrm{p}}_{n+1}-\frac{\gamma_{n+1}}{\alpha_{n+1}\beta_{n+1}}\vec{A}^{n+1}_{n,n+1,n+1+k}\right)^{2}\Psi^{n,n+1,n+1+k}+U^{n,n+1,n+1+k}\Psi^{n,n+1,n+1+k}.$$

Let us consider the second equation from the second group, which has two sources of dissipation (i.22):

$$\frac{\partial f^{n,n+s,n+s+1}}{\partial t}+\frac{\partial}{\partial \xi_{\beta}^{n}}\left[\left\langle \xi_{\beta}^{n+1}\right\rangle_{n,n+s,n+s+1}f^{n,n+s,n+s+1}\right]+\frac{\partial}{\partial \xi_{\beta}^{n+s}}\left[\xi_{\beta}^{n+s+1}f^{n,n+s,n+s+1}\right]+ \\ +\frac{\partial}{\partial \xi_{\beta}^{n+s+1}}\left[f^{n,n+s,n+s+1}\left\langle \xi_{\beta}^{n+s+2}\right\rangle_{n,n+s,n+s+1}\right]=0. \tag{A.32}$$

Representations of average kinematical values are of the form:

$$\left\langle \vec{\xi}^{n+1} \right\rangle_{n,n+s,n+s+1} = i\alpha_n \nabla_{\xi^n} \operatorname{Ln} \frac{\Psi^{n,n+s,n+s+1}}{\bar{\Psi}^{n,n+s,n+s+1}} + \gamma_n \vec{A}^n_{n,n+s,n+s+1}, \tag{A.33}$$
$$\left\langle \vec{\xi}^{n+s+2} \right\rangle_{n,n+s,n+s+1} = i\alpha_{n+s+1} \nabla_{\xi^{n+s+1}} \operatorname{Ln} \frac{\Psi^{n,n+s,n+s+1}}{\bar{\Psi}^{n,n+s,n+s+1}} + \gamma_{n+s+1} \vec{A}^{n+s+1}_{n,n+s,n+s+1}.$$

Let us substitute (A.33) into (A.32)

$$\begin{aligned}
&\bar{\Psi}^{n,n+s,n+s+1}\left[\frac{\partial}{\partial t} + \vec{\xi}^{n+s+1}\nabla_{\xi^{n+s}} + i\alpha_n \Delta_{\xi^n} + \gamma_n \vec{A}^n_{n,n+s,n+s+1}\nabla_{\xi^n} + \right.\\
&\left. + i\alpha_{n+s+1}\Delta_{\xi^{n+s+1}} + \gamma_{n+s+1}\vec{A}^{n+s+1}_{n,n+s,n+s+1}\nabla_{\xi^{n+s+1}}\right]\Psi^{n,n+s,n+s+1} + \\
&\Psi^{n,n+s,n+s+1}\left[\frac{\partial}{\partial t} + \vec{\xi}^{n+s+1}\nabla_{\xi^{n+s}} - i\alpha_n \Delta_{\xi^n} + \gamma_n \vec{A}^n_{n,n+s,n+s+1}\nabla_{\xi^n} - \right.\\
&\left. - i\alpha_{n+s+1}\Delta_{\xi^{n+s+1}} + \gamma_{n+s+1}\vec{A}^{n+s+1}_{n,n+s,n+s+1}\nabla_{\xi^{n+s+1}}\right]\bar{\Psi}^{n,n+s,n+s+1} = 0.
\end{aligned} \tag{A.34}$$

Let us add and subtract summands $\frac{i}{4\alpha_n}\left|\gamma_n \vec{A}^n_{n,n+s,n+s+1}\right|^2 \left|\Psi^{n,n+s,n+s+1}\right|^2$ and $\frac{i}{4\alpha_{n+s+1}}\left|\gamma_{n+s+1} \vec{A}^{n+s+1}_{n,n+s,n+s+1}\right|^2 \left|\Psi^{n,n+s,n+s+1}\right|^2$ in expression (A.34), having it rewritten in operator form, we obtain:

$$L_{n,n+s,n+s+1}\Psi^{n,n+s,n+s+1} = -i\beta_{n+s+1}U^{n,n+s,n+s+1}\Psi^{n,n+s,n+s+1},$$

$$\begin{aligned}
&\frac{\partial \Psi^{n,n+s,n+s+1}}{\partial t} + \vec{\xi}^{n+s+1}\nabla_{\xi^{n+s}}\Psi^{n,n+s,n+s+1} + i\alpha_n \Delta_{\xi^n}\Psi^{n,n+s,n+s+1} + \gamma_n \vec{A}^n_{n,n+s,n+s+1}\nabla_{\xi^n}\Psi^{n,n+s,n+s+1} - \\
&-\frac{i}{4\alpha_n}\left|\gamma_n \vec{A}^n_{n,n+s,n+s+1}\right|^2 \Psi^{n,n+s,n+s+1} + i\alpha_{n+s+1}\Delta_{\xi^{n+s+1}}\Psi^{n,n+s,n+s+1} + \gamma_{n+s+1}\vec{A}^{n+s+1}_{n,n+s,n+s+1}\nabla_{\xi^{n+s+1}}\Psi^{n,n+s,n+s+1} - \\
&-\frac{i}{4\alpha_{n+s+1}}\left|\gamma_{n+s+1}\vec{A}^{n+s+1}_{n,n+s,n+s+1}\right|^2 \Psi^{n,n+s,n+s+1} = -i\beta_{n+s+1}U^{n,n+s,n+s+1}\Psi^{n,n+s,n+s+1}.
\end{aligned} \tag{A.35}$$

Using the notation of operator $\hat{p}_n$ and taking into account (A.10) (A.35), the equation takes the form:

$$\begin{aligned}
&\frac{i}{\beta_{n+s+1}}\left(\frac{\partial}{\partial t} + \vec{\xi}^{n+s+1}\nabla_{\xi^{n+s}}\right)\Psi^{n,n+s,n+s+1} = -\frac{\alpha_n \beta_n^2}{\beta_{n+s+1}}\left(\hat{p}_n - \frac{\gamma_n}{2\alpha_n\beta_n}\vec{A}^n_{n,n+s,n+s+1}\right)^2 \Psi^{n,n+s,n+s+1} - \\
&-\alpha_{n+s+1}\beta_{n+s+1}\left(\hat{p}_{n+s+1} - \frac{\gamma_{n+s+1}}{2\alpha_{n+s+1}\beta_{n+s+1}}\vec{A}^{n+s+1}_{n,n+s,n+s+1}\right)^2 \Psi^{n,n+s,n+s+1} + U^{n,n+s,n+s+1}\Psi^{n,n+s,n+s+1}.
\end{aligned}$$

The third equation from the second group has three sources of dissipations (i.22):

$$\frac{\partial f^{n,n+s,n+s+k}}{\partial t}+\frac{\partial}{\partial \xi_\beta^n}\left[\left\langle \xi_\beta^{n+1}\right\rangle_{n,n+s,n+s+k} f^{n,n+s,n+s+k}\right]+\frac{\partial}{\partial \xi_\beta^{n+s}}\left[\left\langle \xi_\beta^{n+s+1}\right\rangle_{n,n+s,n+s+k} f^{n,n+s,n+s+k}\right]+$$
$$+\frac{\partial}{\partial \xi_\beta^{n+s+k}}\left[f^{n,n+s,n+s+k}\left\langle \xi_\beta^{n+1+s+k}\right\rangle_{n,n+s,n+s+k}\right]=0. \tag{A.36}$$

According to (1.5), average kinematical flows can be represented as:

$$\begin{aligned}
\left\langle \vec{\xi}^{n+1}\right\rangle_{n,n+s,n+s+k} &= i\alpha_n \nabla_{\xi^n} \operatorname{Ln}\frac{\Psi^{n,n+s,n+s+k}}{\bar{\Psi}^{n,n+s,n+s+k}}+\gamma_n \vec{A}^n_{n,n+s,n+s+k},\\
\left\langle \vec{\xi}^{n+s+1}\right\rangle_{n,n+s,n+s+k} &= i\alpha_{n+s} \nabla_{\xi^{n+s}} \operatorname{Ln}\frac{\Psi^{n,n+s,n+s+k}}{\bar{\Psi}^{n,n+s,n+s+k}}+\gamma_{n+s} \vec{A}^{n+s}_{n,n+s,n+s+k},\\
\left\langle \vec{\xi}^{n+1+s+k}\right\rangle_{n,n+s,n+s+k} &= i\alpha_{n+s+k} \nabla_{\xi^{n+s+k}} \operatorname{Ln}\frac{\Psi^{n,n+s,n+s+k}}{\bar{\Psi}^{n,n+s,n+s+k}}+\gamma_{n+s+k} \vec{A}^{n+s+k}_{n,n+s,n+s+k}.
\end{aligned} \tag{A.37}$$

Taking into account representations (A.37), equation (A.36) takes the form:

$$\bar{\Psi}^{n,n+s,n+s+k}\left[\frac{\partial}{\partial t}+i\alpha_n\Delta_{\xi^n}+\gamma_n\vec{A}^n_{n,n+s,n+s+k}\nabla_{\xi^n}+i\alpha_{n+s}\Delta_{\xi^{n+s}}+\gamma_{n+s}\vec{A}^{n+s}_{n,n+s,n+s+k}\nabla_{\xi^{n+s}}+\right.$$
$$\left.+i\alpha_{n+s+k}\Delta_{\xi^{n+s+k}}+\gamma_{n+s+k}\vec{A}^{n+s+k}_{n,n+s,n+s+k}\nabla_{\xi^{n+s+k}}\right]\Psi^{n,n+s,n+s+k}+$$
$$+\Psi^{n,n+s,n+s+k}\left[\frac{\partial}{\partial t}-i\alpha_n\Delta_{\xi^n}+\gamma_n\vec{A}^n_{n,n+s,n+s+k}\nabla_{\xi^n}-i\alpha_{n+s}\Delta_{\xi^{n+s}}+\gamma_{n+s}\vec{A}^{n+s}_{n,n+s,n+s+k}\nabla_{\xi^{n+s}}+\right.$$
$$\left.-i\alpha_{n+s+k}\Delta_{\xi^{n+s+k}}+\gamma_{n+s+k}\vec{A}^{n+s+k}_{n,n+s,n+s+k}\nabla_{\xi^{n+s+k}}\right]\bar{\Psi}^{n,n+s,n+s+k}=0 \tag{A.37}$$

To transform expression (A.37), let us add and subtract three terms:

$$\begin{aligned}
&\frac{i}{4\alpha_n}\left|\gamma_n\vec{A}^n_{n,n+s,n+s+k}\right|^2\left|\Psi^{n,n+s,n+s+k}\right|^2,\\
&\frac{i}{4\alpha_{n+s}}\left|\gamma_{n+s}\vec{A}^{n+s}_{n,n+s,n+s+k}\right|^2\left|\Psi^{n,n+s,n+s+k}\right|^2,\\
&\frac{i}{4\alpha_{n+s+k}}\left|\gamma_{n+s+k}\vec{A}^{n+s+k}_{n,n+s,n+s+k}\right|^2\left|\Psi^{n,n+s,n+s+k}\right|^2.
\end{aligned} \tag{A.38}$$

Using operator representation

$$L_{n,n+s,n+s+k}\Psi^{n,n+s,n+s+k}=-i\beta_{n+s+k}U^{n,n+s,n+s+k}\Psi^{n,n+s,n+s+k}, \tag{A.39}$$

and summands (A.38), expression (A.37) takes the form:

$$\frac{i}{\beta_{n+s+k}}\frac{\partial\Psi^{n,n+s,n+s+k}}{\partial t}=-\frac{\alpha_n\beta_n^2}{\beta_{n+s+k}}\left(\hat{\mathrm{p}}_n-\frac{\gamma_n}{2\alpha_n\beta_n}\vec{A}^n_{n,n+s,n+s+k}\right)^2\Psi^{n,n+s,n+s+k}+$$

$$-\frac{\alpha_{n+s}\beta_{n+s}^2}{\beta_{n+s+k}}\left(\hat{\mathrm{p}}_{n+s}-\frac{\gamma_{n+s}}{2\alpha_{n+s}\beta_{n+s}}\vec{A}^{n+s}_{n,n+s,n+s+k}\right)^2\Psi^{n,n+s,n+s+k}+$$

$$-\alpha_{n+s+k}\beta_{n+s+k}\left(\hat{\mathrm{p}}_{n+s+k}-\frac{\gamma_{n+s+k}}{\alpha_{n+s+k}\beta_{n+s+k}}\vec{A}^{n+s+k}_{n,n+s,n+s+k}\right)^2\Psi^{n,n+s,n+s+k}+U^{n,n+s,n+s+k}\Psi^{n,n+s,n+s+k}.$$

The remaining equations of the dispersion chain for functions $\Psi^{n_1\ldots n_R}$ are constructed similarly. Theorem 1 is proved.

## Appendix B

### *Proof of Theorem 2*

From expression (A.8) it follows that

$$\beta_n U^n\left|\Psi^n\right|^2=i\bar{\Psi}^n\frac{\partial\Psi^n}{\partial t}-\alpha_n\bar{\Psi}^n\Delta_{\xi^n}\Psi^n+i\bar{\Psi}^n\gamma_n\vec{A}^n_n\nabla_{\xi^n}\Psi^n+\frac{1}{4\alpha_n}\left|\gamma_n\vec{A}^n_n\right|^2\left|\Psi^n\right|^2. \tag{B.1}$$

We calculate $\Delta_{\xi^n}\Psi^n$:

$$\nabla_{\xi^n}\Psi^n=\nabla_{\xi^n}\left(\left|\Psi^n\right|e^{i\varphi^n}\right)=e^{i\varphi^n}\nabla_{\xi^n}\left|\Psi^n\right|+ie^{i\varphi^n}\left|\Psi^n\right|\nabla_{\xi^n}\varphi^n, \tag{B.2}$$

$$\Delta_{\xi^n}\Psi^n=2ie^{i\varphi^n}\nabla_{\xi^n}\varphi^n\nabla_{\xi^n}\left|\Psi^n\right|+e^{i\varphi^n}\Delta_{\xi^n}\left|\Psi^n\right|-e^{i\varphi^n}\left|\Psi^n\right|\left|\nabla_{\xi^n}\varphi^n\right|^2+ie^{i\varphi^n}\left|\Psi^n\right|\Delta_{\xi^n}\varphi^n,$$

Let us substitute expressions (B.2) into representation (B.1)

$$\begin{aligned}\beta_n U^n\left|\Psi^n\right|^2&=i\left(\frac{1}{2}\frac{\partial\left|\Psi^n\right|^2}{\partial t}-\alpha_n\nabla_{\xi^n}\varphi^n\nabla_{\xi^n}\left|\Psi^n\right|^2-\alpha_n\left|\Psi^n\right|^2\Delta_{\xi^n}\varphi^n+\frac{\gamma_n\vec{A}^n_n}{2}\nabla_{\xi^n}\left|\Psi^n\right|^2\right)-\\&-\left|\Psi^n\right|^2\left(\frac{\partial\varphi^n}{\partial t}+\alpha_n\frac{\Delta_{\xi^n}\left|\Psi^n\right|}{\left|\Psi^n\right|}-\alpha_n\left|\nabla_{\xi^n}\varphi^n\right|^2+\gamma_n\vec{A}^n_n\nabla_{\xi^n}\varphi^n-\frac{1}{4\alpha_n}\left|\gamma_n\vec{A}^n_n\right|^2\right).\end{aligned} \tag{B.3}$$

We consider the imaginary part of the expression (B.3)

$$\begin{aligned}\mathrm{Im}\left(\beta_n U^n\right)&=\frac{1}{2}\left(\frac{\partial f^n}{\partial t}-\alpha_n\nabla_{\xi^n}\Phi^n\nabla_{\xi^n}f^n-\alpha_n f^n\Delta_{\xi^n}\Phi^n+\gamma_n\vec{A}^n_n\nabla_{\xi^n}f^n\right)=\\&=\frac{1}{2}\left(\frac{\partial f^n}{\partial t}+\left(-\alpha_n\nabla_{\xi^n}\Phi^n+\gamma_n\vec{A}^n_n,\nabla_{\xi^n}f^n\right)+f^n\left(\nabla_{\xi^n},-\alpha_n\nabla_{\xi^n}\Phi^n+\gamma_n\vec{A}^n_n\right)\right)=\\&=\frac{1}{2}\left(\frac{\partial f^n}{\partial t}+\left\langle\vec{\xi}^{n+1}\right\rangle_n\nabla_{\xi^n}f^n+f^n\nabla_{\xi^n}\left\langle\vec{\xi}^{n+1}\right\rangle_n\right)=\frac{1}{2}\left(\frac{\partial f^n}{\partial t}+\mathrm{div}_{\xi^n}\left[f^n\left\langle\vec{\xi}^{n+1}\right\rangle_n\right]\right)=0.\end{aligned} \tag{B.4}$$

By virtue of result (B.4), expression (B.3) has only a real part:

$$-\frac{1}{2\beta_n}\frac{\partial\Phi^n}{\partial t}=-\frac{1}{4\alpha_n\beta_n}\left(-\alpha_n\nabla_{\xi^n}\Phi^n+\gamma_n\vec{A}_n^n\right)^2+U^n+\frac{\alpha_n}{\beta_n}\frac{\Delta_{\xi^n}\left|\Psi^n\right|}{\left|\Psi^n\right|}. \qquad \text{(B.5)}$$

The resulting expression (B.5) corresponds to the Hamilton-Jacobi equation of the first rank.

Let us consider the equation of the second rank belonging to the first group (A.18)

$$\begin{aligned}\beta_{n+1}U^{n,n+1}\left|\Psi^{n,n+1}\right|^2&=i\bar{\Psi}^{n,n+1}\frac{\partial\Psi^{n,n+1}}{\partial t}-\alpha_{n+1}\bar{\Psi}^{n,n+1}\Delta_{\xi^{n+1}}\Psi^{n,n+1}+i\vec{\xi}^{n+1}\bar{\Psi}^{n,n+1}\nabla_{\xi^n}\Psi^{n,n+1}+\\&+i\gamma_{n+1}\bar{\Psi}^{n,n+1}\vec{A}_{n,n+1}^{n+1}\nabla_{\xi^{n+1}}\Psi^{n,n+1}+\frac{1}{4\alpha_{n+1}}\left|\gamma_{n+1}\vec{A}_{n,n+1}^{n+1}\right|^2\left|\Psi^{n,n+1}\right|^2,\end{aligned} \qquad \text{(B.6)}$$

using expression (B.2), one can write the expressions

$$\begin{aligned}\nabla_{\xi^n}\Psi^{n,n+1}&=e^{i\varphi^{n,n+1}}\left(\nabla_{\xi^n}\left|\Psi^{n,n+1}\right|+i\left|\Psi^{n,n+1}\right|\nabla_{\xi^n}\varphi^{n,n+1}\right),\\ \nabla_{\xi^{n+1}}\Psi^{n,n+1}&=e^{i\varphi^{n,n+1}}\left(\nabla_{\xi^{n+1}}\left|\Psi^{n,n+1}\right|+i\left|\Psi^{n,n+1}\right|\nabla_{\xi^{n+1}}\varphi^{n,n+1}\right),\\ e^{-i\varphi^{n,n+1}}\Delta_{\xi^{n+1}}\Psi^{n,n+1}&=\\=2i\nabla_{\xi^{n+1}}\varphi^{n,n+1}\nabla_{\xi^{n+1}}\left|\Psi^{n,n+1}\right|&+\Delta_{\xi^{n+1}}\left|\Psi^{n,n+1}\right|-\left|\Psi^{n,n+1}\right|\left|\nabla_{\xi^{n+1}}\varphi^{n,n+1}\right|^2+i\left|\Psi^{n,n+1}\right|\Delta_{\xi^{n+1}}\varphi^{n,n+1}.\end{aligned} \qquad \text{(B.7)}$$

Substituting (B.7) into (B.6), we obtain

$$\begin{aligned}\beta_{n+1}U^{n,n+1}\left|\Psi^{n,n+1}\right|^2&=\frac{i}{2}\left[\frac{\partial\left|\Psi^{n,n+1}\right|^2}{\partial t}+\left(-\alpha_{n+1}\nabla_{\xi^{n+1}}\Phi^{n,n+1}+\gamma_{n+1}\vec{A}_{n,n+1}^{n+1},\nabla_{\xi^{n+1}}\left|\Psi^{n,n+1}\right|^2\right)+\right.\\&\left.+\vec{\xi}^{n+1}\nabla_{\xi^n}\left|\Psi^{n,n+1}\right|^2-\alpha_{n+1}\left|\Psi^{n,n+1}\right|^2\Delta_{\xi^{n+1}}\Phi^{n,n+1}\right]+\left|\Psi^{n,n+1}\right|^2\left[-\frac{1}{2}\frac{\partial\Phi^{n,n+1}}{\partial t}-\alpha_{n+1}\frac{\Delta_{\xi^{n+1}}\left|\Psi^{n,n+1}\right|}{\left|\Psi^{n,n+1}\right|}+\right.\\&\left.+\frac{\alpha_{n+1}}{4}\left|\nabla_{\xi^{n+1}}\Phi^{n,n+1}\right|^2-\frac{1}{2}\vec{\xi}^{n+1}\nabla_{\xi^n}\Phi^{n,n+1}-\frac{1}{2}\gamma_{n+1}\vec{A}_{n,n+1}^{n+1}\nabla_{\xi^{n+1}}\Phi^{n,n+1}+\frac{1}{4\alpha_{n+1}}\left|\gamma_{n+1}\vec{A}_{n,n+1}^{n+1}\right|^2\right],\end{aligned} \qquad \text{(B.8)}$$

Let us show that the imaginary part of expression (B.8) is equal to zero. Indeed:

$$\begin{aligned}\mathrm{Im}\left(\beta_{n+1}U^{n,n+1}\left|\Psi^{n,n+1}\right|^2\right)&=\frac{\partial f^{n,n+1}}{\partial t}+\left\langle\vec{\xi}^{n+2}\right\rangle_{n,n+1}\nabla_{\xi^{n+1}}f^{n,n+1}+\vec{\xi}^{n+1}\nabla_{\xi^n}f^{n,n+1}+\\&+f^{n,n+1}\left(\nabla_{\xi^{n+1}},-\alpha_{n+1}\nabla_{\xi^{n+1}}\Phi^{n,n+1}+\gamma_{n+1}\vec{A}_{n,n+1}^{n+1}\right)=\frac{\partial f^{n,n+1}}{\partial t}+\vec{\xi}^{n+1}\nabla_{\xi^n}f^{n,n+1}+\mathrm{div}_{\xi^{n+1}}\left[f^{n,n+1}\left\langle\vec{\xi}^{n+2}\right\rangle_{n,n+1}\right]=0.\end{aligned}$$

As a result, expression (B.8) takes the form:

$$
-\frac{1}{2\beta_{n+1}}\frac{\partial\Phi^{n,n+1}}{\partial t}=-\frac{1}{4\alpha_{n+1}\beta_{n+1}}\left(\left|\alpha_{n+1}\nabla_{\xi^{n+1}}\Phi^{n,n+1}\right|^2-2\gamma_{n+1}\vec{A}^{n+1}_{n,n+1}\alpha_{n+1}\nabla_{\xi^{n+1}}\Phi^{n,n+1}+\left|\gamma_{n+1}\vec{A}^{n+1}_{n,n+1}\right|^2\right)+
$$

$$
+\frac{1}{2\alpha_{n+1}\beta_{n+1}}\vec{\xi}^{n+1}\alpha_{n+1}\nabla_{\xi^n}\Phi^{n,n+1}+\frac{\alpha_{n+1}}{\beta_{n+1}}\frac{\Delta_{\xi^{n+1}}\left|\Psi^{n,n+1}\right|}{\left|\Psi^{n,n+1}\right|}+U^{n,n+1}
$$

$$
-\frac{1}{2\beta_{n+1}}\left(\frac{\partial}{\partial t}+\vec{\xi}^{n+1}\nabla_{\xi^n}\right)\Phi^{n,n+1}=
$$

$$
=-\frac{1}{4\alpha_{n+1}\beta_{n+1}}\left|-\alpha_{n+1}\nabla_{\xi^{n+1}}\Phi^{n,n+1}+\gamma_{n+1}\vec{A}^{n+1}_{n,n+1}\right|^2+\frac{\alpha_{n+1}}{\beta_{n+1}}\frac{\Delta_{\xi^{n+1}}\left|\Psi^{n,n+1}\right|}{\left|\Psi^{n,n+1}\right|}+U^{n,n+1}. \qquad \text{(B.9)}
$$

Equation (B.9) corresponds to the Hamilton-Jacobi equation of the second rank belonging to the first group.

We consider equations of the second rank from the second group (A.24).

$$
\beta_{n+k}U^{n,n+k}\left|\Psi^{n,n+k}\right|^2=i\bar{\Psi}^{n,n+k}\frac{\partial\Psi^{n,n+k}}{\partial t}-\alpha_{n+k}\bar{\Psi}^{n,n+k}\Delta_{\xi^{n+k}}\Psi^{n,n+k}+
$$

$$
+i\gamma_{n+k}\vec{A}^{n+k}_{n,n+k}\bar{\Psi}^{n,n+k}\nabla_{\xi^{n+k}}\Psi^{n,n+k}+\frac{1}{4\alpha_{n+k}}\left|\gamma_{n+k}\vec{A}^{n+k}_{n,n+k}\right|^2\left|\Psi^{n,n+k}\right|^2- \qquad \text{(B.10)}
$$

$$
-\alpha_n\bar{\Psi}^{n,n+k}\Delta_{\xi^n}\Psi^{n,n+k}+i\gamma_n\vec{A}^{n}_{n,n+k}\bar{\Psi}^{n,n+k}\nabla_{\xi^n}\Psi^{n,n+k}+\frac{1}{4\alpha_n}\left|\gamma_n\vec{A}^{n}_{n,n+k}\right|^2\left|\Psi^{n,n+k}\right|^2.
$$

The relations are true:

$$
\nabla_{\xi^n}\Psi^{n,n+k}=e^{i\varphi^{n,n+k}}\left(\nabla_{\xi^n}\left|\Psi^{n,n+k}\right|+i\left|\Psi^{n,n+k}\right|\nabla_{\xi^n}\varphi^{n,n+k}\right), \qquad \text{(B.11)}
$$

$$
\nabla_{\xi^{n+k}}\Psi^{n,n+k}=e^{i\varphi^{n,n+k}}\left(\nabla_{\xi^{n+k}}\left|\Psi^{n,n+k}\right|+i\left|\Psi^{n,n+k}\right|\nabla_{\xi^{n+k}}\varphi^{n,n+k}\right),
$$

$$
e^{-i\varphi^{n,n+k}}\Delta_{\xi^n}\Psi^{n,n+k}=
$$

$$
=2i\nabla_{\xi^n}\varphi^{n,n+k}\nabla_{\xi^n}\left|\Psi^{n,n+k}\right|+\Delta_{\xi^n}\left|\Psi^{n,n+k}\right|-\left|\Psi^{n,n+k}\right|\left|\nabla_{\xi^n}\varphi^{n,n+k}\right|^2+i\left|\Psi^{n,n+k}\right|\Delta_{\xi^n}\varphi^{n,n+k},
$$

$$
e^{-i\varphi^{n,n+k}}\Delta_{\xi^{n+k}}\Psi^{n,n+k}=
$$

$$
=2i\nabla_{\xi^{n+k}}\varphi^{n,n+k}\nabla_{\xi^{n+k}}\left|\Psi^{n,n+k}\right|+\Delta_{\xi^{n+k}}\left|\Psi^{n,n+k}\right|-\left|\Psi^{n,n+k}\right|\left|\nabla_{\xi^{n+k}}\varphi^{n,n+k}\right|^2+i\left|\Psi^{n,n+k}\right|\Delta_{\xi^{n+k}}\varphi^{n,n+k}.
$$

We substitute expressions (B.11) into equation (B.10)

$$
\beta_{n+k}U^{n,n+k}\left|\Psi^{n,n+k}\right|^2=
$$

$$
i\frac{1}{2}\frac{\partial\left|\Psi^{n,n+k}\right|^2}{\partial t}-i\alpha_{n+k}\nabla_{\xi^{n+k}}\varphi^{n,n+k}\nabla_{\xi^{n+k}}\left|\Psi^{n,n+k}\right|^2-i\alpha_{n+k}\left|\Psi^{n,n+k}\right|^2\Delta_{\xi^{n+k}}\varphi^{n,n+k}+ \qquad \text{(B.12)}
$$

$$
+i\frac{\gamma_{n+k}}{2}\vec{A}^{n+k}_{n,n+k}\nabla_{\xi^{n+k}}\left|\Psi^{n,n+k}\right|^2-i\alpha_n\nabla_{\xi^n}\varphi^{n,n+k}\nabla_{\xi^n}\left|\Psi^{n,n+k}\right|^2-i\alpha_n\left|\Psi^{n,n+k}\right|^2\Delta_{\xi^n}\varphi^{n,n+k}+
$$

$$
+i\frac{\gamma_n}{2}\vec{A}^{n}_{n,n+k}\nabla_{\xi^n}\left|\Psi^{n,n+k}\right|^2-\left|\Psi^{n,n+k}\right|^2\frac{\partial\varphi^{n,n+k}}{\partial t}-\alpha_{n+k}\left|\Psi^{n,n+k}\right|\Delta_{\xi^{n+k}}\left|\Psi^{n,n+k}\right|+
$$

$$+\alpha_{n+k}\left|\Psi^{n,n+k}\right|^2\left|\nabla_{\xi^{n+k}}\varphi^{n,n+k}\right|^2-\gamma_{n+k}\vec{A}^{n+k}_{n,n+k}\left|\Psi^{n,n+k}\right|^2\nabla_{\xi^{n+k}}\varphi^{n,n+k}+$$
$$+\frac{1}{4\alpha_{n+k}}\left|\gamma_{n+k}\vec{A}^{n+k}_{n,n+k}\right|^2\left|\Psi^{n,n+k}\right|^2-\alpha_n\left|\Psi^{n,n+k}\right|\Delta_{\xi^n}\left|\Psi^{n,n+k}\right|+\alpha_n\left|\Psi^{n,n+k}\right|^2\left|\nabla_{\xi^n}\varphi^{n,n+k}\right|^2-$$
$$-\gamma_n\vec{A}^{n}_{n,n+k}\left|\Psi^{n,n+k}\right|^2\nabla_{\xi^n}\varphi^{n,n+k}+\frac{1}{4\alpha_n}\left|\gamma_n\vec{A}^{n}_{n,n+k}\right|^2\left|\Psi^{n,n+k}\right|^2,$$

from here

$$\begin{gathered}2\operatorname{Im}\left[\beta_{n+k}U^{n,n+k}f^{n,n+k}\right]=\frac{\partial f^{n,n+k}}{\partial t}+\left(-\alpha_{n+k}\nabla_{\xi^{n+k}}\Phi^{n,n+k}+\gamma_{n+k}\vec{A}^{n+k}_{n,n+k}\right)\nabla_{\xi^{n+k}}f^{n,n+k}+\\+\left(\nabla_{\xi^{n+k}},-\alpha_{n+k}\nabla_{\xi^{n+k}\xi^{n+k}}\Phi^{n,n+k}+\gamma_{n+k}\vec{A}^{n+k}_{n,n+k}\right)f^{n,n+k}+\\+\left(-\alpha_n\nabla_{\xi^n}\Phi^{n,n+k}+\gamma_n\vec{A}^{n}_{n,n+k}\right)\nabla_{\xi^n}f^{n,n+k}+\left(\nabla_{\xi^n},-\alpha_n\nabla_{\xi^n},\Phi^{n,n+k}+\gamma_n\vec{A}^{n}_{n,n+k}\right)f^{n,n+k}=\\=\frac{\partial f^{n,n+k}}{\partial t}+\operatorname{div}_{\xi^{n+k}}\left[\left\langle\vec{\xi}^{\,n+k+1}\right\rangle_{n,n+k}f^{n,n+k}\right]+\operatorname{div}_{\xi^n}\left[\left\langle\vec{\xi}^{\,n+1}\right\rangle_{n,n+k}f^{n,n+k}\right]=0.\end{gathered}\tag{B.13}$$

Taking into consideration (B.13), expression (B.12) takes the form:

$$-\frac{1}{\beta_{n+k}}\frac{\partial\varphi^{n,n+k}}{\partial t}=-\frac{1}{4\alpha_{n+k}\beta_{n+k}}\left(-\alpha_{n+k}\nabla_{\xi^{n+k}}\Phi^{n,n+k}+\gamma_{n+k}\vec{A}^{n+k}_{n,n+k}\right)^2$$
$$-\frac{1}{4\alpha_n\beta_{n+k}}\left(-\alpha_n\nabla_{\xi^n}\Phi^{n,n+k}+\gamma_n\vec{A}^{n}_{n,n+k}\right)^2+\frac{\alpha_n}{\beta_{n+k}}\frac{\Delta_{\xi^n}\left|\Psi^{n,n+k}\right|}{\left|\Psi^{n,n+k}\right|}+\frac{\alpha_{n+k}}{\beta_{n+k}}\frac{\Delta_{\xi^{n+k}}\left|\Psi^{n,n+k}\right|}{\left|\Psi^{n,n+k}\right|}+U^{n,n+k},$$

as a result,

$$\begin{gathered}-\frac{1}{\beta_{n+k}}\frac{\partial\varphi^{n,n+k}}{\partial t}=-\frac{1}{4\alpha_{n+k}\beta_{n+k}}\left|\left\langle\vec{\xi}^{\,n+k+1}\right\rangle_{n,n+k}\right|^2-\frac{1}{4\alpha_n\beta_{n+k}}\left|\left\langle\vec{\xi}^{\,n+1}\right\rangle_{n,n+k}\right|^2\\+\frac{\alpha_n}{\beta_{n+k}}\frac{\Delta_{\xi^n}\left|\Psi^{n,n+k}\right|}{\left|\Psi^{n,n+k}\right|}+\frac{\alpha_{n+k}}{\beta_{n+k}}\frac{\Delta_{\xi^{n+k}}\left|\Psi^{n,n+k}\right|}{\left|\Psi^{n,n+k}\right|}+U^{n,n+k}.\end{gathered}\tag{B.14}$$

We consider equations of the third rank from the first group (A.28)

$$\begin{gathered}\beta_{n+2}U^{n,n+1,n+2}\left|\Psi^{n,n+1,n+2}\right|^2=i\bar{\Psi}^{n,n+1,n+2}\frac{\partial\Psi^{n,n+1,n+2}}{\partial t}+i\vec{\xi}^{\,n+1}\bar{\Psi}^{n,n+1,n+2}\nabla_{\xi^n}\Psi^{n,n+1,n+2}+\\+i\vec{\xi}^{\,n+2}\bar{\Psi}^{n,n+1,n+2}\nabla_{\xi^{n+1}}\Psi^{n,n+1,n+2}-\alpha_{n+2}\bar{\Psi}^{n,n+1,n+2}\Delta_{\xi^{n+2}}\Psi^{n,n+1,n+2}+\\+i\gamma_{n+2}\vec{A}^{n+2}_{n,n+1,n+2}\bar{\Psi}^{n,n+1,n+2}\nabla_{\xi^{n+2}}\Psi^{n,n+1,n+2}+\frac{1}{4\alpha_{n+2}}\left|\gamma_{n+2}\vec{A}^{n+2}_{n,n+1,n+2}\right|^2\left|\Psi^{n,n+1,n+2}\right|^2,\end{gathered}\tag{B.15}$$

Taking into account relations like (B.11), expression (B.15) takes the form:

$$\beta_{n+2}U^{n,n+1,n+2}\left|\Psi^{n,n+1,n+2}\right|^2=\frac{1}{2}i\frac{\partial\left|\Psi^{n,n+1,n+2}\right|^2}{\partial t}+i\vec{\xi}^{n+1}\left|\Psi^{n,n+1,n+2}\right|\nabla_{\xi^n}\left|\Psi^{n,n+1,n+2}\right|+$$
$$+i\vec{\xi}^{n+2}\left|\Psi^{n,n+1,n+2}\right|\nabla_{\xi^{n+1}}\left|\Psi^{n,n+1,n+2}\right|-2i\alpha_{n+2}\left|\Psi^{n,n+1,n+2}\right|\nabla_{\xi^{n+2}}\varphi^{n,n+1,n+2}\nabla_{\xi^{n+2}}\left|\Psi^{n,n+1,n+2}\right|-$$
$$-i\alpha_{n+2}\left|\Psi^{n,n+1,n+2}\right|^2\Delta_{\xi^{n+2}}\varphi^{n,n+1,n+2}+i\gamma_{n+2}\vec{A}^{n+2}_{n,n+1,n+2}\left|\Psi^{n,n+1,n+2}\right|\nabla_{\xi^{n+2}}\left|\Psi^{n,n+1,n+2}\right|$$
$$-\left|\Psi^{n,n+1,n+2}\right|^2\frac{\partial\varphi^{n,n+1,n+2}}{\partial t}-\vec{\xi}^{n+1}\left|\Psi^{n,n+1,n+2}\right|^2\nabla_{\xi^n}\varphi^{n,n+1,n+2}-\vec{\xi}^{n+2}\left|\Psi^{n,n+1,n+2}\right|^2\nabla_{\xi^{n+1}}\varphi^{n,n+1,n+2}-$$
$$-\alpha_{n+2}\left|\Psi^{n,n+1,n+2}\right|\Delta_{\xi^{n+2}}\left|\Psi^{n,n+1,n+2}\right|+\alpha_{n+2}\left|\Psi^{n,n+1,n+2}\right|^2\left|\nabla_{\xi^{n+2}}\varphi^{n,n+1,n+2}\right|^2+$$
$$-\gamma_{n+2}\vec{A}^{n+2}_{n,n+1,n+2}\left|\Psi^{n,n+1,n+2}\right|^2\nabla_{\xi^{n+2}}\varphi^{n,n+1,n+2}+\frac{1}{4\alpha_{n+2}}\left|\gamma_{n+2}\vec{A}^{n+2}_{n,n+1,n+2}\right|^2\left|\Psi^{n,n+1,n+2}\right|^2, \qquad \text{(B.16)}$$

or

$$-\frac{1}{\beta_{n+2}}\left(\frac{\partial}{\partial t}+\vec{\xi}^{n+1}\nabla_{\xi^n}+\vec{\xi}^{n+2}\nabla_{\xi^{n+1}}\right)\varphi^{n,n+1,n+2}=$$
$$-\frac{1}{4\alpha_{n+2}\beta_{n+2}}\left(\alpha^2_{n+2}\left|\nabla_{\xi^{n+2}}\Phi^{n,n+1,n+2}\right|^2-2\gamma_{n+2}\vec{A}^{n+2}_{n,n+1,n+2}\alpha_{n+2}\nabla_{\xi^{n+2}}\Phi^{n,n+1,n+2}+\left|\gamma_{n+2}\vec{A}^{n+2}_{n,n+1,n+2}\right|^2\right)+ \qquad \text{(B.17)}$$
$$+\frac{\alpha_{n+2}}{\beta_{n+2}}\frac{\Delta_{\xi^{n+2}}\left|\Psi^{n,n+1,n+2}\right|}{\left|\Psi^{n,n+1,n+2}\right|}+U^{n,n+1,n+2}.$$

it is taken into account here that the imaginary part of (B.16) is equal to zero. Given representation (1.5), equation (B.17) takes the form

$$-\frac{1}{\beta_{n+2}}\left(\frac{\partial}{\partial t}+\vec{\xi}^{n+1}\nabla_{\xi^n}+\vec{\xi}^{n+2}\nabla_{\xi^{n+1}}\right)\varphi^{n,n+1,n+2}=$$
$$-\frac{1}{4\alpha_{n+2}\beta_{n+2}}\left|\left\langle\vec{\xi}^{n+3}\right\rangle_{n,n+1,n+2}\right|^2+\frac{\alpha_{n+2}}{\beta_{n+2}}\frac{\Delta_{\xi^{n+2}}\left|\Psi^{n,n+1,n+2}\right|}{\left|\Psi^{n,n+1,n+2}\right|}+U^{n,n+1,n+2}.$$

We consider the equation of the third rank from the second group with two dissipation sources (A.31)

$$\beta_{n+1+k}U^{n,n+1,n+1+k}\left|\Psi^{n,n+1,n+1+k}\right|^2=i\frac{1}{2}\frac{\partial\left|\Psi^{n,n+1,n+1+k}\right|^2}{\partial t}-\left|\Psi^{n,n+1,n+1+k}\right|^2\frac{\partial\varphi^{n,n+1,n+1+k}}{\partial t}+$$
$$+i\vec{\xi}^{n+1}\left|\Psi^{n,n+1,n+1+k}\right|\nabla_{\xi^n}\left|\Psi^{n,n+1,n+1+k}\right|-\vec{\xi}^{n+1}\left|\Psi^{n,n+1,n+1+k}\right|^2\nabla_{\xi^n}\varphi^{n,n+1,n+1+k}-$$
$$-2i\alpha_{n+1+k}\left|\Psi^{n,n+1,n+1+k}\right|\nabla_{\xi^{n+1+k}}\varphi^{n,n+1,n+1+k}\nabla_{\xi^{n+1+k}}\left|\Psi^{n,n+1,n+1+k}\right|-\alpha_{n+1+k}\left|\Psi^{n,n+1,n+1+k}\right|\Delta_{\xi^{n+1+k}}\left|\Psi^{n,n+1,n+1+k}\right|+$$
$$+\alpha_{n+1+k}\left|\Psi^{n,n+1,n+1+k}\right|^2\left|\nabla_{\xi^{n+1+k}}\varphi^{n,n+1,n+1+k}\right|^2-i\alpha_{n+1+k}\left|\Psi^{n,n+1,n+1+k}\right|^2\Delta_{\xi^{n+1+k}}\varphi^{n,n+1,n+1+k}+$$
$$+i\gamma_{n+1+k}\vec{A}^{n+1+k}_{n,n+1,n+1+k}\left|\Psi^{n,n+1,n+1+k}\right|\nabla_{\xi^{n+1+k}}\left|\Psi^{n,n+1,n+1+k}\right|-\gamma_{n+1+k}\vec{A}^{n+1+k}_{n,n+1,n+1+k}\left|\Psi^{n,n+1,n+1+k}\right|^2\nabla_{\xi^{n+1+k}}\varphi^{n,n+1,n+1+k}+$$
$$\frac{1}{4\alpha_{n+1+k}}\left|\gamma_{n+1+k}\vec{A}^{n+1+k}_{n,n+1,n+1+k}\right|^2\left|\Psi^{n,n+1,n+1+k}\right|^2-\alpha_{n+1}\bar{\Psi}^{n,n+1,n+1+k}\Delta_{\xi^{n+1}}\Psi^{n,n+1,n+1+k}+$$

$$i\gamma_{n+1}\vec{A}^{n+1}_{n,n+1,n+1+k}\overline{\Psi}^{n,n+1,n+1+k}\nabla_{\xi^{n+1}}\Psi^{n,n+1,n+1+k}+\frac{1}{4\alpha_{n+1}}\left|\gamma_{n+1}\vec{A}^{n+1}_{n,n+1,n+1+k}\right|^2\left|\Psi^{n,n+1,n+1+k}\right|^2.$$

Considering that $\operatorname{Im}U^{n,n+1,n+1+k}=0$, we obtain

$$\begin{aligned}
&-\frac{1}{\beta_{n+1+k}}\left(\frac{\partial}{\partial t}+\vec{\xi}^{n+1}\nabla_{\xi^n}\right)\varphi^{n,n+1,n+1+k}=-\frac{\alpha_{n+1+k}}{4\beta_{n+1+k}}\left|\nabla_{\xi^{n+1+k}}\Phi^{n,n+1,n+1+k}\right|^2+\\
&+\frac{\gamma_{n+1+k}}{2\beta_{n+1+k}}\vec{A}^{n+1+k}_{n,n+1,n+1+k}\nabla_{\xi^{n+1+k}}\Phi^{n,n+1,n+1+k}-\frac{1}{4\alpha_{n+1+k}\beta_{n+1+k}}\left|\gamma_{n+1+k}\vec{A}^{n+1+k}_{n,n+1,n+1+k}\right|^2-\\
&-\frac{\alpha_{n+1}}{4\beta_{n+1+k}}\left|\nabla_{\xi^{n+1}}\Phi^{n,n+1,n+1+k}\right|^2+\frac{\gamma_{n+1}}{2\beta_{n+1+k}}\vec{A}^{n+1}_{n,n+1,n+1+k}\nabla_{\xi^{n+1}}\Phi^{n,n+1,n+1+k}-\\
&-\frac{1}{4\alpha_{n+1}\beta_{n+1+k}}\left|\gamma_{n+1}\vec{A}^{n+1}_{n,n+1,n+1+k}\right|^2+\frac{\alpha_{n+1}}{\beta_{n+1+k}}\frac{\Delta_{\xi^{n+1}}\left|\Psi^{n,n+1,n+1+k}\right|}{\left|\Psi^{n,n+1,n+1+k}\right|}+\\
&+\frac{\alpha_{n+1+k}}{\beta_{n+1+k}}\frac{\Delta_{\xi^{n+1+k}}\left|\Psi^{n,n+1,n+1+k}\right|}{\left|\Psi^{n,n+1,n+1+k}\right|}+U^{n,n+1,n+1+k}.
\end{aligned}\tag{B.18}$$

Obtained equation (B.18) transformed into the form

$$\begin{aligned}
&-\frac{1}{\beta_{n+1+k}}\left(\frac{\partial}{\partial t}+\vec{\xi}^{n+1}\nabla_{\xi^n}\right)\varphi^{n,n+1,n+1+k}=\\
&-\frac{1}{4\alpha_{n+1+k}\beta_{n+1+k}}\left|\left\langle\vec{\xi}^{n+k+2}\right\rangle_{n,n+1,n+1+k}\right|^2-\frac{1}{4\alpha_{n+1}\beta_{n+1+k}}\left|\left\langle\vec{\xi}^{n+2}\right\rangle_{n,n+1,n+1+k}\right|^2+\\
&+\frac{\alpha_{n+1}}{\beta_{n+1+k}}\frac{\Delta_{\xi^{n+1}}\left|\Psi^{n,n+1,n+1+k}\right|}{\left|\Psi^{n,n+1,n+1+k}\right|}+\frac{\alpha_{n+1+k}}{\beta_{n+1+k}}\frac{\Delta_{\xi^{n+1+k}}\left|\Psi^{n,n+1,n+1+k}\right|}{\left|\Psi^{n,n+1,n+1+k}\right|}+U^{n,n+1,n+1+k}.
\end{aligned}\tag{B.19}$$

We obtain Hamilton-Jacobi equation for the second equation of the third rank with two sources of dissipation (A.35)

$$\begin{aligned}
&\beta_{n+s+1}U^{n,n+s,n+s+1}\left|\Psi^{n,n+s,n+s+1}\right|^2=i\frac{1}{2}\frac{\partial\left|\Psi^{n,n+s,n+s+1}\right|^2}{\partial t}-\left|\Psi^{n,n+s,n+s+1}\right|^2\frac{\partial\varphi^{n,n+s,n+s+1}}{\partial t}-\\
&+i\vec{\xi}^{n+s+1}\left|\Psi^{n,n+s,n+s+1}\right|\nabla_{\xi^{n+s}}\left|\Psi^{n,n+s,n+s+1}\right|-\vec{\xi}^{n+s+1}\left|\Psi^{n,n+s,n+s+1}\right|^2\nabla_{\xi^{n+s}}\varphi^{n,n+s,n+s+1}-\\
&-i\alpha_n\nabla_{\xi^n}\varphi^{n,n+s,n+s+1}\nabla_{\xi^n}\left|\Psi^{n,n+s,n+s+1}\right|^2-\alpha_n\left|\Psi^{n,n+s,n+s+1}\right|\Delta_{\xi^n}\left|\Psi^{n,n+s,n+s+1}\right|+\\
&+\alpha_n\left|\Psi^{n,n+s,n+s+1}\right|^2\left|\nabla_{\xi^n}\varphi^{n,n+s,n+s+1}\right|^2-i\alpha_n\left|\Psi^{n,n+s,n+s+1}\right|^2\Delta_{\xi^n}\varphi^{n,n+s,n+s+1}+\\
&+i\gamma_n\vec{A}^n_{n,n+s,n+s+1}\left|\Psi^{n,n+s,n+s+1}\right|\nabla_{\xi^n}\left|\Psi^{n,n+s,n+s+1}\right|-\gamma_n\vec{A}^n_{n,n+s,n+s+1}\left|\Psi^{n,n+s,n+s+1}\right|^2\nabla_{\xi^n}\varphi^{n,n+s,n+s+1}+\\
&+\frac{1}{4\alpha_n}\left|\gamma_n\vec{A}^n_{n,n+s,n+s+1}\right|^2\left|\Psi^{n,n+s,n+s+1}\right|^2-i\alpha_{n+s+1}\nabla_{\xi^{n+s+1}}\varphi^{n,n+s,n+s+1}\nabla_{\xi^{n+s+1}}\left|\Psi^{n,n+s,n+s+1}\right|^2-
\end{aligned}$$

$$-\alpha_{n+s+1}\left|\Psi^{n,n+s,n+s+1}\right|\Delta_{\xi^{n+s+1}}\left|\Psi^{n,n+s,n+s+1}\right|+\alpha_{n+s+1}\left|\Psi^{n,n+s,n+s+1}\right|^2\left|\nabla_{\xi^{n+s+1}}\varphi^{n,n+s,n+s+1}\right|^2-$$

$$-i\alpha_{n+s+1}\left|\Psi^{n,n+s,n+s+1}\right|^2\Delta_{\xi^{n+s+1}}\varphi^{n,n+s,n+s+1}+i\frac{\gamma_{n+s+1}}{2}\vec{A}^{n+s+1}_{n,n+s,n+s+1}\nabla_{\xi^{n+s+1}}\left|\Psi^{n,n+s,n+s+1}\right|^2-$$

$$-\gamma_{n+s+1}\vec{A}^{n+s+1}_{n,n+s,n+s+1}\left|\Psi^{n,n+s,n+s+1}\right|^2\nabla_{\xi^{n+s+1}}\varphi^{n,n+s,n+s+1}+\frac{1}{4\alpha_{n+s+1}}\left|\gamma_{n+s+1}\vec{A}^{n+s+1}_{n,n+s,n+s+1}\right|^2\left|\Psi^{n,n+s,n+s+1}\right|^2.$$

or, taking into account that $\operatorname{Im}U^{n,n+s,n+s+1}=0$, we obtain

$$-\frac{1}{\beta_{n+s+1}}\left(\frac{\partial}{\partial t}+\vec{\xi}^{\,n+s+1}\nabla_{\xi^{n+s}}\right)\varphi^{n,n+s,n+s+1}=$$
$$-\frac{1}{4\alpha_n\beta_{n+s+1}}\left|\left\langle\vec{\xi}^{\,n+1}\right\rangle_{n,n+s,n+s+1}\right|^2-\frac{1}{4\alpha_{n+s+1}\beta_{n+s+1}}\left|\left\langle\vec{\xi}^{\,n+s+2}\right\rangle_{n,n+s,n+s+1}\right|^2+ \qquad \text{(B.20)}$$
$$+\frac{\alpha_{n+s+1}}{\beta_{n+s+1}}\frac{\Delta_{\xi^{n+s+1}}\left|\Psi^{n,n+s,n+s+1}\right|}{\left|\Psi^{n,n+s,n+s+1}\right|}+\frac{\alpha_n}{\beta_{n+s+1}}\frac{\Delta_{\xi^n}\left|\Psi^{n,n+s,n+s+1}\right|}{\left|\Psi^{n,n+s,n+s+1}\right|}+U^{n,n+s,n+s+1}.$$

It remains to consider the last type of equations of the third rank with three sources of dissipations (A.39)

$$\beta_{n+s+k}U^{n,n+s,n+s+k}\left|\Psi^{n,n+s,n+s+k}\right|^2=i\left|\Psi^{n,n+s,n+s+k}\right|\frac{\partial\left|\Psi^{n,n+s,n+s+k}\right|}{\partial t}-\left|\Psi^{n,n+s,n+s+k}\right|^2\frac{\partial\varphi^{n,n+s,n+s+k}}{\partial t}$$

$$-i\alpha_n\nabla_{\xi^n}\varphi^{n,n+s,n+s+k}\nabla_{\xi^n}\left|\Psi^{n,n+s,n+s+k}\right|^2-\alpha_n\left|\Psi^{n,n+s,n+s+k}\right|\Delta_{\xi^n}\left|\Psi^{n,n+s,n+s+k}\right|+$$

$$+\alpha_n\left|\Psi^{n,n+s,n+s+k}\right|^2\left|\nabla_{\xi^n}\varphi^{n,n+s,n+s+k}\right|^2-\alpha_n i\left|\Psi^{n,n+s,n+s+k}\right|^2\Delta_{\xi^n}\varphi^{n,n+s,n+s+k}+$$

$$+i\frac{\gamma_n}{2}\vec{A}^{n}_{n,n+s,n+s+k}\nabla_{\xi^n}\left|\Psi^{n,n+s,n+s+k}\right|^2-\gamma_n\vec{A}^{n}_{n,n+s,n+s+k}\left|\Psi^{n,n+s,n+s+k}\right|^2\nabla_{\xi^n}\varphi^{n,n+s,n+s+k}+$$

$$+\frac{1}{4\alpha_n}\left|\gamma_n\vec{A}^{n}_{n,n+s,n+s+k}\right|^2\left|\Psi^{n,n+s,n+s+k}\right|^2-i\alpha_{n+s}\nabla_{\xi^{n+s}}\varphi^{n,n+s,n+s+k}\nabla_{\xi^{n+s}}\left|\Psi^{n,n+s,n+s+k}\right|^2-$$

$$-\alpha_{n+s}\left|\Psi^{n,n+s,n+s+k}\right|\Delta_{\xi^{n+s}}\left|\Psi^{n,n+s,n+s+k}\right|+\alpha_{n+s}\left|\Psi^{n,n+s,n+s+k}\right|^2\left|\nabla_{\xi^{n+s}}\varphi^{n,n+s,n+s+k}\right|^2-$$

$$-i\alpha_{n+s}\left|\Psi^{n,n+s,n+s+k}\right|^2\Delta_{\xi^{n+s}}\varphi^{n,n+s,n+s+k}+i\frac{\gamma_{n+s}}{2}\vec{A}^{n+s}_{n,n+s,n+s+k}\nabla_{\xi^{n+s}}\left|\Psi^{n,n+s,n+s+k}\right|^2-$$

$$-\gamma_{n+s}\vec{A}^{n+s}_{n,n+s,n+s+k}\left|\Psi^{n,n+s,n+s+k}\right|^2\nabla_{\xi^{n+s}}\varphi^{n,n+s,n+s+k}+\frac{1}{4\alpha_{n+s}}\left|\gamma_{n+s}\vec{A}^{n+s}_{n,n+s,n+s+k}\right|^2\left|\Psi^{n,n+s,n+s+k}\right|^2-$$

$$-i\alpha_{n+s+k}\nabla_{\xi^{n+s+k}}\varphi^{n,n+s,n+s+k}\nabla_{\xi^{n+s+k}}\left|\Psi^{n,n+s,n+s+k}\right|^2-\alpha_{n+s+k}\left|\Psi^{n,n+s,n+s+k}\right|\Delta_{\xi^{n+s+k}}\left|\Psi^{n,n+s,n+s+k}\right|+$$

$$+\alpha_{n+s+k}\left|\Psi^{n,n+s,n+s+k}\right|^2\left|\nabla_{\xi^{n+s+k}}\varphi^{n,n+s,n+s+k}\right|^2-i\alpha_{n+s+k}\left|\Psi^{n,n+s,n+s+k}\right|^2\Delta_{\xi^{n+s+k}}\varphi^{n,n+s,n+s+k}+$$

$$+i\frac{\gamma_{n+s+k}}{2}\vec{A}^{n+s+k}_{n,n+s,n+s+k}\nabla_{\xi^{n+s+k}}\left|\Psi^{n,n+s,n+s+k}\right|^2-\gamma_{n+s+k}\vec{A}^{n+s+k}_{n,n+s,n+s+k}\left|\Psi^{n,n+s,n+s+k}\right|^2\nabla_{\xi^{n+s+k}}\varphi^{n,n+s,n+s+k}+ \qquad \text{(B.21)}$$

$$+\frac{1}{4\alpha_{n+s+k}}\left|\gamma_{n+s+k}\vec{A}^{n+s+k}_{n,n+s,n+s+k}\right|^2\left|\Psi^{n,n+s,n+s+k}\right|^2,$$

or

$$-\frac{1}{\beta_{n+s+k}}\frac{\partial\varphi^{n,n+s,n+s+k}}{\partial t}=-\frac{1}{4\alpha_n\beta_{n+s+k}}\left|\left\langle\vec{\xi}^{n+1}\right\rangle_{n,n+s,n+s+k}\right|^2-\frac{1}{4\alpha_{n+s}\beta_{n+s+k}}\left|\left\langle\vec{\xi}^{n+s+1}\right\rangle_{n,n+s,n+s+k}\right|^2-$$
$$-\frac{1}{4\alpha_{n+s+k}\beta_{n+s+k}}\left|\left\langle\vec{\xi}^{n+s+k+1}\right\rangle_{n,n+s,n+s+k}\right|^2+\frac{\alpha_{n+s}}{\beta_{n+s+k}}\frac{\Delta_{\xi^{n+s}}\left|\Psi^{n,n+s,n+s+k}\right|}{\left|\Psi^{n,n+s,n+s+k}\right|}+\frac{\alpha_{n+s+k}}{\beta_{n+s+k}}\frac{\Delta_{\xi^{n+s+k}}\left|\Psi^{n,n+s,n+s+k}\right|}{\left|\Psi^{n,n+s,n+s+k}\right|}+$$
$$+\frac{\alpha_n}{\beta_{n+s+k}}\frac{\Delta_{\xi^n}\left|\Psi^{n,n+s,n+s+k}\right|}{\left|\Psi^{n,n+s,n+s+k}\right|}+U^{n,n+s,n+s+k}. \tag{B.22}$$

Theorem 2 is proved.

## Appendix C

### *Proof of Theorem 3*

We start with equation of the first rank (B.5) for the function $\varphi^n=\varphi^n\left(\vec{\xi}^n,t\right)$. We calculate the value

$$-\frac{1}{\beta_n}\hat{\pi}_n\varphi^n=-\hat{\pi}_n\,\mathrm{S}^n=-\frac{1}{\beta_n}\left(\partial_0\varphi^n+\left\langle\vec{\xi}^{n+1}\right\rangle_n\nabla_{\xi^n}\right)\varphi^n=\mathrm{H}^n+\frac{1}{2\alpha_n\beta_n}\left\langle\vec{\xi}^{n+1}\right\rangle_n\left\langle\vec{\xi}_p^{n+1}\right\rangle_n=$$
$$=-\frac{1}{4\alpha_n\beta_n}\left|\left\langle\vec{\xi}^{n+1}\right\rangle_n\right|^2+\frac{1}{2\alpha_n\beta_n}\left\langle\vec{\xi}^{n+1}\right\rangle_n\left(\left\langle\vec{\xi}^{n+1}\right\rangle_n-\left\langle\vec{\xi}_s^{n+1}\right\rangle_n\right)+\mathrm{V}^n= \tag{C.1}$$
$$=\frac{1}{4\alpha_n\beta_n}\left|\left\langle\vec{\xi}^{n+1}\right\rangle_n\right|^2-\frac{1}{2\alpha_n\beta_n}\left\langle\vec{\xi}^{n+1}\right\rangle_n\left\langle\vec{\xi}_s^{n+1}\right\rangle_n+\mathrm{V}^n=$$
$$=\frac{1}{4\alpha_n\beta_n}\left[\left|\left\langle\vec{\xi}_p^{n+1}\right\rangle_n\right|^2+2\left\langle\vec{\xi}_p^{n+1}\right\rangle_n\left\langle\vec{\xi}_s^{n+1}\right\rangle_n+\left|\left\langle\vec{\xi}_s^{n+1}\right\rangle_n\right|^2-2\left\langle\vec{\xi}_p^{n+1}\right\rangle_n\left\langle\vec{\xi}_s^{n+1}\right\rangle_n-2\left|\left\langle\vec{\xi}_s^{n+1}\right\rangle_n\right|^2\right]+\mathrm{V}^n,$$

$$-\hat{\pi}_n\,\mathrm{S}^n=\frac{1}{4\alpha_n\beta_n}\left[\left|\left\langle\vec{\xi}_p^{n+1}\right\rangle_n\right|^2-\left|\left\langle\vec{\xi}_s^{n+1}\right\rangle_n\right|^2\right]+\mathrm{V}^n=-\mathrm{L}^n. \tag{C.2}$$

The resulting expression (C.2) coincides with equation (2.28), and, from a comparison of (C.1) and (C.2)

$$-\hat{\pi}_n\,\mathrm{S}^n=\mathrm{H}^n+\frac{1}{2\alpha_n\beta_n}\left\langle\vec{\xi}^{n+1}\right\rangle_n\left\langle\vec{\xi}_p^{n+1}\right\rangle_n=-\mathrm{L}^n, \tag{C.3}$$

Legendre transformation (2.29) follows. For equation of the second rank from the first group (B.9)/(2.15), we write

$$-\frac{1}{\beta_{n+1}}\hat{\pi}_{n,n+1}\varphi^{n,n+1}=-\frac{1}{\beta_{n+1}}\left(\partial_n+\left\langle\vec{\xi}^{n+2}\right\rangle_{n,n+1}\nabla_{\xi^{n+1}}\right)\varphi^{n,n+1}=\mathrm{H}^{n,n+1}+\frac{1}{2\alpha_{n+1}\beta_{n+1}}\left\langle\vec{\xi}^{n+2}\right\rangle_{n,n+1}\left\langle\vec{\xi}_p^{n+2}\right\rangle_{n,n+1}=$$
$$=-\frac{1}{4\alpha_{n+1}\beta_{n+1}}\left[\left|\left\langle\vec{\xi}_p^{n+2}\right\rangle_{n,n+1}\right|^2+2\left\langle\vec{\xi}_p^{n+2}\right\rangle_{n,n+1}\left\langle\vec{\xi}_s^{n+2}\right\rangle_{n,n+1}+\left|\left\langle\vec{\xi}_s^{n+2}\right\rangle_{n,n+1}\right|^2\right]+ \tag{C.4}$$
$$+\frac{1}{2\alpha_{n+1}\beta_{n+1}}\left\langle\vec{\xi}^{n+2}\right\rangle_{n,n+1}\left\langle\vec{\xi}_p^{n+2}\right\rangle_{n,n+1}+\mathrm{V}^{n,n+1},$$

$$-\hat{\pi}_{n,n+1}\,\mathrm{S}^{n,n+1}=\frac{1}{4\alpha_{n+1}\beta_{n+1}}\left[\left|\left\langle\vec{\xi}_p^{\,n+2}\right\rangle_{n,n+1}\right|^2-\left|\left\langle\vec{\xi}_s^{\,n+2}\right\rangle_{n,n+1}\right|^2\right]+\mathrm{V}^{n,n+1}=-\mathrm{L}^{n,n+1}. \quad \text{(C.5)}$$

Expression (C.5) proves the validity of equation (2.30), and, from a comparison of (C.4) and (C.5), relation (2.31) follows:

$$-\hat{\pi}_{n,n+1}\,\mathrm{S}^{n,n+1}=\mathrm{H}^{n,n+1}+\frac{1}{2\alpha_{n+1}\beta_{n+1}}\left\langle\vec{\xi}^{\,n+2}\right\rangle_{n,n+1}\left\langle\vec{\xi}_p^{\,n+2}\right\rangle_{n,n+1}=-\mathrm{L}^{n,n+1}. \quad \text{(C.6)}$$

We consider equations of the second rank belonging to the second group (2.16)/(B.14):

$$-\hat{\pi}_{n,n+k}\,\mathrm{S}^{n,n+k}=-\frac{1}{\beta_{n+k}}\left(\frac{\partial}{\partial t}+\left\langle\vec{\xi}^{\,n+1}\right\rangle_{n,n+k}\nabla_{\xi^n}+\left\langle\vec{\xi}^{\,n+1+k}\right\rangle_{n,n+k}\nabla_{\xi^{n+k}}\right)\varphi^{n,n+k}=\mathrm{H}^{n,n+k}- \quad \text{(C.7)}$$
$$-\frac{\alpha_n}{2\alpha_n\beta_{n+k}}\left\langle\vec{\xi}^{\,n+1}\right\rangle_{n,n+k}\nabla_{\xi^n}\Phi^{n,n+k}-\frac{\alpha_{n+k}}{2\alpha_{n+k}\beta_{n+k}}\left\langle\vec{\xi}^{\,n+1+k}\right\rangle_{n,n+k}\nabla_{\xi^{n+k}}\Phi^{n,n+k}=-\frac{1}{4\alpha_{n+k}\beta_{n+k}}\left|\left\langle\vec{\xi}^{\,n+k+1}\right\rangle_{n,n+k}\right|^2-$$
$$-\frac{1}{4\alpha_n\beta_{n+k}}\left|\left\langle\vec{\xi}^{\,n+1}\right\rangle_{n,n+k}\right|^2+\mathrm{V}^{n,n+k}+\frac{1}{2\alpha_n\beta_{n+k}}\left\langle\vec{\xi}^{\,n+1}\right\rangle_{n,n+k}\left\langle\vec{\xi}_p^{\,n+1}\right\rangle_{n,n+k}+$$
$$+\frac{1}{2\alpha_{n+k}\beta_{n+k}}\left\langle\vec{\xi}^{\,n+1+k}\right\rangle_{n,n+k}\left\langle\vec{\xi}_p^{\,n+k+1}\right\rangle_{n,n+k},$$

$$-\hat{\pi}_{n,n+k}\,\mathrm{S}^{n,n+k}=-\frac{1}{4\alpha_{n+k}\beta_{n+k}}\left|\left\langle\vec{\xi}^{\,n+k+1}\right\rangle_{n,n+k}\right|^2+\frac{1}{2\alpha_{n+k}\beta_{n+k}}\left|\left\langle\vec{\xi}^{\,n+k+1}\right\rangle_{n,n+k}\right|^2-$$
$$-\frac{1}{2\alpha_{n+k}\beta_{n+k}}\left\langle\vec{\xi}^{\,n+1+k}\right\rangle_{n,n+k}\left\langle\vec{\xi}_s^{\,n+k+1}\right\rangle_{n,n+k}-\frac{1}{4\alpha_n\beta_{n+k}}\left|\left\langle\vec{\xi}^{\,n+1}\right\rangle_{n,n+k}\right|^2+\frac{1}{2\alpha_n\beta_{n+k}}\left|\left\langle\vec{\xi}^{\,n+1}\right\rangle_{n,n+k}\right|^2- \quad \text{(C.8)}$$
$$-\frac{1}{2\alpha_n\beta_{n+k}}\left\langle\vec{\xi}^{\,n+1}\right\rangle_{n,n+k}\left\langle\vec{\xi}_s^{\,n+1}\right\rangle_{n,n+k}+\mathrm{V}^{n,n+k},$$

from here

$$-\hat{\pi}_{n,n+k}\,\mathrm{S}^{n,n+k}=\frac{1}{4\alpha_{n+k}\beta_{n+k}}\left[\left|\left\langle\vec{\xi}_p^{\,n+k+1}\right\rangle_{n,n+k}\right|^2-\left|\left\langle\vec{\xi}_s^{\,n+k+1}\right\rangle_{n,n+k}\right|^2\right]+$$
$$+\frac{1}{4\alpha_n\beta_{n+k}}\left[\left|\left\langle\vec{\xi}_p^{\,n+1}\right\rangle_{n,n+k}\right|^2-\left|\left\langle\vec{\xi}_s^{\,n+1}\right\rangle_{n,n+k}\right|^2\right]+\mathrm{V}^{n,n+k}=-\mathrm{L}^{n,n+k}. \quad \text{(C.9)}$$

Resulting expression (C.9) coincides with equation (2.32), and, from a comparison of (C.9) and (C.7), relation (2.33) follows.

We consider equations of the third rank from the first group (2.17)/(B.17)

$$-\hat{\pi}_{n,n+1,n+2}\,\mathrm{S}^{n,n+1,n+2}=-\frac{1}{2\beta_{n+2}}\left(\partial_{n,n+1}+\left\langle\vec{\xi}^{\,n+3}\right\rangle_{n,n+1,n+2}\nabla_{\xi^{n+2}}\right)\Phi^{n,n+1,n+2}=\mathrm{H}^{n,n+1,n+2}+$$
$$+\frac{1}{2\alpha_{n+2}\beta_{n+2}}\left\langle\vec{\xi}^{\,n+3}\right\rangle_{n,n+1,n+2}\left\langle\vec{\xi}_p^{\,n+3}\right\rangle_{n,n+1,n+2}=-\frac{1}{4\alpha_{n+2}\beta_{n+2}}\left|\left\langle\vec{\xi}^{\,n+3}\right\rangle_{n,n+1,n+2}\right|^2+\mathrm{V}^{n,n+1,n+2}+ \quad \text{(C.10)}$$
$$+\frac{1}{2\alpha_{n+2}\beta_{n+2}}\left|\left\langle\vec{\xi}^{\,n+3}\right\rangle_{n,n+1,n+2}\right|^2-\frac{1}{2\alpha_{n+2}\beta_{n+2}}\left\langle\vec{\xi}^{\,n+3}\right\rangle_{n,n+1,n+2}\left\langle\vec{\xi}_s^{\,n+3}\right\rangle_{n,n+1,n+2},$$

or

$$-\hat{\pi}_{n,n+1,n+2}\,\mathrm{S}^{n,n+1,n+2}=\frac{1}{4\alpha_{n+2}\beta_{n+2}}\left[\left|\left\langle\vec{\xi}_p^{\,n+3}\right\rangle_{n,n+1,n+2}\right|^2-\left|\left\langle\vec{\xi}_s^{\,n+3}\right\rangle_{n,n+1,n+2}\right|^2\right]+\mathrm{V}^{n,n+1,n+2}=-\mathrm{L}^{n,n+1,n+2}. \quad \text{(C.11)}$$

Equation (C.11) corresponds to equation (2.34), and, from expressions (C.10) and (C.11), transformation (2.35) follows. For equations of the third rank from the second group, we consider equations (2.18)/(B.19)

$$\begin{aligned}
&-\frac{1}{\beta_{n+1+k}}\hat{\pi}_{n,n+1,n+1+k}\varphi^{n,n+1,n+1+k}=\\
&=-\frac{1}{2\beta_{n+1+k}}\left(\partial_n+\left\langle\vec{\xi}^{\,n+2}\right\rangle_{n,n+1,n+1+k}\nabla_{\xi^{n+1}}+\left\langle\vec{\xi}^{\,n+2+k}\right\rangle_{n,n+1,n+1+k}\nabla_{\xi^{n+1+k}}\right)\Phi^{n,n+1,n+1+k}=\\
&=\mathrm{H}^{n,n+1,n+1+k}+\frac{1}{2\alpha_{n+1}\beta_{n+1+k}}\left\langle\vec{\xi}^{\,n+2}\right\rangle_{n,n+1,n+1+k}\left\langle\vec{\xi}_p^{\,n+2}\right\rangle_{n,n+1,n+1+k}+\\
&+\frac{1}{2\alpha_{n+1+k}\beta_{n+1+k}}\left\langle\vec{\xi}^{\,n+2+k}\right\rangle_{n,n+1,n+1+k}\left\langle\vec{\xi}_p^{\,n+2+k}\right\rangle_{n,n+1,n+1+k},
\end{aligned} \quad \text{(C.12)}$$

$$\begin{aligned}
&-\frac{1}{\beta_{n+1+k}}\hat{\pi}_{n,n+1,n+1+k}\varphi^{n,n+1,n+1+k}=\frac{1}{4\alpha_{n+1+k}\beta_{n+1+k}}\left[\left|\left\langle\vec{\xi}_p^{\,n+2+k}\right\rangle_{n,n+1,n+1+k}\right|^2-\left|\left\langle\vec{\xi}_s^{\,n+2+k}\right\rangle_{n,n+1,n+1+k}\right|^2\right]+\\
&+\frac{1}{4\alpha_{n+1}\beta_{n+1+k}}\left[\left|\left\langle\vec{\xi}_p^{\,n+2}\right\rangle_{n,n+1,n+1+k}\right|^2-\left|\left\langle\vec{\xi}_s^{\,n+2}\right\rangle_{n,n+1,n+1+k}\right|^2\right]+\mathrm{V}^{n,n+1,n+1+k}=-\mathrm{L}^{n,n+1,n+1+k}.
\end{aligned} \quad \text{(C.13)}$$

Expressions (C.12), (C.13) show the validity of representations (2.36) and (2.37). For equation (2.18)/(B.20),

$$\begin{aligned}
&-\hat{\pi}_{n,n+s,n+s+1}\,\mathrm{S}^{n,n+s,n+s+1}=\\
&=-\frac{1}{2\beta_{n+s+1}}\left(\partial_{n+s}+\left\langle\vec{\xi}^{\,n+1}\right\rangle_{n,n+s,n+s+1}\nabla_{\xi^{n}}+\left\langle\vec{\xi}^{\,n+s+2}\right\rangle_{n,n+s,n+s+1}\nabla_{\xi^{n+s+1}}\right)\Phi^{n,n+s,n+s+1}=\mathrm{H}^{n,n+s,n+s+1}+\\
&+\frac{1}{2\alpha_n\beta_{n+s+1}}\left\langle\vec{\xi}^{\,n+1}\right\rangle_{n,n+s,n+s+1}\left\langle\vec{\xi}_p^{\,n+1}\right\rangle_{n,n+s,n+s+1}+\frac{1}{2\alpha_{n+s+1}\beta_{n+s+1}}\left\langle\vec{\xi}^{\,n+s+2}\right\rangle_{n,n+s,n+s+1}\left\langle\vec{\xi}_p^{\,n+s+2}\right\rangle_{n,n+s,n+s+1}=\\
&=-\frac{1}{4\alpha_n\beta_{n+s+1}}\left|\left\langle\vec{\xi}^{\,n+1}\right\rangle_{n,n+s,n+s+1}\right|^2-\frac{1}{4\alpha_{n+s+1}\beta_{n+s+1}}\left|\left\langle\vec{\xi}^{\,n+s+2}\right\rangle_{n,n+s,n+s+1}\right|^2+\mathrm{V}^{n,n+s,n+s+1}+\\
&+\frac{1}{2\alpha_n\beta_{n+s+1}}\left|\left\langle\vec{\xi}^{\,n+1}\right\rangle_{n,n+s,n+s+1}\right|^2-\frac{1}{2\alpha_n\beta_{n+s+1}}\left\langle\vec{\xi}^{\,n+1}\right\rangle_{n,n+s,n+s+1}\left\langle\vec{\xi}_s^{\,n+1}\right\rangle_{n,n+s,n+s+1}+\\
&+\frac{1}{2\alpha_{n+s+1}\beta_{n+s+1}}\left|\left\langle\vec{\xi}^{\,n+s+2}\right\rangle_{n,n+s,n+s+1}\right|^2-\frac{1}{2\alpha_{n+s+1}\beta_{n+s+1}}\left\langle\vec{\xi}^{\,n+s+2}\right\rangle_{n,n+s,n+s+1}\left\langle\vec{\xi}_p^{\,n+s+2}\right\rangle_{n,n+s,n+s+1},
\end{aligned} \quad \text{(C.14)}$$

$$\begin{aligned}
&-\hat{\pi}_{n,n+s,n+s+1}\,\mathrm{S}^{n,n+s,n+s+1}=\frac{1}{4\alpha_n\beta_{n+s+1}}\left[\left|\left\langle\vec{\xi}_p^{\,n+1}\right\rangle_{n,n+s,n+s+1}\right|^2-\left|\left\langle\vec{\xi}_s^{\,n+1}\right\rangle_{n,n+s,n+s+1}\right|^2\right]+\\
&+\frac{1}{4\alpha_{n+s+1}\beta_{n+s+1}}\left[\left|\left\langle\vec{\xi}_p^{\,n+s+2}\right\rangle_{n,n+s,n+s+1}\right|^2-\left|\left\langle\vec{\xi}_s^{\,n+s+2}\right\rangle_{n,n+s,n+s+1}\right|^2\right]+\mathrm{V}^{n,n+s,n+s+1}=-\mathrm{L}^{n,n+s,n+s+1}.
\end{aligned} \quad \text{(C.15)}$$

As a result, from expressions (C.14) and (C.15), we obtain representations (2.38) and (2.39). There remains the last type of equations of the third rank with three sources of dissipations (2.19)/(B.22)

$$-\hat{\pi}_{n,n+s,n+s+k}\,\mathrm{S}^{n,n+s,n+s+k}=\mathrm{H}^{n,n+s,n+s+k}+\frac{1}{2\alpha_n\beta_{n+s+k}}\left\langle\vec{\xi}^{\,n+1}\right\rangle_{n,n+s,n+s+k}\left\langle\vec{\xi}_p^{\,n+1}\right\rangle_{n,n+s,n+s+k}+ \tag{C.16}$$
$$+\frac{1}{2\alpha_{n+s}\beta_{n+s+k}}\left\langle\vec{\xi}^{\,n+1+s}\right\rangle_{n,n+s,n+s+k}\left\langle\vec{\xi}_p^{\,n+1+s}\right\rangle_{n,n+s,n+s+k}+\frac{1}{2\alpha_{n+s+k}\beta_{n+s+k}}\left\langle\vec{\xi}^{\,n+1+s+k}\right\rangle_{n,n+s,n+s+k}\left\langle\vec{\xi}_p^{\,n+1+s+k}\right\rangle_{n,n+s,n+s+k},$$

$$-\hat{\pi}_{n,n+s,n+s+k}\,\mathrm{S}^{n,n+s,n+s+k}=\frac{1}{4\alpha_n\beta_{n+s+k}}\left[\left|\left\langle\vec{\xi}_p^{\,n+1}\right\rangle_{n,n+s,n+s+k}\right|^2-\left|\left\langle\vec{\xi}_s^{\,n+1}\right\rangle_{n,n+s,n+s+k}\right|^2\right]+$$
$$+\frac{1}{4\alpha_{n+s}\beta_{n+s+k}}\left[\left|\left\langle\vec{\xi}_p^{\,n+1+s}\right\rangle_{n,n+s,n+s+k}\right|^2-\left|\left\langle\vec{\xi}_s^{\,n+1+s}\right\rangle_{n,n+s,n+s+k}\right|^2\right]+ \tag{C.17}$$
$$+\frac{1}{4\alpha_{n+s+k}\beta_{n+s+k}}\left[\left|\left\langle\vec{\xi}_p^{\,n+1+s+k}\right\rangle_{n,n+s,n+s+k}\right|^2-\left|\left\langle\vec{\xi}_s^{\,n+1+s+k}\right\rangle_{n,n+s,n+s+k}\right|^2\right]+\mathrm{V}^{n,n+s,n+s+k}=-\mathrm{L}^{n,n+s,n+s+k}.$$

Expressions (C.16) and (C.17) prove the validity of equations (2.40) and (2.41). Theorem 3 is proved.

## Appendix D

***Proof of Theorem 4***

Equation (2.40) with the complex Lagrange function $\mathcal{L}^{n_1\dots n_R}$ (2.42), (2.45), (2.48), (2.51), (2.54), (2.57), (2.60) follows directly from definition (2.38) and equation (2.28):

$$\mathcal{L}^{n_1\dots n_R}=\hat{\pi}_{n_1\dots n_R}\mathcal{Z}^{n_1\dots n_R}=\frac{1}{2}\hat{\pi}_{n_1\dots n_R}S^{n_1\dots n_R}+i\beta_{n_R}\hat{\pi}_{n_1\dots n_R}\,\mathrm{S}^{n_1\dots n_R}=\frac{1}{2}\hat{\pi}_{n_1\dots n_R}S^{n_1\dots n_R}+i\beta_{n_R}\,\mathrm{L}^{n_1\dots n_R}, \tag{D.1}$$

where value $\hat{\pi}_{n_1\dots n_R}S^{n_1\dots n_R}$ is determined from dispersion chain of the Vlasov equations (i.3)-(i.5).

We obtain equations (2.41), (2.44), (2.47), (2.50), (2.53), (2.56), (2.59). Let us first take into account the validity of the expression (2.62)

$$f^{n_1\dots n_R}\mathcal{Q}^{\ell}_{n_1\dots n_R}=f^{n_1\dots n_R}\,\mathrm{div}_{\xi^\ell}\left\langle\vec{\xi}^{\,\ell+1}\right\rangle_{n_1\dots n_R}+\left\langle\vec{\xi}^{\,\ell+1}\right\rangle_{n_1\dots n_R}\nabla_{\xi^\ell}f^{n_1\dots n_R},$$
$$-\left\langle\vec{\xi}^{\,\ell+1}\right\rangle_{n_1\dots n_R}\nabla_{\xi^\ell}S^{n_1\dots n_R}=\mathrm{Q}^{\ell}_{n_1\dots n_R}-\mathcal{Q}^{\ell}_{n_1\dots n_R}. \tag{D.2}$$

We start from equation (2.41)

$$-\partial_0\mathcal{Z}^n=-\frac{1}{2}\partial_0 S^n-i\partial_0\varphi^n=-\frac{1}{2}\partial_0 S^n+i\beta_n\,\mathrm{H}^n, \tag{D.2}$$

where, according to equation (i.22) and expression (D.2)

$$\partial_0 S^n=-Q^n_n-\left\langle\vec{\xi}^{\,n+1}\right\rangle_n\nabla_{\xi^n}S^n=-Q^n_n+Q^n_n-\mathcal{Q}^n_n=-\mathcal{Q}^n_n. \tag{D.3}$$

Substituting (D.3) into expression (D.2), we obtain the validity of equation (2.41). For equations of the second rank (2.44), taking into account equation (2.15), we obtain

$$-\partial_n \mathcal{Z}^{n,n+1} = -\frac{1}{2}\partial_n S^{n,n+1} - i\partial_n \varphi^{n,n+1} = -\frac{1}{2}\partial_n S^{n,n+1} + i\beta_{n+1}\,\mathrm{H}^{n,n+1}, \tag{D.4}$$

where, according to equation (i.23) and (D.2)

$$\partial_n S^{n,n+1} = -Q^{n+1}_{n,n+1} - \left\langle \vec{\xi}^{\,n+2} \right\rangle_{n,n+1} \nabla_{\xi^{n+1}} S^{n,n+1} = -\mathcal{Q}^{n+1}_{n,n+1}. \tag{D.5}$$

Equation (2.44) follows from expressions (D.4) and (D.5). Equation (2.47) has two sources of dissipations

$$-\partial_0 \mathcal{Z}^{n,n+k} = -\frac{1}{2}\partial_0 S^{n,n+k} - i\partial_0 \varphi^{n,n+k} = -\frac{1}{2}\partial_0 S^{n,n+k} + i\beta_{n+k}\,\mathrm{H}^{n,n+k}, \tag{D.6}$$

$$\partial_0 S^{n,n+k} = -\left(Q^n_{n,n+k} + Q^{n+k}_{n,n+k}\right) - \left\langle \vec{\xi}^{\,n+1} \right\rangle_{n,n+k} \nabla_{\xi^n} S^{n,n+k} - \left\langle \vec{\xi}^{\,n+1+k} \right\rangle_{n,n+k} \nabla_{\xi^{n+k}} S^{n,n+k},$$

$$\partial_0 S^{n,n+k} = -\left(Q^n_{n,n+k} + Q^{n+k}_{n,n+k}\right) + Q^n_{n,n+k} - \mathcal{Q}^n_{n,n+k} + Q^{n+k}_{n,n+k} - \mathcal{Q}^{n+k}_{n,n+k} = -\mathcal{Q}^n_{n,n+k} - \mathcal{Q}^{n+k}_{n,n+k}, \tag{D.7}$$

where expression (D.2) and equations (i.23), (2.16) are taken into account. Substituting (D.7) into (D.6) gives the validity of equation (2.47).

We consider third-rank equations (2.50), (2.53), (2.56), and (2.59). Equation (2.50) has only one source of dissipations. Equations (2.53) and (2.56) contain two sources each, and equation (2.59) contains three sources of dissipations. The proof for equations (2.50), (2.53) and (2.56) has a structure similar to (D.2)-(D.7), therefore, without loss of generality, we consider the proof for the equation with three sources (2.59).

$$-\partial_0 \mathcal{Z}^{n,n+s,n+s+k} = -\frac{1}{2}\partial_0 S^{n,n+s,n+s+k} + i\beta_{n+s+k}\,\mathrm{H}^{n,n+s,n+s+k}, \tag{D.8}$$

$$\begin{aligned}
\partial_0 S^{n,n+s,n+s+k} &= -\left(Q^n_{n,n+s,n+s+k} + Q^{n+s}_{n,n+s,n+s+k} + Q^{n+s+k}_{n,n+s,n+s+k}\right) - \left\langle \vec{\xi}^{\,n+1} \right\rangle_{n,n+s,n+s+k} \nabla_{\xi^n} S^{n,n+s,n+s+k} - \\
&- \left\langle \vec{\xi}^{\,n+1+s} \right\rangle_{n,n+s,n+s+k} \nabla_{\xi^{n+s}} S^{n,n+s,n+s+k} - \left\langle \vec{\xi}^{\,n+1+s+k} \right\rangle_{n,n+s,n+s+k} \nabla_{\xi^{n+s+k}} S^{n,n+s,n+s+k},
\end{aligned}$$

$$\begin{aligned}
\partial_0 S^{n,n+s,n+s+k} &= -\left(Q^n_{n,n+s,n+s+k} + Q^{n+s}_{n,n+s,n+s+k} + Q^{n+s+k}_{n,n+s,n+s+k}\right) + Q^n_{n,n+s,n+s+k} - \mathcal{Q}^n_{n,n+s,n+s+k} + \\
&+ Q^{n+s}_{n,n+s,n+s+k} - \mathcal{Q}^{n+s}_{n,n+s,n+s+k} + Q^{n+s+k}_{n,n+s,n+s+k} - \mathcal{Q}^{n+s+k}_{n,n+s,n+s+k} = -\mathcal{Q}^n_{n,n+s,n+s+k} - \mathcal{Q}^{n+s}_{n,n+s,n+s+k} - \mathcal{Q}^{n+s+k}_{n,n+s,n+s+k}.
\end{aligned} \tag{D.9}$$

Substituting expression (D.9) into (D.8) proves the validity of equation (2.59). Transformations (2.43), (2.46), (2.49), (2.52), (2.55), (2.58), (2.61) are obtained by adding corresponding expressions (2.42), (2.45), (2.48), (2.51), (2.54), (2.57), (2.60) for functions $\mathcal{L}^{n_1 \ldots n_R}$ and $\mathcal{H}^{n_1 \ldots n_R}$ taking into account transformations (2.29)-(2.35). Theorem 4 is proved.

***Proof of Theorem 5***

Let us differentiate equation (2.14) and take into account (1.10), we obtain

$$-2\alpha_n \partial_0 \nabla_{\xi^n} \varphi^n = \partial_0 \left( \left\langle \vec{\xi}^{\,n+1} \right\rangle_n - \gamma_n \vec{A}^n_n \right) = 2\alpha_n \beta_n \nabla_{\xi^n}\,\mathrm{H}^n. \tag{D.10}$$

Let us transform the right side of equation (D.10). The following relation is true

$$\left[\left\langle \vec{\xi}^{\,n+1}\right\rangle_n, \left[\nabla_{\xi^n}, \left\langle \vec{\xi}^{\,n+1}\right\rangle_n\right]\right] = \frac{1}{2}\nabla_{\xi^n}\left|\left\langle \vec{\xi}^{\,n+1}\right\rangle_n\right|^2 - \left\langle \vec{\xi}^{\,n+1}\right\rangle_n \nabla_{\xi^n}\left\langle \vec{\xi}^{\,n+1}\right\rangle_n,$$

$$\frac{1}{2}\nabla_{\xi^n}\left|\left\langle \vec{\xi}^{\,n+1}\right\rangle_n\right|^2 = \left\langle \vec{\xi}^{\,n+1}\right\rangle_n \nabla_{\xi^n}\left\langle \vec{\xi}^{\,n+1}\right\rangle_n + \gamma_n \left\langle \vec{\xi}^{\,n+1}\right\rangle_n \times \vec{B}_n^n, \tag{D.11}$$

consequently,

$$2\alpha_n\beta_n\nabla_{\xi^n}\mathrm{H}^n = -\left\langle \vec{\xi}^{\,n+1}\right\rangle_n \nabla_{\xi^n}\left\langle \vec{\xi}^{\,n+1}\right\rangle_n - \gamma_n\left\langle \vec{\xi}^{\,n+1}\right\rangle_n \times \vec{B}_n^n + 2\alpha_n\beta_n\nabla_{\xi^n}\mathrm{V}^n. \tag{D.12}$$

We substitute (D.12) into equation (D.10)

$$\left(\partial_0 + \left\langle \vec{\xi}^{\,n+1}\right\rangle_n \nabla_{\xi^n}\right)\left\langle \vec{\xi}^{\,n+1}\right\rangle_n = \gamma_n\partial_0\vec{A}_n^n + 2\alpha_n\beta_n\nabla_{\xi^n}\mathrm{V}^n - \gamma_n\left\langle \vec{\xi}^{\,n+1}\right\rangle_n \times \vec{B}_n^n. \tag{D.13}$$

Expression (D.13) proves the validity of equation (3.13). We consider second-rank equations (2.15) and (2.16). Let us start with equation (2.15).

$$-2\alpha_{n+1}\nabla_{\xi^{n+1}}\partial_n\varphi^{n,n+1} = -\alpha_{n+1}\partial_n\nabla_{\xi^{n+1}}\Phi^{n,n+1} - \alpha_{n+1}\nabla_{\xi^n}\Phi^{n,n+1} = 2\alpha_{n+1}\beta_{n+1}\nabla_{\xi^{n+1}}\mathrm{H}^{n,n+1}, \tag{D.14}$$

where it is taken into account that

$$\begin{aligned}
-\alpha_{n+1}\frac{\partial}{\partial\xi_\lambda^{n+1}}\partial_n\Phi^{n,n+1} &= -\alpha_{n+1}\frac{\partial}{\partial t}\frac{\partial\Phi^{n,n+1}}{\partial\xi_\lambda^{n+1}} - \alpha_{n+1}\frac{\partial}{\partial\xi_\lambda^{n+1}}\left(\xi_\mu^{n+1}\frac{\partial}{\partial\xi_\mu^n}\Phi^{n,n+1}\right) = \\
&= -\alpha_{n+1}\frac{\partial}{\partial t}\frac{\partial\Phi^{n,n+1}}{\partial\xi_\lambda^{n+1}} - \alpha_{n+1}\frac{\partial\xi_\mu^{n+1}}{\partial\xi_\lambda^{n+1}}\frac{\partial}{\partial\xi_\mu^n}\Phi^{n,n+1} - \alpha_{n+1}\left(\xi_\mu^{n+1}\frac{\partial}{\partial\xi_\lambda^{n+1}}\frac{\partial}{\partial\xi_\mu^n}\Phi^{n,n+1}\right) = \\
&= -\alpha_{n+1}\left(\frac{\partial}{\partial t} + \xi_\mu^{n+1}\frac{\partial}{\partial\xi_\mu^n}\right)\frac{\partial\Phi^{n,n+1}}{\partial\xi_\lambda^{n+1}} - \alpha_{n+1}\frac{\partial}{\partial\xi_\lambda^n}\Phi^{n,n+1}.
\end{aligned} \tag{D.15}$$

From expression (D.14), it follows that

$$\partial_n\left(\left\langle \vec{\xi}^{\,n+2}\right\rangle_{n,n+1} - \gamma_{n+1}\vec{A}_{n,n+1}^{n+1}\right) = \alpha_{n+1}\nabla_{\xi^n}\Phi^{n,n+1} + 2\alpha_{n+1}\beta_{n+1}\nabla_{\xi^{n+1}}\mathrm{H}^{n,n+1}, \tag{D.16}$$

where

$$\begin{aligned}
2\alpha_{n+1}\beta_{n+1}\nabla_{\xi^{n+1}}\mathrm{H}^{n,n+1} &= -\left\langle \vec{\xi}^{\,n+2}\right\rangle_{n,n+1}\nabla_{\xi^{n+1}}\left\langle \vec{\xi}^{\,n+2}\right\rangle_{n,n+1} - \\
&\quad -\gamma_{n+1}\left\langle \vec{\xi}^{\,n+2}\right\rangle_{n,n+1} \times \vec{B}_{n,n+1}^{n+1} + 2\alpha_{n+1}\beta_{n+1}\nabla_{\xi^{n+1}}\mathrm{V}^{n,n+1},
\end{aligned}$$

as a result

$$\begin{aligned}
&\left(\partial_n + \left\langle \vec{\xi}^{\,n+2}\right\rangle_{n,n+1}\nabla_{\xi^{n+1}}\right)\left\langle \vec{\xi}^{\,n+2}\right\rangle_{n,n+1} = \\
&= \gamma_{n+1}\partial_n\vec{A}_{n,n+1}^{n+1} + \alpha_{n+1}\nabla_{\xi^n}\Phi^{n,n+1} + 2\alpha_{n+1}\beta_{n+1}\nabla_{\xi^{n+1}}\mathrm{V}^{n,n+1} - \gamma_{n+1}\left\langle \vec{\xi}^{\,n+2}\right\rangle_{n,n+1} \times \vec{B}_{n,n+1}^{n+1}.
\end{aligned} \tag{D.17}$$

Expression (D.17) gives equation (3.14). We consider equation (2.16), which contains two sources:

$$-2\alpha_{n+k}\partial_0\varphi^{n,n+k} = -\alpha_{n+k}\partial_0\Phi^{n,n+k} = 2\alpha_{n+k}\beta_{n+k}\,\mathrm{H}^{n,n+k}, \tag{D.18}$$

where

$$\mathrm{H}^{n,n+k} = -\frac{1}{4\alpha_{n+k}\beta_{n+k}}\left|\left\langle \vec{\xi}^{\,n+k+1}\right\rangle_{n,n+k}\right|^2 - \frac{\beta_n}{\beta_{n+k}}\frac{1}{4\alpha_n\beta_n}\left|\left\langle \vec{\xi}^{\,n+1}\right\rangle_{n,n+k}\right|^2 + \mathrm{V}^{n,n+k}, \tag{D.19}$$

Here, differentiation with respect to variables $n$ and $n+k$ is possible. Let us start with $\nabla_{\xi^{n+k}}$

$$-\alpha_{n+k}\partial_0\nabla_{\xi^{n+k}}\Phi^{n,n+k} = \partial_0\left(\left\langle \vec{\xi}^{\,n+k+1}\right\rangle_{n,n+k} - \gamma_{n+k}\vec{A}^{n+k}_{n,n+k}\right) = 2\alpha_{n+k}\beta_{n+k}\nabla_{\xi^{n+k}}\,\mathrm{H}^{n,n+k},$$

$$\partial_0\left\langle \vec{\xi}^{\,n+k+1}\right\rangle_{n,n+k} = \gamma_{n+k}\partial_0\vec{A}^{n+k}_{n,n+k} + 2\alpha_{n+k}\beta_{n+k}\nabla_{\xi^{n+k}}\tau^{n,n+k}_{n+k} + 2\alpha_{n+k}\beta_{n+k}\nabla_{\xi^{n+k}}\left(\mathrm{V}^{n,n+k} + \frac{\beta_n}{\beta_{n+k}}\tau^{n,n+k}_n\right), \tag{D.20}$$

where expressions (2.1), (2.4) are taken into account. Due to representation (D.11) for $\nabla_{\xi^{n+k}}\tau^{n,n+k}_{n+k}$ the following is true

$$\begin{aligned}-2\alpha_{n+k}\beta_{n+k}\nabla_{\xi^{n+k}}\tau^{n,n+k}_{n+k} &= \frac{1}{2}\nabla_{\xi^{n+k}}\left|\left\langle \vec{\xi}^{\,n+k+1}\right\rangle_{n,n+k}\right|^2 = \\ &= \left\langle \vec{\xi}^{\,n+k+1}\right\rangle_{n,n+k}\nabla_{\xi^{n+k}}\left\langle \vec{\xi}^{\,n+k+1}\right\rangle_{n,n+k} + \gamma_{n+k}\left\langle \vec{\xi}^{\,n+k+1}\right\rangle_{n,n+k}\times\vec{B}^{n+k}_{n,n+k}.\end{aligned} \tag{D.21}$$

Substituting (D.21) into equation (D.20), we obtain

$$\begin{aligned}&\left(\partial_0 + \left\langle \vec{\xi}^{\,n+k+1}\right\rangle_{n,n+k}\nabla_{\xi^{n+k}}\right)\left\langle \vec{\xi}^{\,n+k+1}\right\rangle_{n,n+k} = \gamma_{n+k}\partial_0\vec{A}^{n+k}_{n,n+k} + 2\alpha_{n+k}\beta_{n+k}\nabla_{\xi^{n+k}}\left(\mathrm{V}^{n,n+k} + \frac{\beta_n}{\beta_{n+k}}\tau^{n,n+k}_n\right) - \\ &-\gamma_{n,n+k}\left\langle \vec{\xi}^{\,n+k+1}\right\rangle_{n,n+k}\times\vec{B}^{n+k}_{n,n+k},\end{aligned} \tag{D.22}$$

which coincides with equation (3.15). When differentiating equation (2.16) with respect to variable $n$, we obtain

$$\begin{aligned}&-\alpha_n\partial_0\nabla_{\xi^n}\Phi^{n,n+k} = \partial_0\left(\left\langle \vec{\xi}^{\,n+1}\right\rangle_{n,n+k} - \gamma_n\vec{A}^n_{n,n+k}\right) = 2\alpha_n\beta_{n+k}\nabla_{\xi^n}\,\mathrm{H}^{n,n+k}, \\ &\partial_0\left\langle \vec{\xi}^{\,n+1}\right\rangle_{n,n+k} = \gamma_n\partial_0\vec{A}^n_{n,n+k} + 2\alpha_n\beta_n\nabla_{\xi^n}\tau^{n,n+k}_n + 2\alpha_n\beta_{n+k}\nabla_{\xi^n}\left(\mathrm{V}^{n,n+k} + \tau^{n,n+k}_{n+k}\right).\end{aligned} \tag{D.23}$$

Given representation (D.11),

$$\begin{aligned}-2\alpha_n\beta_n\nabla_{\xi^n}\tau^{n,n+k}_n &= \frac{1}{2}\nabla_{\xi^n}\left|\left\langle \vec{\xi}^{\,n+1}\right\rangle_{n,n+k}\right|^2 = \\ &= \left\langle \vec{\xi}^{\,n+1}\right\rangle_{n,n+k}\nabla_{\xi^n}\left\langle \vec{\xi}^{\,n+1}\right\rangle_{n,n+k} + \gamma_n\left\langle \vec{\xi}^{\,n+1}\right\rangle_{n,n+k}\times\vec{B}^n_{n,n+k},\end{aligned} \tag{D.24}$$

we obtain

$$\left(\partial_0+\left\langle\vec{\xi}^{\,n+1}\right\rangle_{n,n+k}\nabla_{\xi^n}\right)\left\langle\vec{\xi}^{\,n+1}\right\rangle_{n,n+k}=\gamma_n\partial_0\vec{A}^n_{n,n+k}+2\alpha_n\beta_{n+k}\nabla_{\xi^n}\left(\mathrm{V}^{n,n+k}+\tau^{n,n+k}_{n+k}\right)-$$
$$-\gamma_n\left\langle\vec{\xi}^{\,n+1}\right\rangle_{n,n+k}\times\vec{B}^n_{n,n+k}. \tag{D.25}$$

Equation (D.25) corresponds to (3.16). We consider equation of the third rank (2.17). We transform the order of differentiation of $\nabla_{\xi^{n+2}}\partial_{n,n+1}$:

$$\frac{\partial}{\partial\xi^{n+2}_\lambda}\left(\frac{\partial}{\partial t}+\xi^{n+1}_\mu\frac{\partial}{\partial\xi^n_\mu}+\xi^{n+2}_\mu\frac{\partial}{\partial\xi^{n+1}_\mu}\right)=\left(\frac{\partial}{\partial t}+\xi^{n+1}_\mu\frac{\partial}{\partial\xi^n_\mu}+\xi^{n+2}_\mu\frac{\partial}{\partial\xi^{n+1}_\mu}\right)\frac{\partial}{\partial\xi^{n+2}_\lambda}+\frac{\partial}{\partial\xi^{n+1}_\lambda}, \tag{D.26}$$

or

$$\nabla_{\xi^{n+2}}\partial_{n,n+1}=\partial_{n,n+1}\nabla_{\xi^{n+2}}+\nabla_{\xi^{n+1}}.$$

Using relation (D.26), derivative $\nabla_{\xi^{n+2}}$ of equation (2.17) takes the form:

$$-\alpha_{n+2}\partial_{n,n+1}\nabla_{\xi^{n+2}}\Phi^{n,n+1,n+2}=\partial_{n,n+1}\left(\left\langle\vec{\xi}^{\,n+3}\right\rangle_{n,n+1,n+2}-\gamma_{n+2}\vec{A}^{n+2}_{n,n+1,n+2}\right)-\alpha_{n+2}\nabla_{\xi^{n+1}}\Phi^{n,n+1,n+2}=$$
$$=2\alpha_{n+2}\beta_{n+2}\nabla_{\xi^{n+2}}\mathrm{H}^{n,n+1,n+2},$$
$$\partial_{n,n+1}\left\langle\vec{\xi}^{\,n+3}\right\rangle_{n,n+1,n+2}=\gamma_{n+2}\partial_{n,n+1}\vec{A}^{n+2}_{n,n+1,n+2}+\alpha_{n+2}\nabla_{\xi^{n+1}}\Phi^{n,n+1,n+2}+$$
$$+2\alpha_{n+2}\beta_{n+2}\nabla_{\xi^{n+2}}\left(\tau^{n,n+1,n+2}_{n+2}+\mathrm{V}^{n,n+1,n+2}\right), \tag{D.27}$$

where, considering

$$-2\alpha_{n+2}\beta_{n+2}\nabla_{\xi^{n+2}}\tau^{n,n+1,n+2}_{n+2}=\frac{1}{2}\nabla_{\xi^{n+2}}\left|\left\langle\vec{\xi}^{\,n+3}\right\rangle_{n,n+1,n+2}\right|^2=$$
$$=\left\langle\vec{\xi}^{\,n+3}\right\rangle_{n,n+1,n+2}\nabla_{\xi^{n+2}}\left\langle\vec{\xi}^{\,n+3}\right\rangle_{n,n+1,n+2}+\gamma_{n+2}\left\langle\vec{\xi}^{\,n+3}\right\rangle_{n,n+1,n+2}\times\vec{B}^{n+2}_{n,n+1,n+2}, \tag{D.28}$$

we obtain

$$\left(\partial_{n,n+1}+\left\langle\vec{\xi}^{\,n+3}\right\rangle_{n,n+1,n+2}\nabla_{\xi^{n+2}}\right)\left\langle\vec{\xi}^{\,n+3}\right\rangle_{n,n+1,n+2}=\gamma_{n+2}\partial_{n,n+1}\vec{A}^{n+2}_{n,n+1,n+2}+\alpha_{n+2}\nabla_{\xi^{n+1}}\Phi^{n,n+1,n+2}+$$
$$+2\alpha_{n+2}\beta_{n+2}\nabla_{\xi^{n+2}}\mathrm{V}^{n,n+1,n+2}-\gamma_{n+2}\left\langle\vec{\xi}^{\,n+3}\right\rangle_{n,n+1,n+2}\times\vec{B}^{n+2}_{n,n+1,n+2}, \tag{D.29}$$

which coincides with equation (3.17). For the first equation from (2.18), it is possible to take two derivatives with respect to variables $n+1$ and $n+1+k$. Let's start with $\nabla_{\xi^{n+1+k}}$, considering

$$\nabla_{\xi^{n+1+k}}\partial_n=\partial_n\nabla_{\xi^{n+1+k}}, \tag{D.30}$$

we obtain

$$-\alpha_{n+1+k}\partial_n\nabla_{\xi^{n+1+k}}\Phi^{n,n+1,n+1+k}=\partial_n\left(\left\langle\vec{\xi}^{\,n+k+2}\right\rangle_{n,n+1,n+1+k}-\gamma_{n+1+k}\vec{A}^{n+1+k}_{n,n+1,n+1+k}\right)=2\alpha_{n+1+k}\beta_{n+1+k}\nabla_{\xi^{n+1+k}}\mathrm{H}^{n,n+1,n+1+k},$$

$$\begin{aligned}&\partial_n\left\langle\vec{\xi}^{\,n+k+2}\right\rangle_{n,n+1,n+1+k}=\gamma_{n+1+k}\partial_n\vec{A}^{n+1+k}_{n,n+1,n+1+k}+2\alpha_{n+1+k}\beta_{n+1+k}\nabla_{\xi^{n+1+k}}\tau^{n,n+1,n+1+k}_{n+1+k}+\\&+2\alpha_{n+1+k}\beta_{n+1+k}\nabla_{\xi^{n+1+k}}\left(\mathrm{V}^{n,n+1,n+1+k}+\frac{\beta_{n+1}}{\beta_{n+1+k}}\tau^{n,n+1,n+1+k}_{n+1}\right).\end{aligned}\tag{D.31}$$

Since

$$\begin{aligned}&-2\alpha_{n+1+k}\beta_{n+1+k}\nabla_{\xi^{n+1+k}}\tau^{n,n+1,n+1+k}_{n+1+k}=\frac{1}{2}\nabla_{\xi^{n+1+k}}\left|\left\langle\vec{\xi}^{\,n+k+2}\right\rangle_{n,n+1,n+1+k}\right|^2=\\&=\left\langle\vec{\xi}^{\,n+k+2}\right\rangle_{n,n+1,n+1+k}\nabla_{\xi^{n+1+k}}\left\langle\vec{\xi}^{\,n+k+2}\right\rangle_{n,n+1,n+k+1}+\gamma_{n+1+k}\left\langle\vec{\xi}^{\,n+k+2}\right\rangle_{n,n+1,n+k+1}\times\vec{B}^{n+k+1}_{n,n+1,n+k+1},\end{aligned}\tag{D.32}$$

then equation (D.31) takes the form

$$\begin{aligned}&\left(\partial_n+\left\langle\vec{\xi}^{\,n+k+2}\right\rangle_{n,n+1,n+1+k}\nabla_{\xi^{n+1+k}}\right)\left\langle\vec{\xi}^{\,n+k+2}\right\rangle_{n,n+1,n+k+1}=\\&=\gamma_{n+1+k}\partial_n\vec{A}^{n+1+k}_{n,n+1,n+1+k}+2\alpha_{n+1+k}\beta_{n+1+k}\nabla_{\xi^{n+1+k}}\left(\mathrm{V}^{n,n+1,n+1+k}+\frac{\beta_{n+1}}{\beta_{n+1+k}}\tau^{n,n+1,n+1+k}_{n+1}\right)\\&-\gamma_{n+1+k}\left\langle\vec{\xi}^{\,n+k+2}\right\rangle_{n,n+1,n+k+1}\times\vec{B}^{n+k+1}_{n,n+1,n+k+1},\end{aligned}\tag{D.33}$$

resulting expression (D.33) corresponds to equation (3.18). Let us consider derivative $\nabla_{\xi^{n+1}}$

$$\nabla_{\xi^{n+1}}\partial_n=\partial_n\nabla_{\xi^{n+1}}+\nabla_{\xi^n},\tag{D.34}$$

from here

$$\begin{aligned}&-\alpha_{n+1}\nabla_{\xi^{n+1}}\partial_n\Phi^{n,n+1,n+1+k}=-\alpha_{n+1}\partial_n\nabla_{\xi^{n+1}}\Phi^{n,n+1,n+1+k}-\alpha_{n+1}\nabla_{\xi^n}\Phi^{n,n+1,n+1+k}=\\&=\partial_n\left(\left\langle\vec{\xi}^{\,n+2}\right\rangle_{n,n+1,n+1+k}-\gamma_{n+1}\vec{A}^{n+1}_{n,n+1,n+1+k}\right)-\alpha_{n+1}\nabla_{\xi^n}\Phi^{n,n+1,n+1+k}=2\alpha_{n+1}\beta_{n+1+k}\nabla_{\xi^{n+1}}\mathrm{H}^{n,n+1,n+1+k},\\&\partial_n\left\langle\vec{\xi}^{\,n+2}\right\rangle_{n,n+1,n+1+k}=\gamma_{n+1}\partial_n\vec{A}^{n+1}_{n,n+1,n+1+k}+2\alpha_{n+1}\beta_{n+1}\nabla_{\xi^{n+1}}\tau^{n,n+1,n+1+k}_{n+1}+\\&+\alpha_{n+1}\nabla_{\xi^n}\Phi^{n,n+1,n+1+k}+2\alpha_{n+1}\beta_{n+1+k}\nabla_{\xi^{n+1}}\left(\mathrm{V}^{n,n+1,n+1+k}+\tau^{n,n+1,n+1+k}_{n+1+k}\right),\end{aligned}\tag{D.35}$$

or

$$\begin{aligned}&\left(\partial_n+\left\langle\vec{\xi}^{\,n+2}\right\rangle_{n,n+1,n+1+k}\nabla_{\xi^{n+1}}\right)\left\langle\vec{\xi}^{\,n+2}\right\rangle_{n,n+1,n+k+1}=\alpha_{n+1}\nabla_{\xi^n}\Phi^{n,n+1,n+1+k}-\\&-\gamma_{n+1}\partial_n\vec{A}^{n+1}_{n,n+1,n+1+k}+2\alpha_{n+1}\beta_{n+1+k}\nabla_{\xi^{n+1}}\left(\mathrm{V}^{n,n+1,n+1+k}+\tau^{n,n+1,n+1+k}_{n+1+k}\right)-\\&-\gamma_{n+1}\left\langle\vec{\xi}^{\,n+2}\right\rangle_{n,n+1,n+k+1}\times\vec{B}^{n+1}_{n,n+1,n+k+1}\end{aligned}\tag{D.36}$$

where it is taken into consideration that

$$\begin{aligned}&-2\alpha_{n+1}\beta_{n+1}\nabla_{\xi^{n+1}}\tau^{n,n+1,n+1+k}_{n+1}=\frac{1}{2}\nabla_{\xi^{n+1}}\left|\left\langle\vec{\xi}^{\,n+2}\right\rangle_{n,n+1,n+1+k}\right|^2=\\&=\left\langle\vec{\xi}^{\,n+2}\right\rangle_{n,n+1,n+1+k}\nabla_{\xi^{n+1}}\left\langle\vec{\xi}^{\,n+2}\right\rangle_{n,n+1,n+k+1}+\gamma_{n+1}\left\langle\vec{\xi}^{\,n+2}\right\rangle_{n,n+1,n+k+1}\times\vec{B}^{n+1}_{n,n+1,n+k+1},\end{aligned}\tag{D.37}$$

Expression (D.36) coincides with equation (3.19). Consider the second equation in (2.18), which we will differentiate with respect to variables $n$ and $n+s+1$. For the derivative with respect $n$ the relation is true

$$\nabla_{\xi^n}\partial_{n+s} = \partial_{n+s}\nabla_{\xi^n}. \quad \text{(D.38)}$$

Consequently,

$$\begin{aligned}
&-\alpha_n\partial_{n+s}\nabla_{\xi^n}\Phi^{n,n+s,n+s+1} = \partial_{n+s}\left(\left\langle \vec{\xi}^{\,n+1}\right\rangle_{n,n+s,n+s+1} - \gamma_n \vec{A}^n_{n,n+s,n+s+1}\right) = 2\alpha_n\beta_{n+s+1}\nabla_{\xi^n}\mathrm{H}^{n,n+s,n+s+1},\\
&\partial_{n+s}\left\langle \vec{\xi}^{\,n+1}\right\rangle_{n,n+s,n+s+1} = \gamma_n\partial_{n+s}\vec{A}^n_{n,n+s,n+s+1} + 2\alpha_n\beta_n\nabla_{\xi^n}\tau_n^{n,n+s,n+s+1} +\\
&+2\alpha_n\beta_{n+s+1}\nabla_{\xi^n}\left(\tau_{n+s+1}^{n,n+s,n+s+1} + \mathrm{V}^{n,n+s,n+s+1}\right).
\end{aligned} \quad \text{(D.39)}$$

Let us calculate $\nabla_{\xi^n}\tau_n^{n,n+s,n+s+1}$, we obtain

$$\begin{aligned}
&-2\alpha_n\beta_n\nabla_{\xi^n}\tau_n^{n,n+s,n+s+1} = \frac{1}{2}\nabla_{\xi^n}\left|\left\langle \vec{\xi}^{\,n+1}\right\rangle_{n,n+s,n+s+1}\right|^2 =\\
&=\left\langle \vec{\xi}^{\,n+1}\right\rangle_{n,n+s,n+s+1}\nabla_{\xi^n}\left\langle \vec{\xi}^{\,n+1}\right\rangle_{n,n+s,n+s+1} + \gamma_n\left\langle \vec{\xi}^{\,n+1}\right\rangle_{n,n+s,n+s+1}\times\vec{B}^n_{n,n+s,n+s+1}.
\end{aligned} \quad \text{(D.40)}$$

Substituting (D.40) into equation (D.39), we obtain equation (3.20)

$$\begin{aligned}
&\left(\partial_{n+s} + \left\langle \vec{\xi}^{\,n+1}\right\rangle_{n,n+s,n+s+1}\nabla_{\xi^n}\right)\left\langle \vec{\xi}^{\,n+1}\right\rangle_{n,n+s,n+s+1} = \gamma_n\partial_{n+s}\vec{A}^n_{n,n+s,n+s+1} +\\
&+2\alpha_n\beta_{n+s+1}\nabla_{\xi^n}\left(\tau_{n+s+1}^{n,n+s,n+s+1} + \mathrm{V}^{n,n+s,n+s+1}\right) - \gamma_n\left\langle \vec{\xi}^{\,n+1}\right\rangle_{n,n+s,n+s+1}\times\vec{B}^n_{n,n+s,n+s+1}.
\end{aligned} \quad \text{(D.41)}$$

Let us calculate the derivative with respect to variable $n+s+1$ for equation (2.18). Mixed derivative $\nabla_{\xi^{n+s+1}}\partial_{n+s}$ has the form

$$\nabla_{\xi^{n+s+1}}\partial_{n+s} = \partial_{n+s}\nabla_{\xi^{n+s+1}} + \nabla_{\xi^{n+s}}. \quad \text{(D.42)}$$

As a result,

$$\begin{aligned}
&-\alpha_{n+s+1}\nabla_{\xi^{n+s+1}}\partial_{n+s}\Phi^{n,n+s,n+s+1} = \partial_{n+s}\left(\left\langle \vec{\xi}^{\,n+s+2}\right\rangle_{n,n+s,n+s+1} - \gamma_{n+s+1}\vec{A}^{n+s+1}_{n,n+s,n+s+1}\right) - \alpha_{n+s+1}\nabla_{\xi^{n+s}}\Phi^{n,n+s,n+s+1} =\\
&= 2\alpha_{n+s+1}\beta_{n+s+1}\nabla_{\xi^{n+s+1}}\mathrm{H}^{n,n+s,n+s+1},\\
&\partial_{n+s}\left\langle \vec{\xi}^{\,n+s+2}\right\rangle_{n,n+s,n+s+1} = \gamma_{n+s+1}\partial_{n+s}\vec{A}^{n+s+1}_{n,n+s,n+s+1} + \alpha_{n+s+1}\nabla_{\xi^{n+s}}\Phi^{n,n+s,n+s+1} +\\
&+2\alpha_{n+s+1}\beta_{n+s+1}\nabla_{\xi^{n+s+1}}\tau_{n+s+1}^{n,n+s,n+s+1} + 2\alpha_{n+s+1}\beta_{n+s+1}\nabla_{\xi^{n+s+1}}\left(\frac{\beta_n}{\beta_{n+s+1}}\tau_n^{n,n+s,n+s+1} + \mathrm{V}^{n,n+s,n+s+1}\right).
\end{aligned} \quad \text{(D.43)}$$

Since

$$-2\alpha_{n+s+1}\beta_{n+s+1}\nabla_{\xi^{n+s+1}}\tau_{n+s+1}^{n,n+s,n+s+1} = \frac{1}{2}\nabla_{\xi^{n+s+1}}\left|\left\langle \vec{\xi}^{n+s+2}\right\rangle_{n,n+s,n+s+1}\right|^2 =$$
$$= \left\langle \vec{\xi}^{n+s+2}\right\rangle_{n,n+s,n+s+1}\nabla_{\xi^{n+s+1}}\left\langle \vec{\xi}^{n+s+2}\right\rangle_{n,n+s,n+s+1} + \gamma_{n+s+1}\left\langle \vec{\xi}^{n+s+2}\right\rangle_{n,n+s,n+s+1}\times \vec{B}_{n,n+s,n+s+1}^{n+s+1}, \tag{D.44}$$

then equation (D.43) takes the form (3.21):

$$\left(\partial_{n+s} + \left\langle \vec{\xi}^{n+s+2}\right\rangle_{n,n+s,n+s+1}\nabla_{\xi^{n+s+1}}\right)\left\langle \vec{\xi}^{n+s+2}\right\rangle_{n,n+s,n+s+1} = \gamma_{n+s+1}\partial_{n+s}\vec{A}_{n,n+s,n+s+1}^{n+s+1} + \alpha_{n+s+1}\nabla_{\xi^{n+s}}\Phi^{n,n+s,n+s+1} +$$
$$+2\alpha_{n+s+1}\beta_{n+s+1}\nabla_{\xi^{n+s+1}}\left(\frac{\beta_n}{\beta_{n+s+1}}\tau_n^{n,n+s,n+s+1} + \mathrm{V}^{n,n+s,n+s+1}\right) - \gamma_{n+s+1}\left\langle \vec{\xi}^{n+s+2}\right\rangle_{n,n+s,n+s+1}\times \vec{B}_{n,n+s,n+s+1}^{n+s+1}. \tag{D.45}$$

For equation (2.19) with three sources of dissipations, we calculate three derivatives with respect to variables $n, n+s, n+s+k$. Let us start in order with variable $n$:

$$-\alpha_n\partial_0\nabla_{\xi^n}\Phi^{n,n+s,n+s+k} = \partial_0\left(\left\langle \vec{\xi}^{n+1}\right\rangle_{n,n+s,n+s+k} - \gamma_n\vec{A}_{n,n+s,n+s+k}^{n}\right) = 2\alpha_n\beta_{n+s+k}\nabla_{\xi^n}\mathrm{H}^{n,n+s,n+s+k},$$
$$\partial_0\left\langle \vec{\xi}^{n+1}\right\rangle_{n,n+s,n+s+k} = \gamma_n\partial_0\vec{A}_{n,n+s,n+s+k}^{n} + 2\alpha_n\beta_n\nabla_{\xi^n}\tau_n^{n,n+s,n+s+k} +$$
$$+2\alpha_n\beta_{n+s+k}\nabla_{\xi^n}\left(\tau_{n+s+k}^{n,n+s,n+s+k} + \frac{\beta_{n+s}}{\beta_{n+s+k}}\tau_{n+s}^{n,n+s,n+s+k} + \mathrm{V}^{n,n+s,n+s+k}\right), \tag{D.45}$$

$$-2\alpha_n\beta_n\nabla_{\xi^n}\tau_n^{n,n+s,n+s+k} = \frac{1}{2}\nabla_{\xi^n}\left|\left\langle \vec{\xi}^{n+1}\right\rangle_{n,n+s,n+s+k}\right|^2 =$$
$$= \left\langle \vec{\xi}^{n+1}\right\rangle_{n,n+s,n+s+k}\nabla_{\xi^n}\left\langle \vec{\xi}^{n+1}\right\rangle_{n,n+s,n+s+k} + \gamma_n\left\langle \vec{\xi}^{n+1}\right\rangle_{n,n+s,n+s+k}\times \vec{B}_{n,n+s,n+s+k}^{n} \tag{D.46}$$

Substituting (D.46) into (D.45), we obtain equation (3.22)

$$\left(\partial_0 + \left\langle \vec{\xi}^{n+1}\right\rangle_{n,n+s,n+s+k}\nabla_{\xi^n}\right)\left\langle \vec{\xi}^{n+1}\right\rangle_{n,n+s,n+s+k} = \gamma_n\partial_0\vec{A}_{n,n+s,n+s+k}^{n} +$$
$$+2\alpha_n\beta_{n+s+k}\nabla_{\xi^n}\left(\tau_{n+s+k}^{n,n+s,n+s+k} + \frac{\beta_{n+s}}{\beta_{n+s+k}}\tau_{n+s}^{n,n+s,n+s+k} + \mathrm{V}^{n,n+s,n+s+k}\right) - \gamma_n\left\langle \vec{\xi}^{n+1}\right\rangle_{n,n+s,n+s+k}\times \vec{B}_{n,n+s,n+s+k}^{n}. \tag{D.47}$$

Similarly for the derivative with respect to variable $n+s$:

$$-\alpha_{n+s}\partial_0\nabla_{\xi^{n+s}}\Phi^{n,n+s,n+s+k} = \partial_0\left(\left\langle \vec{\xi}^{n+s+1}\right\rangle_{n,n+s,n+s+k} - \gamma_{n+s}\vec{A}_{n,n+s,n+s+k}^{n+s}\right) = 2\alpha_{n+s}\beta_{n+s+k}\nabla_{\xi^{n+s}}\mathrm{H}^{n,n+s,n+s+k},$$
$$\partial_0\left\langle \vec{\xi}^{n+s+1}\right\rangle_{n,n+s,n+s+k} = \gamma_{n+s}\partial_0\vec{A}_{n,n+s,n+s+k}^{n+s} + 2\alpha_{n+s}\beta_{n+s}\nabla_{\xi^{n+s}}\tau_{n+s}^{n,n+s,n+s+k}$$
$$+2\alpha_{n+s}\beta_{n+s+k}\nabla_{\xi^{n+s}}\left(\tau_{n+s+k}^{n,n+s,n+s+k} + \frac{\beta_n}{\beta_{n+s+k}}\tau_n^{n,n+s,n+s+k} + \mathrm{V}^{n,n+s,n+s+k}\right), \tag{D.48}$$

$$-2\alpha_{n+s}\beta_{n+s}\nabla_{\xi^{n+s}}\tau_{n+s}^{n,n+s,n+s+k} = \frac{1}{2}\nabla_{\xi^{n+s}}\left|\left\langle \vec{\xi}^{n+s+1}\right\rangle_{n,n+s,n+s+k}\right|^2 =$$
$$= \left\langle \vec{\xi}^{n+s+1}\right\rangle_{n,n+s,n+s+k}\nabla_{\xi^{n+s}}\left\langle \vec{\xi}^{n+s+1}\right\rangle_{n,n+s,n+s+k} + \gamma_{n+s}\left\langle \vec{\xi}^{n+s+1}\right\rangle_{n,n+s,n+s+k}\times \vec{B}_{n,n+s,n+s+k}^{n+s} \tag{D.49}$$

From expressions (D.48)-(D.49), equation (3.23) follows

$$\left(\partial_0+\left\langle\vec{\xi}^{n+s+1}\right\rangle_{n,n+s,n+s+k}\nabla_{\xi^{n+s}}\right)\left\langle\vec{\xi}^{n+s+1}\right\rangle_{n,n+s,n+s+k}=-\gamma_{n+s}\left\langle\vec{\xi}^{n+s+1}\right\rangle_{n,n+s,n+s+k}\times\vec{B}^{n+s}_{n,n+s,n+s+k}+$$
$$+2\alpha_{n+s}\beta_{n+s+k}\nabla_{\xi^{n+s}}\left(\tau^{n,n+s,n+s+k}_{n+s+k}+\frac{\beta_n}{\beta_{n+s+k}}\tau^{n,n+s,n+s+k}_{n}+\mathrm{V}^{n,n+s,n+s+k}\right)+\gamma_{n+s}\partial_0\vec{A}^{n+s}_{n,n+s,n+s+k}. \tag{D.50}$$

The last equation (3.24) is obtained by differentiating equation (2.19) with respect to variable $n+s+k$. Indeed,

$$-\alpha_{n+s+k}\partial_0\nabla_{\xi^{n+s+k}}\Phi^{n,n+s,n+s+k}=\partial_0\left(\left\langle\vec{\xi}^{n+s+k+1}\right\rangle_{n,n+s,n+s+k}-\gamma_{n+s+k}\vec{A}^{n+s+k}_{n,n+s,n+s+k}\right)=2\alpha_{n+s+k}\beta_{n+s+k}\nabla_{\xi^{n+s+k}}\mathrm{H}^{n,n+s,n+s+k},$$

$$\partial_0\left\langle\vec{\xi}^{n+s+k+1}\right\rangle_{n,n+s,n+s+k}=\gamma_{n+s+k}\partial_0\vec{A}^{n+s+k}_{n,n+s,n+s+k}+2\alpha_{n+s+k}\beta_{n+s+k}\nabla_{\xi^{n+s+k}}\tau^{n,n+s,n+s+k}_{n+s+k}+$$
$$+2\alpha_{n+s+k}\beta_{n+s+k}\nabla_{\xi^{n+s+k}}\left(\frac{\beta_{n+s}}{\beta_{n+s+k}}\tau^{n,n+s,n+s+k}_{n+s}+\frac{\beta_n}{\beta_{n+s+k}}\tau^{n,n+s,n+s+k}_{n}+\mathrm{V}^{n,n+s,n+s+k}\right), \tag{D.51}$$

$$-2\alpha_{n+s+k}\beta_{n+s+k}\nabla_{\xi^{n+s+k}}\tau^{n,n+s,n+s+k}_{n+s+k}=\frac{1}{2}\nabla_{\xi^{n+s+k}}\left|\left\langle\vec{\xi}^{n+s+k+1}\right\rangle_{n,n+s,n+s+k}\right|^2=$$
$$=\left\langle\vec{\xi}^{n+s+k+1}\right\rangle_{n,n+s,n+s+k}\nabla_{\xi^{n+s+k}}\left\langle\vec{\xi}^{n+s+k+1}\right\rangle_{n,n+s,n+s+k}+\gamma_{n+s+k}\left\langle\vec{\xi}^{n+s+k+1}\right\rangle_{n,n+s,n+s+k}\times\vec{B}^{n+s+k}_{n,n+s,n+s+k}. \tag{D.52}$$

Taking into account expression (D.52) in equation (D.51), we obtain

$$\left(\partial_0+\left\langle\vec{\xi}^{n+s+k+1}\right\rangle_{n,n+s,n+s+k}\nabla_{\xi^{n+s+k}}\right)\left\langle\vec{\xi}^{n+s+k+1}\right\rangle_{n,n+s,n+s+k}=\gamma_{n+s+k}\partial_0\vec{A}^{n+s+k}_{n,n+s,n+s+k}-$$
$$-\gamma_{n+s+k}\left\langle\vec{\xi}^{n+s+k+1}\right\rangle_{n,n+s,n+s+k}\times\vec{B}^{n+s+k}_{n,n+s,n+s+k}+$$
$$+2\alpha_{n+s+k}\beta_{n+s+k}\nabla_{\xi^{n+s+k}}\left(\frac{\beta_{n+s}}{\beta_{n+s+k}}\tau^{n,n+s,n+s+k}_{n+s}+\frac{\beta_n}{\beta_{n+s+k}}\tau^{n,n+s,n+s+k}_{n}+\mathrm{V}^{n,n+s,n+s+k}\right). \tag{D.53}$$

Equation (D.53) is the same as equation (3.24). Theorem 5 is proved.

## Appendix E

### *Proof of Theorem 6*

Equation (3.32) follows directly from the definition of field $\vec{B}^{\ell}_{n_1\ldots n_R}=\mathrm{curl}_{\xi^\ell}\vec{A}^{\ell}_{n_1\ldots n_R}$. Let us consider the equations of the first rank ($R=1$). We calculate $\mathrm{curl}_{\xi^n}\vec{E}^n_n$ using definition (3.1), we obtain the expression

$$\mathrm{curl}_{\xi^n}\vec{E}^n_n=-\partial_0\,\mathrm{curl}_{\xi^n}\vec{A}^n_n-\frac{2\alpha_n\beta_n}{\gamma_n}\mathrm{curl}_{\xi^n}\nabla_{\xi^n}\mathrm{V}^n=-\partial_0\vec{B}^n_n, \tag{E.1}$$

which coincides with equation (3.33). Equation (3.34) is obtained from dispersion chain of the Vlasov equations of the first rank (i.3)-(i.5):

$$\frac{\partial}{\partial t} f^n + \mathrm{div}_{\xi^n}\left[ f^n \left\langle \vec{\xi}^{\,n+1} \right\rangle_n \right] = 0, \tag{E.2}$$

$$\partial_0 \,\mathrm{div}_{\xi^n} \vec{D}_n^n + \mathrm{div}_{\xi^n} \vec{J}_n^n = 0 \;\Rightarrow\; \mathrm{div}_{\xi^n}\left(\partial_0 \vec{D}_n^n + \vec{J}_n^n\right) = 0,$$

$$\partial_0 \vec{D}_n^n + \vec{J}_n^n = \mathrm{curl}_{\xi^n} \vec{H}_n^n, \tag{E.3}$$

where representation (3.28) is taken into account and $\vec{H}_n^n$ is some field.

Let us consider equations of the second rank. We calculate $\mathrm{curl}_{\xi^{n+1}} \vec{E}_{n,n+1}^{n+1}$ using definition (3.2):

$$\mathrm{curl}_{\xi^{n+1}} \vec{E}_{n,n+1}^{n+1} = -\mathrm{curl}_{\xi^{n+1}} \partial_n \vec{A}_{n,n+1}^{n+1} - \frac{\alpha_{n+1}}{\gamma_{n+1}} \mathrm{curl}_{\xi^{n+1}} \nabla_{\xi^n} \Phi^{n,n+1}, \tag{E.4}$$

where the right side of (E.4) requires additional transformations. Let us transform each summand on the right side of (E.4) separately. Using the definition of derivative $\partial_n$ for $\mathrm{curl}_{\xi^{n+1}} \partial_n \vec{A}_{n,n+1}^{n+1}$, we obtain

$$\mathrm{curl}_{\xi^{n+1}} \partial_n \vec{A}_{n,n+1}^{n+1} = \partial_0 \vec{B}_{n,n+1}^{n+1} + \mathrm{curl}_{\xi^{n+1}} \left( \vec{\xi}^{\,n+1} \nabla_{\xi^n} \vec{A}_{n,n+1}^{n+1} \right), \tag{E.5}$$

where value $\mathrm{curl}_{\xi^{n+1}} \left( \vec{\xi}^{\,n+1} \nabla_{\xi^n} \vec{A}_{n,n+1}^{n+1} \right)$ can be represented as:

$$\mathrm{curl}_{\xi^{n+1}} \left( \vec{\xi}^{\,n+1} \nabla_{\xi^n} \vec{A}_{n,n+1}^{n+1} \right) = \left[ \nabla_{\xi^{n+1}}, \left( \vec{\xi}^{\,n+1}, \nabla_{\xi^n} \right) \vec{A}_{n,n+1}^{n+1} \right] = \left[ \nabla_{\xi^{n+1}} \underline{\left( \vec{\xi}^{\,n+1}, \nabla_{\xi^n} \right)}, \vec{A}_{n,n+1}^{n+1} \right] +$$

$$+ \left( \vec{\xi}^{\,n+1}, \nabla_{\xi^n} \right) \left[ \nabla_{\xi^{n+1}}, \underline{\vec{A}_{n,n+1}^{n+1}} \right] = \left[ \nabla_{\xi^n}, \vec{A}_{n,n+1}^{n+1} \right] + \left( \vec{\xi}^{\,n+1}, \nabla_{\xi^n} \right) \vec{B}_{n,n+1}^{n+1},$$

$$\mathrm{curl}_{\xi^{n+1}} \left( \vec{\xi}^{\,n+1} \nabla_{\xi^n} \vec{A}_{n,n+1}^{n+1} \right) = \mathrm{curl}_{\xi^n} \vec{A}_{n,n+1}^{n+1} + \left( \vec{\xi}^{\,n+1}, \nabla_{\xi^n} \right) \vec{B}_{n,n+1}^{n+1}, \tag{E.6}$$

where

$$\nabla_{\xi^{n+1}} \left( \vec{\xi}^{\,n+1}, \nabla_{\xi^n} \right) = \nabla_{\xi^n}. \tag{E.7}$$

Using the result of (E.6), expression (E.5) takes the form:

$$\mathrm{curl}_{\xi^{n+1}} \partial_n \vec{A}_{n,n+1}^{n+1} = \partial_0 \vec{B}_{n,n+1}^{n+1} + \left( \vec{\xi}^{\,n+1}, \nabla_{\xi^n} \right) \vec{B}_{n,n+1}^{n+1} + \mathrm{curl}_{\xi^n} \vec{A}_{n,n+1}^{n+1} = \partial_n \vec{B}_{n,n+1}^{n+1} + \mathrm{curl}_{\xi^n} \vec{A}_{n,n+1}^{n+1}. \tag{E.8}$$

We transform the second summand on the right side of expression (E.4):

$$\mathrm{curl}_{\xi^{n+1}} \nabla_{\xi^n} \Phi^{n,n+1} = \left[ \nabla_{\xi^{n+1}}, \nabla_{\xi^n} \Phi^{n,n+1} \right] = -\left[ \nabla_{\xi^n}, \nabla_{\xi^{n+1}} \Phi^{n,n+1} \right] = -\mathrm{curl}_{\xi^n} \nabla_{\xi^{n+1}} \Phi^{n,n+1}. \tag{E.9}$$

Substituting expressions (E.8) and (E.9) into equation (E.4), we obtain

$$\operatorname{curl}_{\xi^{n+1}} \vec{E}^{n+1}_{n,n+1} = -\partial_n \vec{B}^{n+1}_{n,n+1} - \frac{1}{\gamma_{n+1}} \operatorname{curl}_{\xi^n} \left( -\alpha_{n+1} \nabla_{\xi^{n+1}} \Phi^{n,n+1} + \gamma_{n+1} \vec{A}^{n+1}_{n,n+1} \right) =$$

$$= -\partial_n \vec{B}^{n+1}_{n,n+1} - \frac{1}{\gamma_{n+1}} \operatorname{curl}_{\xi^n} \left\langle \vec{\xi}^{\,n+2} \right\rangle_{n,n+1},$$

$$\operatorname{curl}_{\xi^{n+1}} \vec{E}^{n+1}_{n,n+1} = -\partial_n \vec{B}^{n+1}_{n,n+1}, \tag{E.10}$$

where condition $\operatorname{curl}_{\xi^n} \left\langle \vec{\xi}^{\,n+2} \right\rangle_{n,n+1} = 0$ (3.27) is taken into account. Resulting equation (E.10) coincides with equation (3.35). Equations (3.37) and (3.38) are obtained by direct differentiation of expressions (3.3) and (3.4), respectively. Let us proceed to the derivation of equation (3.36). The equation for function $f^{n,n+1}$ has the form:

$$\partial_0 f^{n,n+1} + \vec{\xi}^{\,n+1} \nabla_{\xi^n} f^{n,n+1} + \operatorname{div}_{\xi^{n+1}} \left[ f^{n,n+1} \left\langle \vec{\xi}^{\,n+2} \right\rangle_{n,n+1} \right] = 0,$$

$$\partial_n f^{n,n+1} + \operatorname{div}_{\xi^{n+1}} \vec{J}^{n+1}_{n,n+1} = 0,$$

$$\partial_n \operatorname{div}_{\xi^{n+1}} \vec{D}^{n+1}_{n,n+1} + \partial_n \operatorname{div}_{\xi^n} \vec{D}^{n}_{n,n+1} + \operatorname{div}_{\xi^{n+1}} \vec{J}^{n+1}_{n,n+1} = 0, \tag{E.11}$$

where $\vec{J}^{n+1}_{n,n+1} \overset{\text{det}}{=} f^{n,n+1} \left\langle \vec{\xi}^{\,n+2} \right\rangle_{n,n+1}$. Let us transform expression $\partial_n \operatorname{div}_{\xi^{n+1}} \vec{D}^{n+1}_{n,n+1}$:

$$\partial_n \operatorname{div}_{\xi^{n+1}} \vec{D}^{n+1}_{n,n+1} = \operatorname{div}_{\xi^{n+1}} \partial_0 \vec{D}^{n+1}_{n,n+1} + \vec{\xi}^{\,n+1} \nabla_{\xi^n} \operatorname{div}_{\xi^{n+1}} \vec{D}^{n+1}_{n,n+1}. \tag{E.12}$$

We calculate the summand $\vec{\xi}^{\,n+1} \nabla_{\xi^n} \operatorname{div}_{\xi^{n+1}} \vec{D}^{n+1}_{n,n+1}$ on the right side of equation (E.12):

$$\frac{\partial}{\partial \xi^{n+1}_\lambda} \left( \xi^{n+1}_\mu \frac{\partial}{\partial \xi^n_\mu} D_\lambda \right) = \left( \frac{\partial}{\partial \xi^{n+1}_\lambda} \xi^{n+1}_\mu \right) \frac{\partial}{\partial \xi^n_\mu} D_\lambda + \xi^{n+1}_\mu \frac{\partial}{\partial \xi^n_\mu} \frac{\partial}{\partial \xi^{n+1}_\lambda} D_\lambda =$$

$$= \frac{\partial}{\partial \xi^n_\lambda} D_\lambda + \xi^{n+1}_\mu \frac{\partial}{\partial \xi^n_\mu} \frac{\partial}{\partial \xi^{n+1}_\lambda} D_\lambda,$$

from here

$$\xi^{n+1}_\mu \frac{\partial}{\partial \xi^n_\mu} \frac{\partial}{\partial \xi^{n+1}_\lambda} D_\lambda = \frac{\partial}{\partial \xi^{n+1}_\lambda} \left( \xi^{n+1}_\mu \frac{\partial}{\partial \xi^n_\mu} D_\lambda \right) - \frac{\partial}{\partial \xi^n_\lambda} D_\lambda. \tag{E.13}$$

Consequently,

$$\left( \vec{\xi}^{\,n+1}, \nabla_{\xi^n} \right) \left( \nabla_{\xi^{n+1}}, \vec{D}^{n+1}_{n,n+1} \right) = \left( \nabla_{\xi^{n+1}}, \left( \vec{\xi}^{\,n+1}, \nabla_{\xi^n} \right) \vec{D}^{n+1}_{n,n+1} \right) - \left( \nabla_{\xi^n}, \vec{D}^{n+1}_{n,n+1} \right). \tag{E.14}$$

Substituting expression (E.14) into (E.12), we obtain

$$\partial_n \operatorname{div}_{\xi^{n+1}} \vec{D}^{n+1}_{n,n+1} = \operatorname{div}_{\xi^{n+1}} \partial_0 \vec{D}^{n+1}_{n,n+1} + \operatorname{div}_{\xi^{n+1}} \left[ \left( \vec{\xi}^{\,n+1}, \nabla_{\xi^n} \right) \vec{D}^{n+1}_{n,n+1} \right] - \operatorname{div}_{\xi^n} \vec{D}^{n+1}_{n,n+1} =$$

$$= \operatorname{div}_{\xi^{n+1}} \left[ \partial_0 \vec{D}^{n+1}_{n,n+1} + \left( \vec{\xi}^{\,n+1}, \nabla_{\xi^n} \right) \vec{D}^{n+1}_{n,n+1} \right] - \operatorname{div}_{\xi^n} \vec{D}^{n+1}_{n,n+1},$$

$$\partial_n \operatorname{div}_{\xi^{n+1}} \vec{D}^{n+1}_{n,n+1} = \operatorname{div}_{\xi^{n+1}} \partial_n \vec{D}^{n+1}_{n,n+1} - \operatorname{div}_{\xi^n} \vec{D}^{n+1}_{n,n+1}. \tag{E.15}$$

Let us find expression $\partial_n \operatorname{div}_{\xi^n} \vec{D}^n_{n,n+1}$

$$\partial_n \operatorname{div}_{\xi^n} \vec{D}^n_{n,n+1} = \operatorname{div}_{\xi^n} \partial_0 \vec{D}^n_{n,n+1} + \left(\vec{\xi}^{\,n+1}, \nabla_{\xi^n}\right) \operatorname{div}_{\xi^n} \vec{D}^n_{n,n+1} = \operatorname{div}_{\xi^n} \partial_0 \vec{D}^n_{n,n+1} + \operatorname{div}_{\xi^n} \left(\vec{\xi}^{\,n+1}, \nabla_{\xi^n}\right) \vec{D}^n_{n,n+1},$$

$$\partial_n \operatorname{div}_{\xi^n} \vec{D}^n_{n,n+1} = \operatorname{div}_{\xi^n} \partial_n \vec{D}^n_{n,n+1}. \tag{E.16}$$

As a result, equation (E.11), taking into account (E.15) and (E.16), takes the form:

$$\operatorname{div}_{\xi^{n+1}} \left(\partial_n \vec{D}^{n+1}_{n,n+1} + \vec{J}^{\,n+1}_{n,n+1}\right) + \operatorname{div}_{\xi^n} \left(\partial_n \vec{D}^n_{n,n+1} - \vec{D}^{n+1}_{n,n+1}\right) = 0,$$

from here, taking into account the independence of the expressions in brackets and condition (3.29), we obtain:

$$\partial_n \vec{D}^{n+1}_{n,n+1} + \vec{J}^{\,n+1}_{n,n+1} = \operatorname{curl}_{\xi^{n+1}} \vec{H}^{n+1}_{n,n+1}, \tag{E.17}$$

where $\vec{H}^{n+1}_{n,n+1}$ is some field.

Equation for function $f^{\,n,n+k}$ has the form:

$$\partial_0 f^{\,n,n+k} + \operatorname{div}_{\xi^n} \vec{J}^{\,n}_{n,n+k} + \operatorname{div}_{\xi^{n+k}} \vec{J}^{\,n+k}_{n,n+k} = 0,$$

$$\operatorname{div}_{\xi^n} \left(\partial_0 \vec{D}^n_{n,n+k} + \vec{J}^{\,n}_{n,n+k}\right) + \operatorname{div}_{\xi^{n+k}} \left(\partial_0 \vec{D}^{n+k}_{n,n+k} + \vec{J}^{\,n+k}_{n,n+k}\right) = 0. \tag{E.18}$$

Due to the independence of the expressions in the brackets of expression (E.18), we obtain the equations

$$\partial_0 \vec{D}^n_{n,n+k} + \vec{J}^{\,n}_{n,n+k} = \operatorname{curl}_{\xi^n} \vec{H}^n_{n,n+k}, \qquad \partial_0 \vec{D}^{n+k}_{n,n+k} + \vec{J}^{\,n+k}_{n,n+k} = \operatorname{curl}_{\xi^{n+k}} \vec{H}^{n+k}_{n,n+k}, \tag{E.19}$$

which coincide with (3.39) and (3.40).

Let us consider the equations of the third rank for function $f^{\,n,n+1,n+2}$:

$$\frac{\partial f^{\,n,n+1,n+2}}{\partial t} + \xi^{n+1}_\beta \frac{\partial f^{\,n,n+1,n+2}}{\partial \xi^n_\beta} + \xi^{n+2}_\beta \frac{\partial f^{\,n,n+1,n+2}}{\partial \xi^{n+1}_\beta} + \frac{\partial}{\partial \xi^{n+2}_\beta}\left[ f^{\,n,n+1,n+2} \left\langle \xi^{n+3}_\beta \right\rangle_{n,n+1,n+2} \right] = 0,$$

$$\partial_{n,n+1} f^{\,n,n+1,n+2} + \operatorname{div}_{\xi^{n+2}} \vec{J}^{\,n+2}_{n,n+1,n+2} = 0, \tag{E.20}$$

$$\partial_{n,n+1} \operatorname{div}_{\xi^{n+2}} \vec{D}^{n+2}_{n,n+1,n+2} + \partial_{n,n+1} \operatorname{div}_{\xi^{n+1}} \vec{D}^{n+1}_{n,n+1,n+2} + \partial_{n,n+1} \operatorname{div}_{\xi^n} \vec{D}^n_{n,n+1,n+2} + \operatorname{div}_{\xi^{n+2}} \vec{J}^{\,n+2}_{n,n+1,n+2} = 0, \tag{E.21}$$

where $\vec{J}^{\,n+2}_{n,n+1,n+2} \overset{\text{det}}{=} f^{\,n,n+1,n+2} \left\langle \xi^{n+3}_\beta \right\rangle_{n,n+1,n+2}$. Let us transform the first summand in equation (E.21):

$$\begin{gathered}\partial_{n,n+1} \operatorname{div}_{\xi^{n+2}} \vec{D}^{n+2}_{n,n+1,n+2} = \operatorname{div}_{\xi^{n+2}} \partial_0 \vec{D}^{n+2}_{n,n+1,n+2} + \\ + \operatorname{div}_{\xi^{n+2}} \left( \xi^{n+1}_\beta \frac{\partial}{\partial \xi^n_\beta} \vec{D}^{n+2}_{n,n+1,n+2} \right) + \xi^{n+2}_\beta \frac{\partial}{\partial \xi^{n+1}_\beta} \operatorname{div}_{\xi^{n+2}} \vec{D}^{n+2}_{n,n+1,n+2}.\end{gathered} \tag{E.22}$$

The relation is true

$$\xi_\beta^{n+2}\frac{\partial}{\partial\xi_\beta^{n+1}}\frac{\partial}{\partial\xi_\lambda^{n+2}}\vec{D}_\lambda=\frac{\partial}{\partial\xi_\lambda^{n+2}}\left(\xi_\beta^{n+2}\frac{\partial}{\partial\xi_\beta^{n+1}}\vec{D}_\lambda\right)-\frac{\partial}{\partial\xi_\lambda^{n+1}}\vec{D}_\lambda,$$

$$\xi_\beta^{n+2}\frac{\partial}{\partial\xi_\beta^{n+1}}\operatorname{div}_{\xi^{n+2}}\vec{D}_{n,n+1,n+2}^{n+2}=\operatorname{div}_{\xi^{n+2}}\left(\xi_\beta^{n+2}\frac{\partial}{\partial\xi_\beta^{n+1}}\vec{D}_{n,n+1,n+2}^{n+2}\right)-\operatorname{div}_{\xi^{n+1}}\vec{D}_{n,n+1,n+2}^{n+2}. \quad \text{(E.23)}$$

Let us substitute (E.23) into expression (E.22), we obtain

$$\partial_{n,n+1}\operatorname{div}_{\xi^{n+2}}\vec{D}_{n,n+1,n+2}^{n+2}=-\operatorname{div}_{\xi^{n+1}}\vec{D}_{n,n+1,n+2}^{n+2}+$$
$$+\operatorname{div}_{\xi^{n+2}}\left(\partial_0\vec{D}_{n,n+1,n+2}^{n+2}+\xi_\beta^{n+1}\frac{\partial}{\partial\xi_\beta^{n}}\vec{D}_{n,n+1,n+2}^{n+2}+\xi_\beta^{n+2}\frac{\partial}{\partial\xi_\beta^{n+1}}\vec{D}_{n,n+1,n+2}^{n+2}\right),$$
$$\partial_{n,n+1}\operatorname{div}_{\xi^{n+2}}\vec{D}_{n,n+1,n+2}^{n+2}=\operatorname{div}_{\xi^{n+2}}\partial_{n,n+1}\vec{D}_{n,n+1,n+2}^{n+2}-\operatorname{div}_{\xi^{n+1}}\vec{D}_{n,n+1,n+2}^{n+2}. \quad \text{(E.24)}$$

Let us transform the second and third summands in equation (E.21). For the second summand $\partial_{n,n+1}\operatorname{div}_{\xi^{n+1}}\vec{D}_{n,n+1,n+2}^{n+1}$, we obtain:

$$\partial_{n,n+1}\operatorname{div}_{\xi^{n+1}}\vec{D}_{n,n+1,n+2}^{n+1}=\operatorname{div}_{\xi^{n+1}}\partial_0\vec{D}_{n,n+1,n+2}^{n+1}+\operatorname{div}_{\xi^{n+1}}\left(\xi_\beta^{n+2}\frac{\partial}{\partial\xi_\beta^{n+1}}\vec{D}_{n,n+1,n+2}^{n+1}\right)+$$
$$+\xi_\beta^{n+1}\frac{\partial}{\partial\xi_\beta^{n}}\operatorname{div}_{\xi^{n+1}}\vec{D}_{n,n+1,n+2}^{n+1}=\operatorname{div}_{\xi^{n+1}}\left(\partial_0\vec{D}_{n,n+1,n+2}^{n+1}+\xi_\beta^{n+2}\frac{\partial}{\partial\xi_\beta^{n+1}}\vec{D}_{n,n+1,n+2}^{n+1}\right)+$$
$$+\operatorname{div}_{\xi^{n+1}}\left(\xi_\beta^{n+1}\frac{\partial}{\partial\xi_\beta^{n}}\vec{D}_{n,n+1,n+2}^{n+1}\right)-\operatorname{div}_{\xi^{n}}\vec{D}_{n,n+1,n+2}^{n+1},$$
$$\partial_{n,n+1}\operatorname{div}_{\xi^{n+1}}\vec{D}_{n,n+1,n+2}^{n+1}=\operatorname{div}_{\xi^{n+1}}\left(\partial_{n,n+1}\vec{D}_{n,n+1,n+2}^{n+1}\right)-\operatorname{div}_{\xi^{n}}\vec{D}_{n,n+1,n+2}^{n+1}, \quad \text{(E.25)}$$

The third summand takes the form:

$$\partial_{n,n+1}\operatorname{div}_{\xi^{n}}\vec{D}_{n,n+1,n+2}^{n}=\operatorname{div}_{\xi^{n}}\partial_0\vec{D}_{n,n+1,n+2}^{n}+\operatorname{div}_{\xi^{n}}\left(\xi_\beta^{n+2}\frac{\partial}{\partial\xi_\beta^{n+1}}\vec{D}_{n,n+1,n+2}^{n}\right)+\operatorname{div}_{\xi^{n}}\left(\xi_\beta^{n+1}\frac{\partial}{\partial\xi_\beta^{n}}\vec{D}_{n,n+1,n+2}^{n}\right),$$
$$\partial_{n,n+1}\operatorname{div}_{\xi^{n}}\vec{D}_{n,n+1,n+2}^{n}=\operatorname{div}_{\xi^{n}}\partial_{n,n+1}\vec{D}_{n,n+1,n+2}^{n}. \quad \text{(E.26)}$$

Taking into account relations (E.24)-(E.26), equation (E.21) takes the form:

$$\operatorname{div}_{\xi^{n+2}}\left(\partial_{n,n+1}\vec{D}_{n,n+1,n+2}^{n+2}+\vec{J}_{n,n+1,n+2}^{n+2}\right)+\operatorname{div}_{\xi^{n+1}}\left(\partial_{n,n+1}\vec{D}_{n,n+1,n+2}^{n+1}-\vec{D}_{n,n+1,n+2}^{n+2}\right)+$$
$$+\operatorname{div}_{\xi^{n}}\left(\partial_{n,n+1}\vec{D}_{n,n+1,n+2}^{n}-\vec{D}_{n,n+1,n+2}^{n+1}\right)=0,$$

from here, due to (3.30) and (3.31):

$$\partial_{n,n+1}\vec{D}_{n,n+1,n+2}^{n+2}+\vec{J}_{n,n+1,n+2}^{n+2}=\operatorname{curl}_{\xi^{n+2}}\vec{H}_{n,n+1,n+2}^{n+2}, \quad \text{(E.27)}$$

where $\vec{H}^{n+2}_{n,n+1,n+2}$ is some field. Let us calculate $\operatorname{curl}_{\xi^{n+2}} \vec{E}^{n+2}_{n,n+1,n+2}$ using expression (3.5)

$$\operatorname{curl}_{\xi^{n+2}} \vec{E}^{n+2}_{n,n+1,n+2} \overset{\det}{=} -\operatorname{curl}_{\xi^{n+2}} \left(\partial_{n,n+1}\vec{A}^{n+2}_{n,n+1,n+2}\right) - \frac{\alpha_{n+2}}{\gamma_{n+2}} \operatorname{curl}_{\xi^{n+2}} \left(\nabla_{\xi^{n+1}} \Phi^{n,n+1,n+2}\right). \tag{E.28}$$

We simplify separately each summand in expression (E.28):

$$\operatorname{curl}_{\xi^{n+2}} \left(\partial_{n,n+1}\vec{A}^{n+2}_{n,n+1,n+2}\right) = \partial_0 \vec{B}^{n+2}_{n,n+1,n+2} + \xi^{n+1}_\beta \frac{\partial}{\partial \xi^n_\beta} \vec{B}^{n+2}_{n,n+1,n+2} + \operatorname{curl}_{\xi^{n+2}} \left( \xi^{n+2}_\beta \frac{\partial}{\partial \xi^{n+1}_\beta} \vec{A}^{n+2}_{n,n+1,n+2} \right). \tag{E.29}$$

Let us transform the last summand in (E.29), we obtain

$$\left[\nabla_{\xi^{n+2}}, \left(\vec{\xi}^{\,n+2}, \nabla_{\xi^{n+1}}\right) \vec{A}^{n+2}_{n,n+1,n+2}\right] = \left[\nabla_{\xi^{n+2}}, \underline{\left(\vec{\xi}^{\,n+2}, \nabla_{\xi^{n+1}}\right)} \vec{A}^{n+2}_{n,n+1,n+2}\right] + \left[\nabla_{\xi^{n+2}}, \left(\vec{\xi}^{\,n+2}, \nabla_{\xi^{n+1}}\right) \underline{\vec{A}^{n+2}_{n,n+1,n+2}}\right] =$$
$$= \left[\nabla_{\xi^{n+1}}, \vec{A}^{n+2}_{n,n+1,n+2}\right] + \left(\vec{\xi}^{\,n+2}, \nabla_{\xi^{n+1}}\right)\left[\nabla_{\xi^{n+2}}, \vec{A}^{n+2}_{n,n+1,n+2}\right],$$
$$\operatorname{curl}_{\xi^{n+2}} \left(\vec{\xi}^{\,n+2} \nabla_{\xi^{n+1}} \vec{A}^{n+2}_{n,n+1,n+2}\right) = \operatorname{curl}_{\xi^{n+1}} \vec{A}^{n+2}_{n,n+1,n+2} + \left(\vec{\xi}^{\,n+2}, \nabla_{\xi^{n+1}}\right) \vec{B}^{n+2}_{n,n+1,n+2}. \tag{E.30}$$

Taking into account (E.30), expression (E.29) takes the form

$$\operatorname{curl}_{\xi^{n+2}} \left(\partial_{n,n+1}\vec{A}^{n+2}_{n,n+1,n+2}\right) = \partial_{n,n+1} \vec{B}^{n+2}_{n,n+1,n+2} + \operatorname{curl}_{\xi^{n+1}} \vec{A}^{n+2}_{n,n+1,n+2}. \tag{E.31}$$

By analogy with (E.9), we transform the second summand in equation (E.28)

$$\operatorname{curl}_{\xi^{n+2}} \left(\nabla_{\xi^{n+1}} \Phi^{n,n+1,n+2}\right) = -\operatorname{curl}_{\xi^{n+1}} \nabla_{\xi^{n+2}} \Phi^{n,n+1,n+2}. \tag{E.32}$$

Substituting expressions (E.31) and (E.32) into equation (E.28), we obtain:

$$\operatorname{curl}_{\xi^{n+2}} \vec{E}^{n+2}_{n,n+1,n+2} \overset{\det}{=} -\partial_{n,n+1} \vec{B}^{n+2}_{n,n+1,n+2} - \frac{1}{\gamma_{n+2}} \operatorname{curl}_{\xi^{n+1}} \left(\gamma_{n+2} \vec{A}^{n+2}_{n,n+1,n+2} - \alpha_{n+2} \nabla_{\xi^{n+2}} \Phi^{n,n+1,n+2}\right),$$
$$\operatorname{curl}_{\xi^{n+2}} \vec{E}^{n+2}_{n,n+1,n+2} \overset{\det}{=} -\partial_{n,n+1} \vec{B}^{n+2}_{n,n+1,n+2} - \frac{1}{\gamma_{n+2}} \operatorname{curl}_{\xi^{n+1}} \left\langle \vec{\xi}^{\,n+3} \right\rangle_{n,n+1,n+2}. \tag{E.33}$$

Equation (E.33) under condition (3.27) $\operatorname{curl}_{\xi^{n+1}} \left\langle \vec{\xi}^{\,n+3} \right\rangle_{n,n+1,n+2} = 0$ goes into equation (3.41).

Let us consider the equation of the third rank for function $f^{\,n,n+1,n+1+k}$ with two sources of dissipations:

$$\partial_n f^{\,n,n+1,n+1+k} + \frac{\partial}{\partial \xi^{n+1}_\beta} \left[ f^{\,n,n+1,n+1+k} \left\langle \xi^{n+2}_\beta \right\rangle_{n,n+1,n+1+k} \right] +$$
$$+ \frac{\partial}{\partial \xi^{n+1+k}_\beta} \left[ f^{\,n,n+1,n+1+k} \left\langle \xi^{n+2+k}_\beta \right\rangle_{n,n+1,n+1+k} \right] = 0,$$

$$
\begin{aligned}
&\partial_n \operatorname{div}_{\xi^n} \vec{D}^{n}_{n,n+1,n+1+k} + \partial_n \operatorname{div}_{\xi^{n+1}} \vec{D}^{n+1}_{n,n+1,n+1+k} + \partial_n \operatorname{div}_{\xi^{n+1+k}} \vec{D}^{n+1+k}_{n,n+1,n+1+k} + \\
&+ \operatorname{div}_{\xi^{n+1}} \vec{J}^{n+1}_{n,n+1,n+1+k} + \operatorname{div}_{\xi^{n+1+k}} \vec{J}^{n+1+k}_{n,n+1,n+1+k} = 0,
\end{aligned}
\tag{E.34}
$$

where representation (3.28) is taken into account. We transform the first three summands in equation (E.34). Let us start with expression $\partial_n \operatorname{div}_{\xi^n} \vec{D}^{n}_{n,n+1,n+1+k}$:

$$
\partial_n \operatorname{div}_{\xi^n} \vec{D}^{n}_{n,n+1,n+1+k} = \operatorname{div}_{\xi^n} \partial_0 \vec{D}^{n}_{n,n+1,n+1+k} + \xi^{n+1}_\beta \frac{\partial}{\partial \xi^n_\beta} \operatorname{div}_{\xi^n} \vec{D}^{n}_{n,n+1,n+1+k} = \operatorname{div}_{\xi^n} \partial_n \vec{D}^{n}_{n,n+1,n+1+k}. \tag{E.35}
$$

Let us transform the second summand $\partial_n \operatorname{div}_{\xi^{n+1}} \vec{D}^{n+1}_{n,n+1,n+1+k}$:

$$
\partial_n \operatorname{div}_{\xi^{n+1}} \vec{D}^{n+1}_{n,n+1,n+1+k} = \operatorname{div}_{\xi^{n+1}} \left( \partial_0 \vec{D}^{n+1}_{n,n+1,n+1+k} + \xi^{n+1}_\beta \frac{\partial}{\partial \xi^n_\beta} \vec{D}^{n+1}_{n,n+1,n+1+k} \right) - \operatorname{div}_{\xi^n} \vec{D}^{n+1}_{n,n+1,n+1+k},
$$

$$
\partial_n \operatorname{div}_{\xi^{n+1}} \vec{D}^{n+1}_{n,n+1,n+1+k} = \operatorname{div}_{\xi^{n+1}} \left( \partial_n \vec{D}^{n+1}_{n,n+1,n+1+k} \right) - \operatorname{div}_{\xi^n} \vec{D}^{n+1}_{n,n+1,n+1+k}, \tag{E.36}
$$

where it is taken into account that

$$
\xi^{n+1}_\beta \frac{\partial}{\partial \xi^n_\beta} \operatorname{div}_{\xi^{n+1}} \vec{D}^{n+1}_{n,n+1,n+1+k} = \operatorname{div}_{\xi^{n+1}} \left[ \xi^{n+1}_\beta \frac{\partial}{\partial \xi^n_\beta} \vec{D}^{n+1}_{n,n+1,n+1+k} \right] - \operatorname{div}_{\xi^n} \vec{D}^{n+1}_{n,n+1,n+1+k}. \tag{E.37}
$$

Similarly, for the third summand from equation (E.34), we obtain

$$
\partial_n \operatorname{div}_{\xi^{n+1+k}} \vec{D}^{n+1+k}_{n,n+1,n+1+k} = \operatorname{div}_{\xi^{n+1+k}} \left( \partial_0 \vec{D}^{n+1+k}_{n,n+1,n+1+k} + \xi^{n+1}_\beta \frac{\partial}{\partial \xi^n_\beta} \vec{D}^{n+1+k}_{n,n+1,n+1+k} \right),
$$

$$
\partial_n \operatorname{div}_{\xi^{n+1+k}} \vec{D}^{n+1+k}_{n,n+1,n+1+k} = \operatorname{div}_{\xi^{n+1+k}} \left( \partial_n \vec{D}^{n+1+k}_{n,n+1,n+1+k} \right). \tag{E.38}
$$

We substitute expressions (E.35)-(E.37) into equation (E.34)

$$
\begin{aligned}
&\operatorname{div}_{\xi^{n+1}} \left( \partial_n \vec{D}^{n+1}_{n,n+1,n+1+k} + \vec{J}^{n+1}_{n,n+1,n+1+k} \right) + \operatorname{div}_{\xi^{n+1+k}} \left( \partial_n \vec{D}^{n+1+k}_{n,n+1,n+1+k} + \vec{J}^{n+1+k}_{n,n+1,n+1+k} \right) + \\
&+ \operatorname{div}_{\xi^n} \left( \partial_n \vec{D}^{n}_{n,n+1,n+1+k} - \vec{D}^{n+1}_{n,n+1,n+1+k} \right) = 0,
\end{aligned}
$$

$$
\begin{aligned}
\partial_n \vec{D}^{n+1}_{n,n+1,n+1+k} + \vec{J}^{n+1}_{n,n+1,n+1+k} &= \operatorname{curl}_{\xi^{n+1}} \vec{H}^{n+1}_{n,n+1,n+1+k}, \\
\partial_n \vec{D}^{n+1+k}_{n,n+1,n+1+k} + \vec{J}^{n+1+k}_{n,n+1,n+1+k} &= \operatorname{curl}_{\xi^{n+1+k}} \vec{H}^{n+1+k}_{n,n+1,n+1+k},
\end{aligned}
\tag{E.39}
$$

where the independence of expressions in brackets and condition (3.29) are taken into account. We calculate $\operatorname{curl}_{\xi^{n+1+k}} \vec{E}^{n+1+k}_{n,n+1,n+1+k}$ using expression (3.6)

$$
\operatorname{curl}_{\xi^{n+1+k}} \vec{E}^{n+1+k}_{n,n+1,n+1+k} = -\partial_n \operatorname{curl}_{\xi^{n+1+k}} \vec{A}^{n+1+k}_{n,n+1,n+1+k} = -\partial_n \vec{B}^{n+1+k}_{n,n+1,n+1+k}. \tag{E.40}
$$

Equation (E.40) coincides with equation (3.43). Then, we find $\operatorname{curl}_{\xi^{n+1}} \vec{E}^{n+1}_{n,n+1,n+1+k}$ and $\operatorname{curl}_{\xi^{n}} \vec{E}^{n}_{n,n+1,n+1+k}$

$$\operatorname{curl}_{\xi^{n+1}} \vec{E}^{n+1}_{n,n+1,n+1+k} = -\operatorname{curl}_{\xi^{n+1}} \partial_n \vec{A}^{n+1}_{n,n+1,n+1+k} - \frac{\alpha_{n+1}}{\gamma_{n+1}} \operatorname{curl}_{\xi^{n+1}} \nabla_{\xi^n} \Phi^{n,n+1,n+1+k}. \tag{E.41}$$

We calculate separately each summand in expression (E.41):

$$\operatorname{curl}_{\xi^{n+1}} \partial_n \vec{A}^{n+1}_{n,n+1,n+1+k} = \operatorname{curl}_{\xi^{n+1}} \partial_0 \vec{A}^{n+1}_{n,n+1,n+1+k} + \operatorname{curl}_{\xi^{n+1}} \left(\vec{\xi}^{\,n+1}, \nabla_{\xi^n}\right) \vec{A}^{n+1}_{n,n+1,n+1+k} =$$
$$= \partial_0 \operatorname{curl}_{\xi^{n+1}} \vec{A}^{n+1}_{n,n+1,n+1+k} + \left[\nabla_{\xi^n}, \vec{A}^{n+1}_{n,n+1,n+1+k}\right] + \left(\vec{\xi}^{\,n+1}, \nabla_{\xi^n}\right)\left[\nabla_{\xi^{n+1}}, \vec{A}^{n+1}_{n,n+1,n+1+k}\right],$$
$$\operatorname{curl}_{\xi^{n+1}} \partial_n \vec{A}^{n+1}_{n,n+1,n+1+k} = \partial_n \vec{B}^{n+1}_{n,n+1,n+1+k} + \operatorname{curl}_{\xi^n} \vec{A}^{n+1}_{n,n+1,n+1+k}. \tag{E.42}$$

By analogy with expression (E.9), we can write as follows

$$\operatorname{curl}_{\xi^{n+1}} \nabla_{\xi^n} \Phi^{n,n+1,n+1+k} = -\operatorname{curl}_{\xi^n} \nabla_{\xi^{n+1}} \Phi^{n,n+1,n+1+k}. \tag{E.43}$$

Substituting expressions (E.42) and (E.43) into equation (E.41), we obtain

$$\operatorname{curl}_{\xi^{n+1}} \vec{E}^{n+1}_{n,n+1,n+1+k} = -\partial_n \vec{B}^{n+1}_{n,n+1,n+1+k} - \frac{1}{\gamma_{n+1}} \operatorname{curl}_{\xi^n} \left(\gamma_{n+1} \vec{A}^{n+1}_{n,n+1,n+1+k} - \alpha_{n+1} \nabla_{\xi^{n+1}} \Phi^{n,n+1,n+1+k}\right),$$
$$\operatorname{curl}_{\xi^{n+1}} \vec{E}^{n+1}_{n,n+1,n+1+k} = -\partial_n \vec{B}^{n+1}_{n,n+1,n+1+k} - \frac{1}{\gamma_{n+1}} \operatorname{curl}_{\xi^n} \left\langle \vec{\xi}^{\,n+2} \right\rangle_{n,n+1,n+1+k}, \tag{E.44}$$

which, taking into account condition (3.27) $\operatorname{curl}_{\xi^n} \left\langle \vec{\xi}^{\,n+2} \right\rangle_{n,n+1,n+1+k} = 0$, transforms into equation (3.44).

Let us consider the second equation with two sources of dissipation for function $f^{n,n+s,n+s+1}$:

$$\begin{aligned}
&\partial_{n+s} f^{n,n+s,n+s+1} + \operatorname{div}_{\xi^n} \vec{J}^{n}_{n,n+s,n+s+1} + \operatorname{div}_{\xi^{n+s+1}} \vec{J}^{n+s+1}_{n,n+s,n+s+1} = 0,\\
&\partial_{n+s} \operatorname{div}_{\xi^n} \vec{D}^{n}_{n,n+s,n+s+1} + \partial_{n+s} \operatorname{div}_{\xi^{n+s+1}} \vec{D}^{n+s+1}_{n,n+s,n+s+1} + \partial_{n+s} \operatorname{div}_{\xi^{n+s}} \vec{D}^{n+s}_{n,n+s,n+s+1}\\
&+ \operatorname{div}_{\xi^n} \vec{J}^{n}_{n,n+s,n+s+1} + \operatorname{div}_{\xi^{n+s+1}} \vec{J}^{n+s+1}_{n,n+s,n+s+1} = 0.
\end{aligned} \tag{E.45}$$

where representations (3.28) and (3.60) are taken into account. We transform the first three summands in equation (E.45):

$$\partial_{n+s} \operatorname{div}_{\xi^{n+s}} \vec{D}^{n+s}_{n,n+s,n+s+1} = \operatorname{div}_{\xi^{n+s}} \partial_{n+s} \vec{D}^{n+s}_{n,n+s,n+s+1}, \tag{E.46}$$

$$\partial_{n+s} \operatorname{div}_{\xi^{n}} \vec{D}^{n}_{n,n+s,n+s+1} = \operatorname{div}_{\xi^{n}} \partial_{n+s} \vec{D}^{n}_{n,n+s,n+s+1}, \tag{E.47}$$

$$\partial_{n+s} \operatorname{div}_{\xi^{n+s+1}} \vec{D}^{n+s+1}_{n,n+s,n+s+1} = \operatorname{div}_{\xi^{n+s+1}} \partial_0 \vec{D}^{n+s+1}_{n,n+s,n+s+1} + \vec{\xi}^{\,n+s+1} \nabla_{\xi^{n+s}} \operatorname{div}_{\xi^{n+s+1}} \vec{D}^{n+s+1}_{n,n+s,n+s+1} =$$
$$= \operatorname{div}_{\xi^{n+s+1}} \left[\partial_0 \vec{D}^{n+s+1}_{n,n+s,n+s+1} + \left(\vec{\xi}^{\,n+s+1}, \nabla_{\xi^{n+s}}\right) \vec{D}^{n+s+1}_{n,n+s,n+s+1}\right] - \operatorname{div}_{\xi^{n+s}} \vec{D}^{n+s+1}_{n,n+s,n+s+1},$$

$$\partial_{n+s}\,\mathrm{div}_{\xi^{n+s+1}}\,\vec{D}^{n+s+1}_{n,n+s,n+s+1}=\mathrm{div}_{\xi^{n+s+1}}\,\partial_{n+s}\vec{D}^{n+s+1}_{n,n+s,n+s+1}-\mathrm{div}_{\xi^{n+s}}\,\vec{D}^{n+s+1}_{n,n+s,n+s+1}, \tag{E.48}$$

where it is taken into account that

$$\xi^{n+s+1}_{\mu}\frac{\partial}{\partial\xi^{n+s}_{\mu}}\left(\frac{\partial}{\partial\xi^{n+s+1}_{\lambda}}D_{\lambda}\right)=\frac{\partial}{\partial\xi^{n+s+1}_{\lambda}}\left(\xi^{n+s+1}_{\mu}\frac{\partial}{\partial\xi^{n+s}_{\mu}}D_{\lambda}\right)-\frac{\partial}{\partial\xi^{n+s}_{\lambda}}D_{\lambda},$$

$$\vec{\xi}^{\,n+s+1}\nabla_{\xi^{n+s}}\,\mathrm{div}_{\xi^{n+s+1}}\,\vec{D}^{n+s+1}_{n,n+s,n+s+1}=\mathrm{div}_{\xi^{n+s+1}}\left[\left(\vec{\xi}^{\,n+s+1},\nabla_{\xi^{n+s}}\right)\vec{D}^{n+s+1}_{n,n+s,n+s+1}\right]-\mathrm{div}_{\xi^{n+s}}\,\vec{D}^{n+s+1}_{n,n+s,n+s+1}.$$

Substituting expressions (E.46)-(E.48) into equation (E.45), we obtain:

$$\begin{gathered}\mathrm{div}_{\xi^{n}}\left(\partial_{n+s}\vec{D}^{n}_{n,n+s,n+s+1}+\vec{J}^{\,n}_{n,n+s,n+s+1}\right)+\mathrm{div}_{\xi^{n+s+1}}\left(\partial_{n+s}\vec{D}^{n+s+1}_{n,n+s,n+s+1}+\vec{J}^{\,n+s+1}_{n,n+s,n+s+1}\right)+\\+\mathrm{div}_{\xi^{n+s}}\left(\partial_{n+s}\vec{D}^{n+s}_{n,n+s,n+s+1}-\vec{D}^{n+s+1}_{n,n+s,n+s+1}\right)=0,\\ \partial_{n+s}\vec{D}^{n}_{n,n+s,n+s+1}+\vec{J}^{\,n}_{n,n+s,n+s+1}=\mathrm{curl}_{\xi^{n}}\,\vec{H}^{n}_{n,n+s,n+s+1},\\ \partial_{n+s}\vec{D}^{n+s+1}_{n,n+s,n+s+1}+\vec{J}^{\,n+s+1}_{n,n+s,n+s+1}=\mathrm{curl}_{\xi^{n+s+1}}\,\vec{H}^{n+s+1}_{n,n+s,n+s+1},\end{gathered} \tag{E.49}$$

where the condition of independence of the expressions in brackets and expression (3.29) are taken into account. Let us calculate $\mathrm{curl}_{\xi^{n+s+1}}\,\vec{E}^{n+s+1}_{n,n+s,n+s+1}$:

$$\begin{gathered}\mathrm{curl}_{\xi^{n+s+1}}\,\vec{E}^{n+s+1}_{n,n+s,n+s+1}=-\mathrm{curl}_{\xi^{n+s+1}}\,\partial_{n+s}\vec{A}^{n+s+1}_{n,n+s,n+s+1}-\frac{\alpha_{n+s+1}}{\gamma_{n+s+1}}\mathrm{curl}_{\xi^{n+s+1}}\,\nabla_{\xi^{n+s}}\Phi^{n,n+s,n+s+1},\\ \mathrm{curl}_{\xi^{n+s+1}}\,\vec{E}^{n+s+1}_{n,n+s,n+s+1}=-\partial_{n+s}\vec{B}^{n+s+1}_{n,n+s,n+s+1}-\frac{1}{\gamma_{n+s+1}}\mathrm{curl}_{\xi^{n+s}}\left(\gamma_{n+s+1}\vec{A}^{n+s+1}_{n,n+s,n+s+1}-\alpha_{n+s+1}\nabla_{\xi^{n+s+1}}\Phi^{n,n+s,n+s+1}\right),\\ \mathrm{curl}_{\xi^{n+s+1}}\,\vec{E}^{n+s+1}_{n,n+s,n+s+1}=-\partial_{n+s}\vec{B}^{n+s+1}_{n,n+s,n+s+1}-\frac{1}{\gamma_{n+s+1}}\mathrm{curl}_{\xi^{n+s}}\left\langle\vec{\xi}^{\,n+s+2}\right\rangle_{n,n+s,n+s+1},\end{gathered} \tag{E.50}$$

where expressions (E.51) and (E.52) are taken into account:

$$\begin{gathered}\mathrm{curl}_{\xi^{n+s+1}}\,\partial_{n+s}\vec{A}^{n+s+1}_{n,n+s,n+s+1}=\mathrm{curl}_{\xi^{n+s+1}}\,\partial_{0}\vec{A}^{n+s+1}_{n,n+s,n+s+1}+\mathrm{curl}_{\xi^{n+s+1}}\left(\vec{\xi}^{\,n+s+1},\nabla_{\xi^{n+s}}\right)\vec{A}^{n+s+1}_{n,n+s,n+s+1}=\\=\partial_{0}\,\mathrm{curl}_{\xi^{n+s+1}}\,\vec{A}^{n+s+1}_{n,n+s,n+s+1}+\left[\nabla_{\xi^{n+s}},\vec{A}^{n+s+1}_{n,n+s,n+s+1}\right]+\left(\vec{\xi}^{\,n+s+1},\nabla_{\xi^{n+s}}\right)\left[\nabla_{\xi^{n+s+1}},\vec{A}^{n+s+1}_{n,n+s,n+s+1}\right],\\ \mathrm{curl}_{\xi^{n+s+1}}\,\partial_{n+s}\vec{A}^{n+s+1}_{n,n+s,n+s+1}=\partial_{n+s}\vec{B}^{n+s+1}_{n,n+s,n+s+1}+\mathrm{curl}_{\xi^{n+s}}\,\vec{A}^{n+s+1}_{n,n+s,n+s+1},\end{gathered} \tag{E.51}$$

$$\mathrm{curl}_{\xi^{n+s+1}}\,\nabla_{\xi^{n+s}}\Phi^{n,n+s,n+s+1}=-\mathrm{curl}_{\xi^{n+s}}\,\nabla_{\xi^{n+s+1}}\Phi^{n,n+s,n+s+1}. \tag{E.52}$$

Equation (E.50) under condition (3.27) $\mathrm{curl}_{\xi^{n+s}}\left\langle\vec{\xi}^{\,n+s+2}\right\rangle_{n,n+s,n+s+1}=0$ transforms into equation (3.47).

The direct calculation of $\mathrm{curl}_{\xi^{n}}\,\vec{E}^{n}_{n,n+s,n+s+1}$, $\mathrm{curl}_{\xi^{n}}\,\vec{E}^{n}_{n,n+s,n+s+k}$, $\mathrm{curl}_{\xi^{n+s}}\,\vec{E}^{n+s}_{n,n+s,n+s+k}$ and $\mathrm{curl}_{\xi^{n+s+k}}\,\vec{E}^{n+s+k}_{n,n+s,n+s+k}$ gives equations (3.48), (3.51)-(3.53), respectively.

Let us consider an equation with three sources of dissipations for distribution function $f^{\,n,n+s,n+s+k}$:

$$\mathrm{div}_{\xi^n}\left(\partial_0\vec{D}^n_{n,n+s,n+s+k}+\vec{J}^n_{n,n+s,n+s+k}\right)+\mathrm{div}_{\xi^{n+s}}\left(\partial_0\vec{D}^{n+s}_{n,n+s,n+s+k}+\vec{J}^{n+s}_{n,n+s,n+s+k}\right)+ \\ +\mathrm{div}_{\xi^{n+s+k}}\left(\partial_0\vec{D}^{n+s+k}_{n,n+s,n+s+k}+\vec{J}^{n+s+k}_{n,n+s,n+s+k}\right)=0, \tag{E.53}$$

from here

$$\begin{aligned}
\partial_0\vec{D}^n_{n,n+s,n+s+k}+\vec{J}^n_{n,n+s,n+s+k} &= \mathrm{curl}_{\xi^n}\,\vec{H}^n_{n,n+s,n+s+k}, \\
\partial_0\vec{D}^{n+s}_{n,n+s,n+s+k}+\vec{J}^{n+s}_{n,n+s,n+s+k} &= \mathrm{curl}_{\xi^{n+s}}\,\vec{H}^{n+s}_{n,n+s,n+s+k}, \\
\partial_0\vec{D}^{n+s+k}_{n,n+s,n+s+k}+\vec{J}^{n+s+k}_{n,n+s,n+s+k} &= \mathrm{curl}_{\xi^{n+s+k}}\,\vec{H}^{n+s+k}_{n,n+s,n+s+k},
\end{aligned} \tag{E.54}$$

where representations (3.28) and (3.57) and the independence of the expressions in brackets are taken into account. Theorem 6 is proved.

## Appendix F

***Proof of Theorem 7***

Let us prove the validity of relation (3.60), indeed, using definition (3.57) and the rule for calculating the average kinematical values, we obtain:

$$\int\limits_{(\infty)}\vec{J}^{\ell}_{n_1\dots n_R,n_R+k}d^3\xi^{n_R+k}=\int\limits_{(\infty)}f^{\,n_1\dots n_R,n_R+k}\left\langle\vec{\xi}^{\,\ell+1}\right\rangle_{n_1\dots n_R,n_R+k}d^3\xi^{n_R+k}=f^{\,n_1\dots n_R}\left\langle\vec{\xi}^{\,\ell+1}\right\rangle_{n_1\dots n_R}=\vec{J}^{\ell}_{n_1\dots n_R}. \tag{F.1}$$

Let us consider the properties of fields $\vec{D}^{\ell}_{n_1\dots n_R}$ (3.61) and $\vec{H}^{\ell}_{n_1\dots n_R}$ (3.62) using the example of equations (3.39), (3.40), (3.45), (3.54) and (3.55). Let us integrate each of these equation over the corresponding phase subspace

$$\int\limits_{(\infty)}\left(\partial_0\vec{D}^n_{n,n+k}+\vec{J}^n_{n,n+k}\right)d^3\xi^{n+k}=\partial_0\vec{D}^n_n+\vec{J}^n_n=\mathrm{curl}_{\xi^n}\int\limits_{(\infty)}\vec{H}^n_{n,n+k}d^3\xi^{n+k}=\mathrm{curl}_{\xi^n}\,\vec{H}^n_n, \tag{F.2}$$

where the properties of (3.61) and (3.62) are used. Thus, integrating equation of the second rank (3.39) for functions $\vec{D}^n_{n,n+k}$ and $\vec{H}^n_{n,n+k}$ over subspace $\xi^{n+k}$ gives equation of the first rank (3.34) for functions $\vec{D}^n_n$ and $\vec{H}^n_n$. Similarly for equations of the third rank (3.45), (3.54) and (3.55):

$$\partial_n\int\limits_{(\infty)}\vec{D}^{n+1}_{n,n+1,n+1+k}d^3\xi^{n+1+k}+\int\limits_{(\infty)}\vec{J}^{n+1}_{n,n+1,n+1+k}d^3\xi^{n+1+k}=\partial_n\vec{D}^{n+1}_{n,n+1,n+1+k}+\vec{J}^{n+1}_{n,n+1}= \\ =\mathrm{curl}_{\xi^{n+1}}\int\limits_{(\infty)}\vec{H}^{n+1}_{n,n+1,n+1+k}d^3\xi^{n+1+k}=\mathrm{curl}_{\xi^{n+1}}\,\vec{H}^{n+1}_{n,n+1}, \tag{F.3}$$

$$\int\limits_{(\infty)}\left(\partial_0\vec{D}^n_{n,n+s,n+s+k}+\vec{J}^n_{n,n+s,n+s+k}\right)d^3\xi^{n+s+k}=\partial_0\vec{D}^n_{n,n+s}+\vec{J}^n_{n,n+s}= \\ =\int\limits_{(\infty)}\mathrm{curl}_{\xi^n}\,\vec{H}^n_{n,n+s,n+s+k}d^3\xi^{n+s+k}=\mathrm{curl}_{\xi^n}\int\limits_{(\infty)}\vec{H}^n_{n,n+s,n+s+k}d^3\xi^{n+s+k}=\mathrm{curl}_{\xi^n}\,\vec{H}^n_{n,n+s}, \tag{F.4}$$

$$\partial_0 \int_{(\infty)} \vec{D}^{n+s}_{n,n+s,n+s+k} d^3\xi^{n+s+k} + \int_{(\infty)} \vec{J}^{n+s}_{n,n+s,n+s+k} d^3\xi^{n+s+k} = \partial_0 \vec{D}^{n+s}_{n,n+s} + \vec{J}^{n+s}_{n,n+s} =$$
$$= \mathrm{curl}_{\xi^{n+s}} \int_{(\infty)} \vec{H}^{n+s}_{n,n+s,n+s+k} d^3\xi^{n+s+k} = \mathrm{curl}_{\xi^{n+s}} \vec{H}^{n+s}_{n,n+s}. \quad \text{(F.5)}$$

Equations (F.3), (F.4) and (F.5) coincide with equations of the second rank (3.36), (3.39) and (3.40), respectively.

Property (3.61) for field $\vec{E}^{\ell}_{n_1 \dots n_R}$ follows from the direct integration of equations (3.4), (3.7), (3.10) and (3.11). Let us integrate equation (3.4):

$$\int_{(\infty)} \vec{E}^{n}_{n,n+k} d^3\xi^{n+k} = -\partial_0 \int_{(\infty)} \vec{A}^{n}_{n,n+k} d^3\xi^{n+k} - \frac{2\alpha_n \beta_{n+k}}{\gamma_n} \nabla_{\xi^n} \int_{(\infty)} \left( \tau^{n,n+k}_{n+k} + \mathrm{V}^{n,n+k} \right) d^3\xi^{n+k} =$$
$$= -\partial_0 \vec{A}^{n}_{n} - \frac{2\alpha_n \beta_n}{\gamma_n} \nabla_{\xi^n} \mathrm{V}^n = \vec{E}^{n}_{n}, \quad \text{(F.6)}$$

where the conditions of theorem (3.58) and (3.59) are taken into account:

$$\vec{A}^{n}_{n} = \int_{(\infty)} \vec{A}^{n}_{n,n+k} d^3\xi^{n+k}, \ \mathrm{V}^n = \frac{\beta_{n+k}}{\beta_n} \int_{(\infty)} \left( \tau^{n,n+k}_{n+k} + \mathrm{V}^{n,n+k} \right) d^3\xi^{n+k}. \quad \text{(F.7)}$$

Similarly, for equations (3.7), (3.10) and (3.11), we obtain:

$$\int_{(\infty)} \vec{E}^{n+1}_{n,n+1,n+1+k} d^3\xi^{n+1+k} = -\partial_n \int_{(\infty)} \vec{A}^{n+1}_{n,n+1,n+1+k} d^3\xi^{n+1+k} -$$
$$-\frac{\alpha_{n+1}}{\gamma_{n+1}} \left[ \nabla_{\xi^n} \int_{(\infty)} \Phi^{n,n+1,n+1+k} d^3\xi^{n+1+k} + 2\beta_{n+1+k} \nabla_{\xi^{n+1}} \int_{(\infty)} \left( \tau^{n,n+1,n+1+k}_{n+1+k} + \mathrm{V}^{n,n+1,n+1+k} \right) d^3\xi^{n+1+k} \right] = \quad \text{(F.8)}$$
$$= -\partial_n \vec{A}^{n+1}_{n,n+1} - \frac{\alpha_{n+1}}{\gamma_{n+1}} \left[ \nabla_{\xi^n} \Phi^{n,n+1} + 2\beta_{n+1} \nabla_{\xi^{n+1}} \mathrm{V}^{n,n+1} \right] = \vec{E}^{n+1}_{n,n+1},$$

where, according to the theorem:

$$\vec{A}^{n+1}_{n,n+1} = \int_{(\infty)} \vec{A}^{n+1}_{n,n+1,n+1+k} d^3\xi^{n+1+k}, \ \Phi^{n,n+1} = \int_{(\infty)} \Phi^{n,n+1,n+1+k} d^3\xi^{n+1+k},$$
$$\mathrm{V}^{n,n+1} = \frac{\beta_{n+1+k}}{\beta_{n+1}} \int_{(\infty)} \left( \tau^{n,n+1,n+1+k}_{n+1+k} + \mathrm{V}^{n,n+1,n+1+k} \right) d^3\xi^{n+1+k}. \quad \text{(F.9)}$$

$$\int_{(\infty)} \vec{E}^{n}_{n,n+s,n+s+k} d^3\xi^{n+s+k} = -\partial_0 \int_{(\infty)} \vec{A}^{n}_{n,n+s,n+s+k} d^3\xi^{n+s+k} -$$
$$-\frac{2\alpha_n \beta_{n+s+k}}{\gamma_n} \nabla_{\xi^n} \int_{(\infty)} \left( \tau^{n,n+s,n+s+k}_{n+s+k} + \frac{\beta_{n+s}}{\beta_{n+s+k}} \tau^{n,n+s,n+s+k}_{n+s} + \mathrm{V}^{n,n+s,n+s+k} \right) d^3\xi^{n+s+k} = \quad \text{(F.10)}$$
$$= -\partial_0 \vec{A}^{n}_{n,n+s} - \frac{2\alpha_n \beta_{n+s}}{\gamma_n} \nabla_{\xi^n} \left( \tau^{n,n+s}_{n+s} + \mathrm{V}^{n,n+s} \right) = \vec{E}^{n}_{n,n+s},$$

where, according to conditions (3.58):

$$\vec{A}^{n}_{n,n+s}=\int\limits_{(\infty)}\vec{A}^{n}_{n,n+s,n+s+k}d^3\xi^{n+s+k},\ \mathrm{V}^{n,n+s}=\frac{\beta_{n+s+k}}{\beta_{n+s}}\int\limits_{(\infty)}\left(\tau^{n,n+s,n+s+k}_{n+s+k}+\mathrm{V}^{n,n+s,n+s+k}\right)d^3\xi^{n+s+k},\quad \text{(F.11)}$$

$$\tau^{n,n+s}_{n+s}=\int\limits_{(\infty)}\tau^{n,n+s,n+s+k}_{n+s}d^3\xi^{n+s+k}.$$

$$\begin{aligned}&\int\limits_{(\infty)}\vec{E}^{n+s}_{n,n+s,n+s+k}d^3\xi^{n+s+k}=-\partial_0\int\limits_{(\infty)}\vec{A}^{n+s}_{n,n+s,n+s+k}d^3\xi^{n+s+k}-\\&-\frac{2\alpha_{n+s}\beta_{n+s+k}}{\gamma_{n+s}}\nabla_{\xi^{n+s}}\int\limits_{(\infty)}\left(\tau^{n,n+s,n+s+k}_{n+s+k}+\frac{\beta_n}{\beta_{n+s+k}}\tau^{n,n+s,n+s+k}_{n}+\mathrm{V}^{n,n+s,n+s+k}\right)d^3\xi^{n+s+k}=\\&=-\partial_0\vec{A}^{n+s}_{n,n+s}-\frac{2\alpha_{n+s}\beta_{n+s}}{\gamma_{n+s}}\nabla_{\xi^{n+s}}\left(\frac{\beta_n}{\beta_{n+s}}\tau^{n,n+s}_{n}+\mathrm{V}^{n,n+s}\right)=\vec{E}^{n+s}_{n,n+s},\end{aligned}\quad \text{(F.12)}$$

where

$$\vec{A}^{n+s}_{n,n+s}=\int\limits_{(\infty)}\vec{A}^{n+s}_{n,n+s,n+s+k}d^3\xi^{n+s+k},\mathrm{V}^{n,n+k}=\frac{\beta_{n+s+k}}{\beta_{n+s}}\int\limits_{(\infty)}\left(\tau^{n,n+s,n+s+k}_{n+s+k}+\mathrm{V}^{n,n+s,n+s+k}\right)d^3\xi^{n+s+k},\ \text{(F.13)}$$

$$\tau^{n,n+s}_{n}=\int\limits_{(\infty)}\tau^{n,n+s,n+s+k}_{n}d^3\xi^{n+s+k}.$$

Expressions (F.6), (F.8), (F.10) and (F.12) confirm the validity of property (3.61) for field $\vec{E}^{\ell}_{n_1\ldots n_R}$.

Property (3.62) for field $\vec{B}^{\ell}_{n_1\ldots n_R}$ follows directly from the integration of the field equations (3.38), (3.44), (3.48), (3.51), and (3.52). Indeed, let us integrate equation (3.38), we obtain

$$\int\limits_{(\infty)}\mathrm{curl}_{\xi^n}\vec{E}^{n}_{n,n+k}d^3\xi^{n+k}=\mathrm{curl}_{\xi^n}\int\limits_{(\infty)}\vec{E}^{n}_{n,n+k}d^3\xi^{n+k}=\mathrm{curl}_{\xi^n}\vec{E}^{n}_{n}=-\partial_0\vec{B}^{n}_{n}=-\partial_0\int\limits_{(\infty)}\vec{B}^{n}_{n,n+k}d^3\xi^{n+k},\quad \text{(F.14)}$$

where property (F.6) and equation (3.33) are used. Therefore, based on expression (F.14), we can assume $\vec{B}^{n}_{n}=\int\limits_{(\infty)}\vec{B}^{n}_{n,n+k}d^3\xi^{n+k}$. Similar calculations are valid for the remaining equations (3.44), (3.48), (3.51) and (3.52). The integration of the equations of the third rank (3.44), (3.51) and (3.52) will give the equations of the second rank (3.55), (3.38) and (3.37), respectively. Theorem 7 is proved.